\documentclass[sigconf,table,xcdraw,9pt]{acmart}

\usepackage{graphicx}
\usepackage{amsmath}
\usepackage[]{subfig}
\usepackage{float}
\usepackage{caption}
\usepackage[T1]{fontenc}
\usepackage{array}
\newcolumntype{L}[1]{>{\centering\arraybackslash}m{#1}}
\usepackage{etoolbox}
\usepackage{xspace}
\usepackage{makecell}
\usepackage{lipsum}
\usepackage[ruled]{algorithm2e}

\setcopyright{none} 
\acmYear{2019}

\begin{document}
\title{The Maestro Attack: \\ Orchestrating Malicious Flows with BGP}

\author{Tyler McDaniel}
 \affiliation{%
   \institution{UT Computer Security Lab \\ University of Tennessee, Knoxville}} \email{bmcdan16@utk.edu}

\author{Jared M. Smith}
\affiliation{%
   \institution{UT Computer Security Lab \\ University of Tennessee, Knoxville}}
 \email{jms@vols.utk.edu}

 \author{Max Schuchard}
 \affiliation{%
   \institution{UT Computer Security Lab \\ University of Tennessee, Knoxville}}
 \email{mschucha@utk.edu}

\begin{abstract}
We present the \textbf{Maestro} attack, a novel Link Flooding Attack (LFA) that leverages control-plane traffic engineering techniques to concentrate botnet-sourced Distributed Denial of Service flows on transit links. Executed from a compromised or malicious Autonomous System (AS), Maestro advertises specific-prefix routes poisoned for selected ASes to collapse inbound traffic paths onto a single target link. A greedy heuristic fed by publicly available AS relationship data iteratively builds the set of ASes to poison. Given a compromised BGP speaker with advantageous positioning relative to the target link in the Internet topology, an adversary can expect to enhance total flow density by \textit{more than 30\%}. For a large botnet (e.g., Mirai), that translates to augmenting a DDoS by more than a million additional infected hosts. Interestingly, the size of the adversary-controlled AS plays little role in this amplification effect. Devastating attacks on core links can be executed by small, resource-limited ASes. To understand the scope of the attack, we evaluate widespread Internet link vulnerability across several metrics, including BGP betweenness and botnet flow density. We then assess where an adversary must be positioned to execute the attack most successfully. Finally, we present effective mitigations for network operators seeking to insulate themselves from this attack.
\end{abstract}



\keywords{Routing Security; DDoS; BGP Poisoning; Adversarial Autonomous Systems (ASes)}

\maketitle

\section{Introduction}{\label{intro}}
Distributed denial of service (DDoS) attacks direct traffic from many distinct sources on the Internet to overwhelm the capacity of links or end hosts. These attacks proliferate despite extensive academic and economic investment in mitigation, and intensify with the Internet's expansion to new devices and services. A reflection attack fueled by unprotected memcached servers, for example, temporarily disabled Github~\cite{majkowski2018memcrashed}. \textit{Link Flooding Attacks} or \textit{LFAs} are DDoS attacks on infrastructure links~\cite{kang2013crossfire, studer2009coremelt}. LFAs may have moved from proposed attack to present threat: in 2016, a Mirai botnet-sourced attack directed over 500 Gbps of traffic to a Liberian infrastructure provider in what may have been a real-world LFA~\cite{scott2016rise}.

A botnet can only launch an LFA if there exists some set of destinations such that bot traffic addressed to those hosts traverses the target link.  As a result, the reach of an LFA attacker is constrained by Internet routing choices. It turns out that these route selection constraints protect the majority of Internet links from large volumes of botnet traffic. Even highly trafficked links in the Internet's dense core, the most plausible targets for an LFA~\cite{studer2009coremelt,kang2013crossfire}, are not universally exposed to bot traffic.

This LFA limitation results from bots' inability to choose the links their traffic transits to a given destination. That choice rests with the Autonomous Systems, or ASes, the bots reside within. In this work, we examine how an adversary's ability to influence routing decisions - i.e., access to a compromised Boarder Gateway Protocol, or BGP speaker - can shape remote networks' path selection process to their advantage. Our novel attack, Maestro, orchestrates path selection of remote ASes and bot traffic destinations to steer malicious flows onto links otherwise unreachable by the botnet. Our attack requires an adversary to have two tools: an edge router in some compromised AS and a botnet. 

The Maestro attack utilizes a BGP traffic engineering technique called BGP poisoning to adjust the path from remote networks to the compromised BGP speaker's network. By adjusting these inbound paths to flow over the targeted link, the adversary gains a destination for bot traffic that traverses the target. The attacker then launches a traditional DDoS attack against the compromised BGP speaker's network, resulting in attack flows that congest the otherwise unreachable link. Maestro relies on a greedy algorithm to efficiently compute which ASes to poison given a botnet distribution, targeted link, and compromised BGP speaker location in order to force botnet traffic over the victim link. Our algorithm on average maneuvers 80\% of the bots in a botnet over links vulnerable to the Maestro attack with just five poisons.

We evaluate both the attacker's need for the Maestro attack, as well as its effectiveness. Our simulations join botnet models derived from real-world bot distributions with CAIDA's Internet topology data to quantify the vulnerability of Internet links to LFAs. The goal is to measure how Internet routing characteristics limit Link Flooding Attacks. These initial experiments will motivate our attack by illustrating the prevalence of likely LFA targets that are unreachable by real world botnets.  For one of our link sample sets, we found that between 18\% to 23\% of links are traversable by the majority of hosts in three major botnets. Fewer than 10\% of sampled links were vulnerable to 75\% or more bots. 

In the same simulation framework, we demonstrate the Maestro attack's ability to both amplify Link Flooding Attacks for already-vulnerable links, and to extend a botmaster's reach to previously unexposed targets. After executing the Maestro attack from a well-positioned adversary, more than 90\% of the previous set of links are exposed to a majority of bots across each botnets, and 85\% to 87\% of links are exposed to 75\% of bots.  Our analysis explores a number of different target link/compromised BGP speaker selection methods in an effort to explore how these properties factor into attack success. Additionally, we explore how a compromised AS can "leak" valley paths to expose Tier 1 peering links to Link Flooding Attacks.

We also consider defenses to mitigate the Maestro attack. The relative effectiveness of potential defenses are explored via simulation, with in-depth discussion of results. Our goal is to give a first look into mitigation techniques network operators can individually deploy to protect their links from the Maestro attack without global coordination. Feedback from outreach to the network operator community is also be presented.

We make the following key contributions throughout the rest of the paper:
\begin{itemize}
    \item We measure how Internet routing properties limit Link Flooding Attacks in Section~\ref{motivation}.
    \item We develop a technique to overcome these limitations: the Maestro attack. We explore the setup and execution of this attack in Section~\ref{attack}.
    \item We evaluate our attack, Maestro, via realistic simulations on an up-to-date Internet model in Section~\ref{evaluation}. There we will also present and evaluate a valley path leak attack.
    \item We develop an understanding of the Internet's large-scale vulnerability to Maestro in Section~\ref{formalization}.
    \item We propose and evaluate mitigations to our attack, as well as seek operator feedback from mailing lists in Section~\ref{mitigation}.
\end{itemize}
\section{Background}{\label{background}}
\subsection{Border Gateway Protocol}{\label{bgp}}

The \textit{Border Gateway Protocol} (BGP)~\cite{Rekhter:1995:BGP:RFC1771} is the de facto routing protocol of the Internet. BGP enables over 60,000 Autonomous Systems (ASes)~\cite{as-rfc-source} to exchange routing information and connect disparate parts of the Internet's infrastructure. Routes in BGP are defined by a destination IP prefix and a collection of attributes, including the AS PATH or AS-level hops to reach the destination. ASes originate routes to hosted IP prefixes via BGP advertisements to neighboring ASes. An AS's routers store received paths and make decisions about which paths to use for each destination prefix. Each AS chooses paths per prefix based on attributes of stored paths, most notably AS PATH length and LOCAL PREF. LOCAL PREF represents the AS operator's local policy choices regarding path qualities. LOCAL PREF holds precedence over AS PATH length in the decision process. Of all available paths, the \textit{longest prefix matching} rule dictates that the stored path with the longest (most specific) IP prefix match is used to forward packets when received. 

Because the BGP decision process draws on path and policy attributes in route selection, BGP is a path-vector algorithm \textit{with policies}. These policies often manifest themselves as a result of the unique \textit{business relationships} on the Internet. ASes can have peers, customers, and providers. Peers exchange traffic for free, customers pay to exchange traffic through a provider, and providers gain economic incentives for traffic exchange provided to customers. Due to this economic aspect, the Internet topology is shaped by behaviors dictated by the valley-free routing model~\cite{gao2001inferring} and shown in Figure~\ref{fig:bgp_rel} in the appendix. In simple terms, the model states that BGP routes will not transit from a customer to a provider after transiting from a provider to a customer, which ensures ASes do not incur monetary costs whenever possible.


\begin{figure}
	\centering
	\includegraphics[width=1.0\columnwidth]{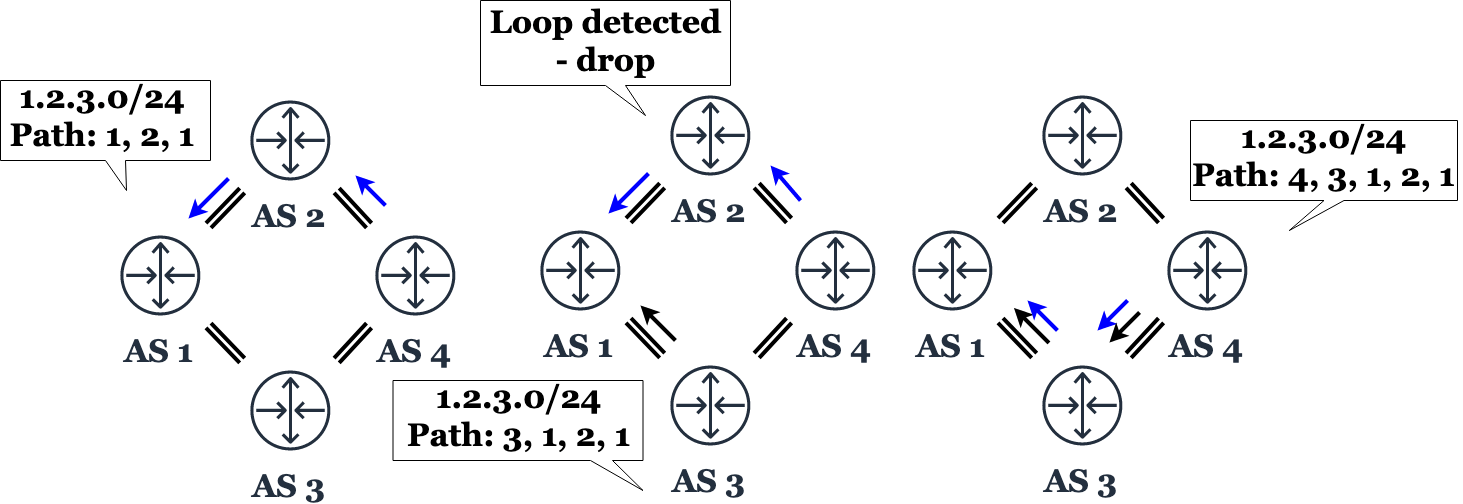}
	\caption{BGP poisoning. AS 1 advertises a specific prefix (thicker arrow). AS 4's traffic to AS 1 (blue) is moved to the more specific route. AS 2 is said to have been \textit{poisoned}.}
	{\label{fig:bgp_poison}}
	\vspace{-10pt}
\end{figure}

\subsection{BGP Poisoning}{\label{poisoning}}

The BGP decision process gives local operators control over outbound paths. Unfortunately, operators have relatively little influence on inbound traffic paths. Techniques do exist for next-hop inbound path control, including the MULTI EXIT DISC (exit discriminator) attribute~\cite{bgp-med} and BGP communities~\cite{bgp-comm}, but both are subject to the source AS's policies. This means inbound path control cannot be exerted by a destination AS arbitrarily on the broader Internet. Fortunately, \textit{BGP poisoning}, a traffic engineering technique growing in use in academic and operator communities~\cite{smith2018routing,smith2019internet,tran2019feasibility,katz2012lifeguard,Anwar:2015tk,RAD}, allows for the manipulation of an AS's inbound traffic routes \textit{without coordination} from other ASes.

BGP poisoning relies on two characteristics of BGP: loop detection and longest-prefix matching. Longest-prefix matching was discussed earlier, but loop detection is a specified BGP behavior where an AS will drop paths which already contain its own AS number (ASN). This prevents loops, but it also allows BGP poisoning. An illustration of BGP poisoning is shown in Fig.~\ref{fig:bgp_poison}. The advertising or poisoning AS advertises a more specific (longer) prefix for the traffic it wishes to move. Longest prefix matching means that ASes directing traffic to included IPs will switch on to the new route (see AS 2). However,
some set of ASes are included in the AS PATH for the advertisement, sandwitched between copies of the originator's ASN. Because they are notionally "on" the AS PATH, these ASes are \textit{poisoned}; that is, they will detect a loop and drop the advertisement. While these
poisoned ASes still have connectivity to the advertising AS's other prefixes, their traffic flows are unchanged by the advertisement. The poisoning AS has adjusted inbound traffic paths without reliance on remote AS policies. Notably, poisoning functions on a \textit{per-prefix} basis.


\subsection{Distributed Denial of Service}{\label{lfas}}

\textit{Distributed Denial of Service} (DDoS) describes a coordinated attack on a target link or end host using traffic from multiple sources. Often the traffic sources for these attacks are botnets, or networks of compromised end hosts (bots) under an attacker's control. Botnets can include PCs, IoT devices, and/or SCADA systems, and are freely available for rent as attack sources from illicit online marketplaces. The most common form of DDoS is volumetric DDoS, where an attacker uses the sheer magnitude of malicious traffic flows to overwhelm a target. These attacks have been carried out by nation-states~\cite{case2016analysis, herzog2011revisiting}, and can also be used to isolate or degrade Internet performance for large geographic regions~\cite{DynDDoS} by targeting core DNS or network infrastructure providers.

A more recent class of DDoS attacks, \textit{Link Flooding Attacks} (LFAs), targets network infrastructure rather than end hosts. One of the first such attacks in the literature is Coremelt~\cite{studer2009coremelt}. Coremelt specifies that a botmaster 1) map which links are present on paths on routes between bots, 2) target a specific link used on paths between many bots, and 3) direct bot traffic \textit{between bots} over the link. The resulting $n^2$ flows (for $n$ bots with paths over the link) overwhelms benign traffic on the target link. The bot traffic is especially difficult to classify/filter as it is ``wanted" by the destination host and therefore appears legitimate. The Crossfire attack, like Coremelt, targets Internet links, but has the more ambitious goal of isolating an entire region (military installation, university, geographic region, etc.) by targeting key links~\cite{kang2013crossfire}. Rather than directing traffic to one another, bots map paths to publicly available web services (decoys) that transit target links. In the Crossfire evaluation, the decoys were PlanetLab servers. Bots then use sustained, low-intensity flows to the decoys to execute the attack, a pattern that makes Crossfire extremely difficult to detect and counter.

\subsection{Adversarial BGP}{\label{adv-bgp}}
Beyond DDoS, adversaries continue to exploit long-known vulnerabilities in the Internetâs routing architecture in increasingly sophisticated control-plane attacks. In 2014, researchers discovered a Canadian ISP surreptitiously hijacking bitcoin mining related traffic to steal victim miners' computational work, netting over \$80,000~\cite{hijack}. Academic research has explored this space with recent papers on hijacking bitcoin traffic and the Tor privacy system with BGP~\cite{Apostolaki:2017ud,Sun:2015tp}. On a larger scale, fraudulent networks designed to deceive advertisers into paying for automated ad views have raked in multi-million dollar hauls~\cite{ops2016methbot}. One such operation, 3ve, persisted for several years, even registering their own ASes, and earned nearly \$30 million~\cite{ops20183ve}. These examples serve to demonstrate that attackers are capable and willing of leveraging the control-plane to accomplish their goals.

\subsection{Key Terminology}{\label{terms}}
\noindent Before we proceed, we now summarize key terminology in the paper, which we will cover in more detail in the next section:\\

\noindent \textbf{Poisoned/poisoning advertisement/route/path:} A route originated with selected ASes included between copies of the originator's ASN. This technique will be exploited to prevent route installation/propagation by the selected ASes.

\noindent \textbf{Adversary/Compromised AS:} A compromised AS used to issue poison advertisements for a prefix. Botnet flows are then directed to the advertised prefix. The path of these flows is manipulated via BGP poisoning to steer them onto a target link. 

\noindent \textbf{Poisoned AS:} An AS included in a poisoned advertisement. Loop detection prevents these ASes from installing the route.

\noindent \textbf{Link Flooding Attack (LFA):} DDoS targeting a link in the Internet topology rather than an end host.

\noindent \textbf{Target link:} The link targeted by the adversary AS executing Maestro.

\noindent \textbf{From AS/To AS:} The endpoint ASes of the target link. The \textit{From AS} is the target link source; the \textit{To AS} is the target link destination.

\noindent \textbf{Flow density:} Our primary attack success metric. \textit{Bot-to-bot} flow density is the percentage of infected hosts in a botnet with paths to another infected host that transit the target link; \textit{(bot-to-any} flow density is the percentage of infected hosts with a path to any destination that transits the target link.

\noindent \textbf{Link betweenness:} The number of times a link appears on best paths between ASes.

\noindent \textbf{Link relationship:} The economic relationship between link endpoints. A \textit{provider to customer} is a link from a provider to one of its customers; a \textit{customer to provider} link is the same link in the reverse direction. A \textit{peer/peering} link connects peer ASes.

\noindent \textbf{Customer cone:} The set of all ASes reachable from an AS via only customer links. That is, direct/indirect customers of an AS.

\noindent \textbf{Adversarial region:} A measure of the adversary AS's positioning relative to the target link. The \textit{customer} region includes the customer cone of the To AS. The \textit{peer} region is all ASes reached from a To AS peer, while the \textit{provider} region includes any AS reached from a provider of the To AS.

\noindent \textbf{Attack path:} A valley-free path from some malicious flow source (e.g., an AS containing infected hosts) to the adversary AS that transits the target link.
\section{Measuring Botnet Reach}{\label{motivation}}
\begin{figure*}
    \centering
    \subfloat[AS-level edge betweenness of Internet links based on the CAIDA topology]{\label{fig:cdf_between}\includegraphics[width=0.29\textwidth]{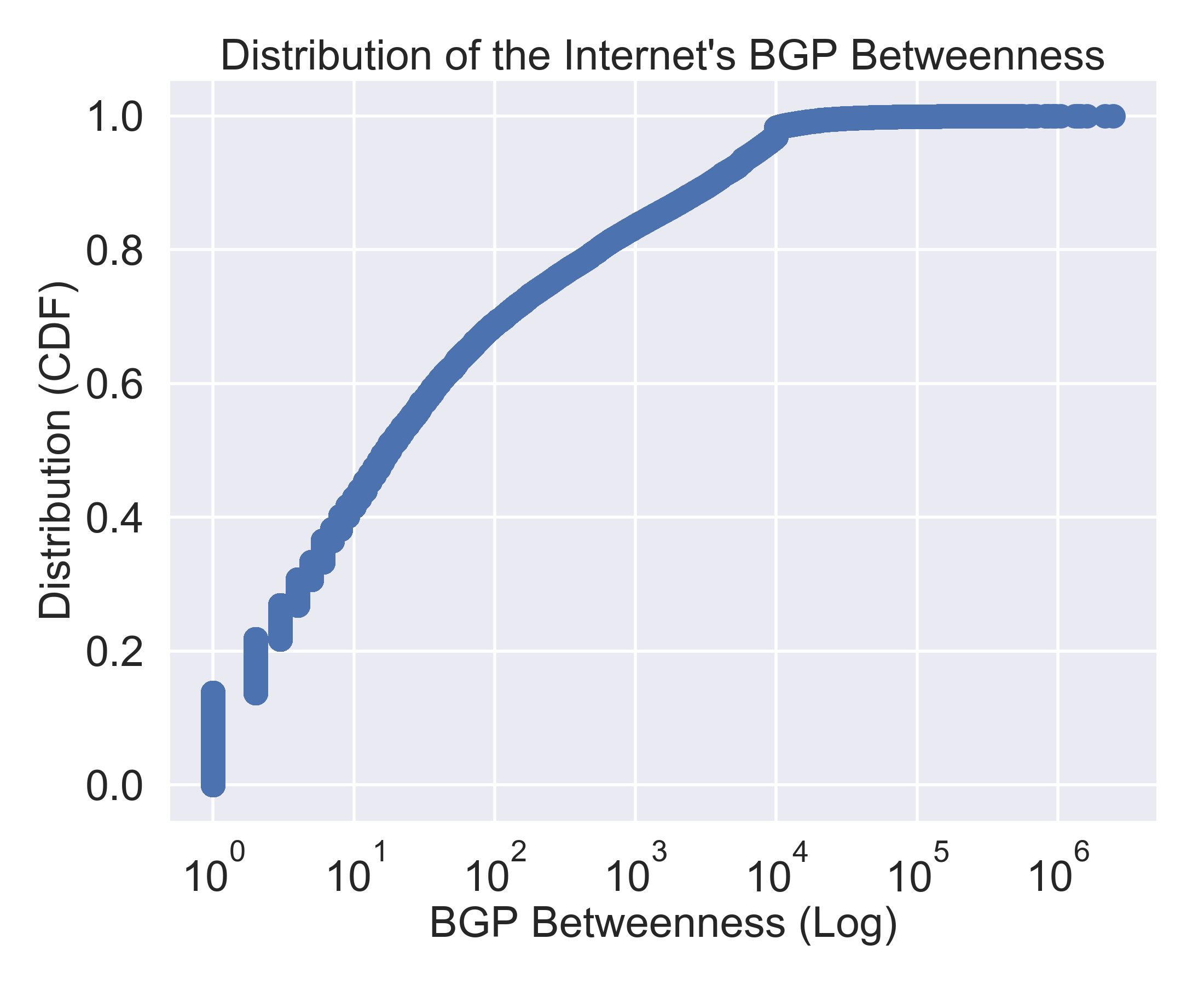}}
    \subfloat[Flow density (bot-to-bot) by betweenness]{\label{fig:bottobot_density}\includegraphics[width=0.29\textwidth]{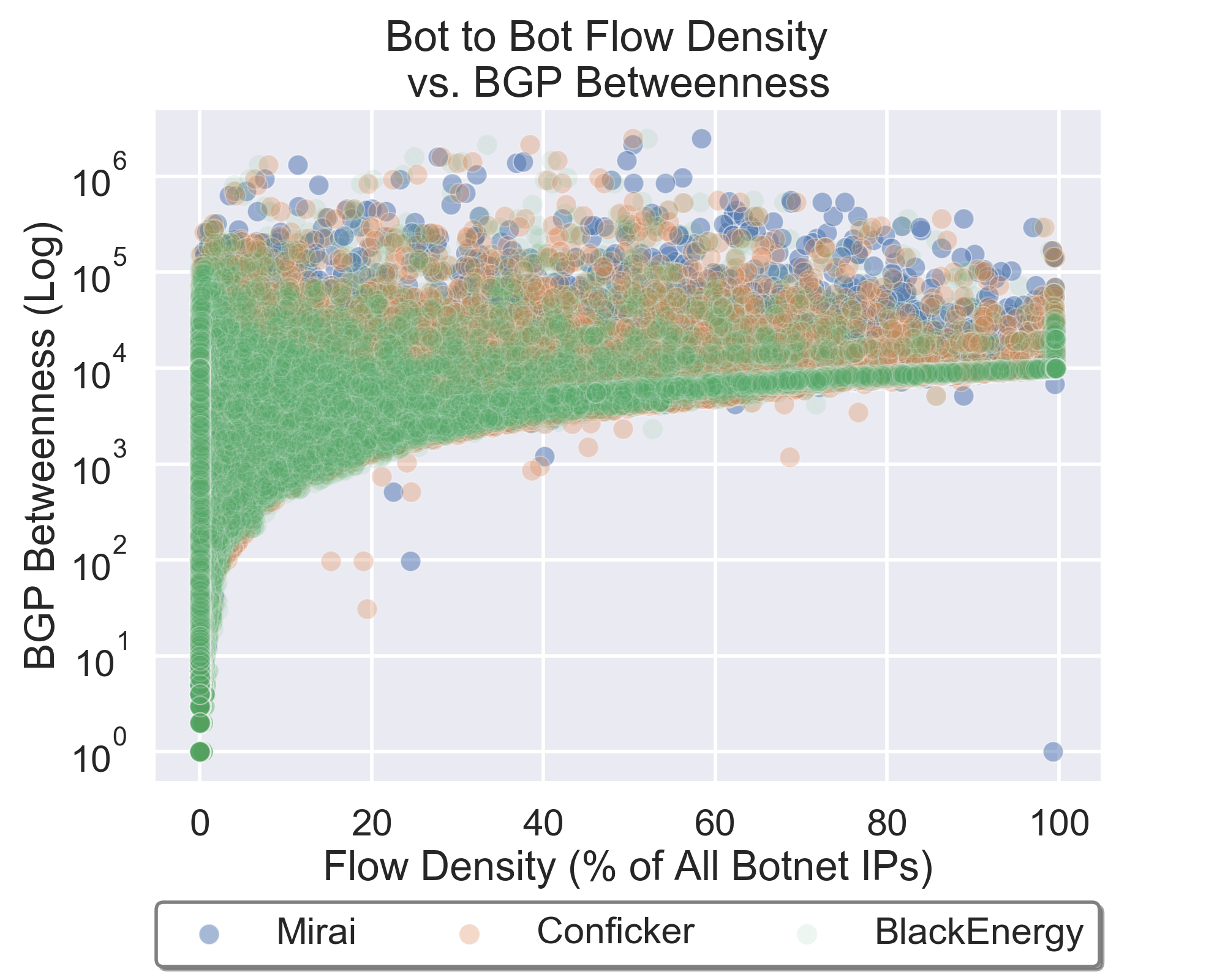}}
    \subfloat[Extended flow density (bot-to-any) by betweenness]{\label{fig:bottoany_density}\includegraphics[width=0.29\textwidth]{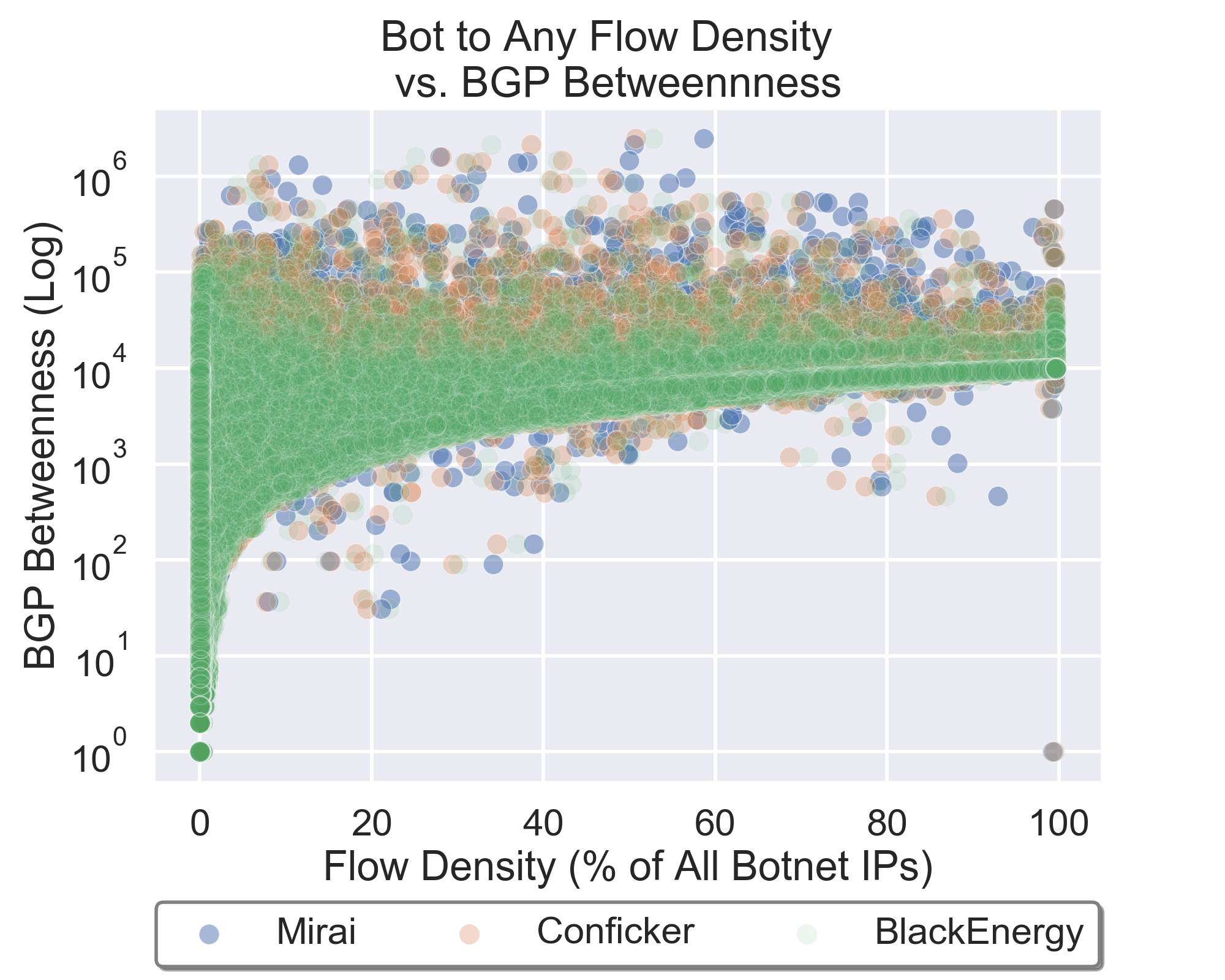}}
    \vspace{-10pt}
    \caption{Exploring Internet link betweenness and LFA vulnerability}
    \label{fig:bet-vs-density}
    \vspace{-10pt}
\end{figure*}

To execute a successful Link Flooding Attack (see Section~\ref{lfas} earlier) on an Internet link, a botnet must be able to drive traffic over it. To accomplish this, the botmaster must find paths between a sufficient number of bots and other destinations that cross the target link. This allows the traffic to be aggregated at the target, which is degraded by overwhelming flows and fails to transit benign traffic. This failure realizes the initial goal of the DDoS. 

However, prior work on link-flooding DDoS attacks often 1) do not perform their measurements with distribution data from a real botnet, notably Kang et al.'s Crossfire~\cite{kang2013crossfire} (recall their approach uses PlanetLab servers which are traffic and location-limited as research platforms), 2) assume botnets can direct significant flows over arbitrary links on the Internet, including Tran et al.'s examination of the feasibility of LFA defense mechanisms using re-routing~\cite{tran2019feasibility}, or 3) choose specific links based on botnet flows, including Coremelt~\cite{studer2009coremelt, schuchard2010losing}. With these limitations in mind, we set out to examine the limitations of LFAs based on the topological positioning of available botnets. We seek to illustrate the critical nature of select core Internet links by examining their relative usage.

For this purpose, we classify and examine links by \textit{betweenness}, defined as the number of times a link appears on the currently-used (best) path between any pair of ASes. High betweenness indicates that a link carries traffic between many ASes; a low betweenness link serves relatively few ASes. Fig.~\ref{fig:cdf_between} shows the cumulative distribution of Internet links on the y-axis by betweenness, based on CAIDA's AS relationship inference~\cite{CAIDA}. This figure shows that clearly not all links have the same importance. The majority of links appear on 10 or fewer paths, but select links have a betweenness of more than \textit{1 million}. These high betweenness links connect more than 1,000 AS source/destination pairs. While we do not expect the low betweenness links to have high amounts of bot-to-bot paths over them, we do expect high amounts of viable paths for the high betweenness links. Accordingly, attacks on these critical links would wreak havoc with upstream and downstream networks, threatening entire regions.

Next, we set out to measure the reach of botnets using real botnet models, built from IP mappings collected for three of the largest botnet families: Conficker, Mirai, and BlackEnergy. These models allow us to measure the relative exposure of Internet links to attacks based on real-world botnet distribution data. We then quantify link vulnerability via \textit{flow density}: the percentage of a botnet's infected hosts with paths over the target link to either 1) another bot, called bot-to-bot flow density, or 2) any destination, called bot-to-any flow density. These two metrics were both covered in the prior section on LFAs. The bot-to-bot flow density models the effectiveness of the Coremelt attack, and the bot-to-any flow density models the Crossfire attack. Fig.~\ref{fig:bottobot_density} depicts the results of our second experiment, measuring bot-to-bot flow density as a function of betweenness for Internet links. Note that some low betweenness (peripheral) links are, not unexpectedly, wholly outside an LFA attacker's reach. \textbf{Critically, some moderate to high betweenness (core) links are also partially or completely devoid of paths between bots}. We note that relaxing our attack technique by allowing bots to send traffic to any AS destination does not significantly alleviate these limitations, as shown in Fig.~\ref{fig:bottoany_density}. This means that both Coremelt and Crossfire as measured previously \textit{make false assumptions} based on the real distributions of botnets available to those systems. 

Given that these highly trafficked links within the Internet's core could be both high-value targets to an adversary and outside the reach of existing LFAs, it is intuitive to investigate whether a routing-capable adversary can manipulate the control-plane to expose them. Next, we will introduce the Maestro attack, a novel combination of traffic engineering techniques with LFAs that 1) increases the flow density a botmaster can drive onto target links, and 2) enables an adversary to steer flows onto previously unreachable targets. This combination is the first known attack to combine inbound traffic influence mechanisms such as BGP poisoning \textit{and} powerful botnet-based DDoS attacks from both research and practice.
\section{The Maestro Attack}{\label{attack}}
We have demonstrated that most links, including many likely LFA targets, are not vulnerable to the full force of a botnet. This condition arises from the lack of end host control over traffic routes; bots cannot always find a destination for their traffic that crosses a target link. The Maestro attack, introduced in this section, is designed to alter the control-plane to expose these links. In this section, we present the threat model for a Maestro attacker, followed by a high-level description of the attack and implementation details.

\subsection{Threat Model}{\label{threat-model}}
To execute the attack, an adversary requires 1) command of a botnet and 2) control of a BGP speaker, i.e., an AS's edge router. The first item is trivially obtainable, as botmasters routinely monetize their networks by renting them out in an attack-as-a-service model on the dark web~\cite{putman2018business}. Recent events demonstrate that multiple feasible avenues exist for adversaries to gain routing capability. The 3ve fraud operation~\cite{ops20183ve} demonstrated the most straightforward route - simply registering a new AS. Network operators could also be compromised by an insider, as may have occurred in the Canadian bitcoin hijack~\cite{hijack}. Recent Cisco router zero-days demonstrate the ongoing possibility of remote attackers, as well~\cite{cisco_exploit}. Finally, BGP has previously been weaponized for intelligence gathering~\cite{demchak2018china} and censorship~\cite{dainotti2011analysis} by nation states. While powerful nation-state adversaries have many other tools, they certainly have the leverage to execute Maestro.

Of course, the degree to which a compromised AS can exert control for the attack may vary with its topological position relative to the target. We will address this question in Section~\ref{evaluation}. Note that a Maestro attacker does not need to control every end host in the network - they only need to issue advertisements from an edge router.

\subsection{Maestro Concept}{\label{attack-overview}}

\begin{figure}[!ht]
    \centering
    \includegraphics[width=0.60\columnwidth]{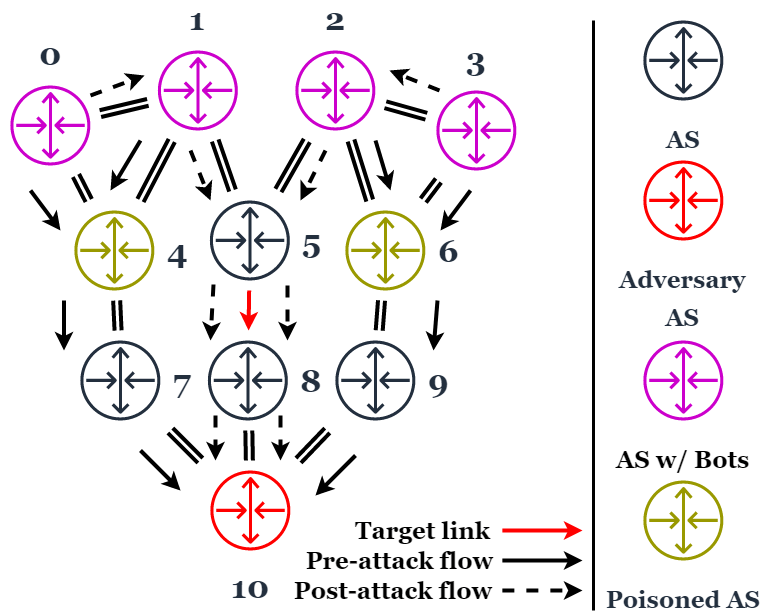}
    \caption{Demonstration of the Maestro Attack: utilizing BGP Poisoning to collapse botnet traffic onto a single link.}
    \label{fig:maestro-concept}
    \vspace{-8pt}
\end{figure}


Bots are located in disparate ASes; to dictate \textit{outbound} bot traffic paths would require that an adversary control every network. Such an adversary is far more powerful the one in our threat model. Maestro's central insight is that a routing-capable adversary can issue poisoned advertisements to alter \textit{inbound} paths to themselves. If an adversary first directs bot traffic to an AS/prefix they control (the \textit{compromised AS} or \textit{adversary AS}), the adversary can orchestrate those flows onto a target link using poisoned BGP advertisements (like a conductor, or \textit{maestro}). We call the origin endpoint of the target link the \textit{From AS} and destination endpoint of the target link the \textit{To AS}. Note that the adversary cannot directly influence route selection in the ASes hosting bots; rather, BGP poisoning essentially \textit{bypasses} route selection by presenting a more specific prefix to those ASes.

In effect, this also executes a traditional DDoS against the adversary AS. This is of little concern to an adversary who compromises an AS they do not own, like those in Section~\ref{adv-bgp}. Fig.~\ref{fig:maestro-concept} shows the attack in abstract, with link $5 \mapsto 6$ as the target. Before the attack, traffic from bot-infected ASes (ASes 0-3) to the adversary (AS 10) flows around (and not over) the target link. AS 10 first issues specific prefix advertisements with ASes 4 and 6 poisoned. This causes inbound flows from the bot-infected ASes to the adversary to concentrate over the target link. After altering these paths, the adversary AS (AS 10) directs bot traffic to itself. The result is a channeled DDoS flowing over $5 \mapsto 6$.

\subsection{Poison Selection Algorithm}{\label{poison-selection}}
The attacker's core utility is an algorithm we developed to determine which ASes to poison to maximize inbound bot traffic over the target link. We call this set of ASes the \textit{poison set}. These ASes will be sandwiched between the compromised ASN in the poisoned advertisement (see Section~\ref{poisoning} on BGP poisoning). Finding a poison set that successfully steers bot traffic is a non-trivial task, because poison sets can conflict; that is, the poisons required to steer one bot-containing AS (or \textit{source AS}) onto the target link will disconnect from the advertised prefix both the poisoned ASes and all ASes requiring an AS from the poison set to reach the adversary AS. Additionally, each poison increments the poisoning advertisement's AS PATH by 1. Because excessively long paths are often filtered, the number of ASes we can poison in practice is limited.

We developed a MAX SAT formulation of the optimal poison choice problem, but its high runtime complexity and failure to exploit the specific structure of our problem led to the heuristic explained below. For completeness, the optimal choice formulation is included in the appendix, Section~\ref{npcomplete-formalization}. 

\subsubsection{Iterative Poison Choice Heuristic}{\label{poison-heuristic}}
We begin from the observation that the adversary wants to selectively poison ASes on source AS paths to the adversary that \textit{do not} cross the target link in an attempt to force source ASes to switch onto paths that \textit{do} contain the target link. Intuitively, the adversary wants to form a bottleneck to the poisoning prefix over the target link. 

Our poison choice heuristic represents the core of the Maestro attack. Once it is used to determine the poison set, the adversary only has to issue the poisoned advertisement and direct bot traffic to the poisoned prefix. We will build this poison set iteratively. At each iteration, we begin by partitioning ASes into four sets:

1) \textit{Sacred} ASes. This set is initialized with the From AS, the To AS, and the adversary AS. It will be updated at each iteration with every AS that appears on all paths from the To AS to the adversary AS. Naturally, we must have a path for traffic from the target link to the poisoning prefix, so these ASes should never be poisoned.

2) \textit{Disconnected} ASes, which includes those poisoned and those without a route to the advertising prefix that does not transit a poisoned AS.

3) \textit{Successful} ASes are those already transiting the target link to the adversary.

4) \textit{Source} ASes who are not yet sacred, disconnected, or successful. 

After these sets are updated, we select an AS to poison from the source ASes and add it to the poison set. Finally, we simulate the remaining source AS's path changes in response to the new poison set, and move to the next iteration. 

We will terminate iteration when no ASes remain in the source set. An additional termination condition is reached if the poison set (which is included in the AS PATH as described in Section~\ref{poisoning}) causes the AS PATH to exceed the size AS operators will almost certainly filter in practice: around 254 hops~\cite{tran2019feasibility, cisco}. We will show in Section~\ref{mitigation} that this condition is rarely required.

At every iteration, the adversary must select one of the source ASes to poison. To accomplish this, we select the AS with the highest vertex betweenness on the set of paths from remaining source ASes to the adversary. Intuitively, this is the poison that invalidates the maximum number of source paths not containing the target link. While no guarantee exists that source ASes will select a path that \textit{contains} the target link, we at least remove a common hop used to \textit{avoid} the link. Other techniques could weight the scoring based on source AS CAIDA rank, customer cone size, or any number of other factors. The entire algorithm, including this poison scoring technique, is shown in the algorithm block below. A full example of poison scoring is shown in the appendix, Section~\ref{example-subsection}.

\renewcommand{\thealgocf}{} 
{\footnotesize 
\begin{algorithm}
    \SetKwInOut{Input}{Input}
    \SetKwInOut{Output}{Output}

    \underline{function ChoosePoisons} $(From,To,Adv,Sources,n)$\:\\
    \Input{From AS $From$, To AS $To$, adversary AS $Adv$, source ASes $Sources$, poison limit $n$}
    \Output{poison set $Poisons$}
    $Poisons = \emptyset $\\
    \While{$ Sources \neq \emptyset \, \text{and}\, |Poisons| < n$} {
    $B = \{ b \mid b \text{ is a bgp path}\, To \mapsto Adv \}$\\
    $Sacred = \{From, To, Adv\} + \bigcup_{i=1}^{|B|} B_{i}$\\
    $Success = \{ s \in Sources \mid \{From, To\} \in s \mapsto Adv$\}\\
    $Disconn = \{s \in Sources \mid \nexists \text{ a specific-prefix path}\,s \mapsto Adv \}$\\
    $Sources \mathrel{-}= Sacred \cup Success \cup Disconn$\\
    $Score = [0]*|Sources|$\\
    \ForEach{$s_{i} \in Sources$} {
       \ForEach{$s_{j} \in s_{i} \mapsto Adv$}{
         $Score_{j} \mathrel{+}=1$
       }
    }
    $Poisons \mathrel{+}= max(Score)$\\
    $Adv$ \text{sends advertisement to poison $Poisons$}
    }
    \caption{Poison Choice Heuristic}
\end{algorithm}{\label{poison-algo}}
}

In the following section, we discuss our experimental setup and present the results of thousands of simulated Maestro attacks. Additionally, we describe a related valley path \textit{leak} attack that can be used to expose Tier 1 to Tier 1 peer links to devastating attack flows.

\section{Evaluation}{\label{evaluation}}
\subsection{Simulation Methodology}{\label{simulator}}
Realistic, active measurement of an attack like Maestro on the Internet would present serious ethical challenges in both control- and data-plane disruption. Instead, we evaluate Maestro by extending the Chaos BGP simulator used in previous related work~\cite{tran2019feasibility,smith2018routing,RAD,schuchard2010losing}. This Internet-scale simulator builds a BGP topology based on publicly available, state-of-the-art inferred AS relationship data from CAIDA (February 2019 dataset)~\cite{CAIDA}. In the simulator, ASes perform a simplified BGP decision process for path selection that includes longest-prefix matching, shortest AS PATH, and abbreviated local policy. As true local AS policies are private, this is the most accurate simulation of AS behavior we can devise. Recent work from Smith et al.~\cite{smith2019internet} suggests that short poisoned paths are rarely filtered by ASes, and we will demonstrate that Maestro does not require a long list of poisons. We model our simulator's poison mechanics based on the live Internet's treatment of BGP poisoning~\cite{smith2019internet} and other work that employs poisoning~\cite{katz2012lifeguard,Anwar:2015tk}.

For each attack, we use three botnet models based on Mirai, Blackenergy, and Conficker botnet IP measurements. With these models, we can measure pre-attack \textit{flow density} for a target, which represents the present vulnerability of the link to LFAs. Next, we execute the Maestro attack using the technique detailed in the previous section in an attempt to bring additional bot traffic to bear on the target. Finally, we measure post-attack flow density to determine how well we steered bot-containing ASes onto the target link. For most experiments, we make bot-to-bot (Coremelt-style) flow density measurements; when using bot-to-any target link sample set, we will instead measure bot-to-any flow density.

Our botnet models are built from passive and active measurements of infected hosts from a variety of sources. The \textit{Mirai} botnet model includes 2.29 million IP addresses in 11,633 ASes. These addresses were recorded by a Chinese CDN as they attempted to spread the malware, a process with a unique signature~\cite{Netlab360}. Our \textit{Conficker} model is composed of 2.28 million bots from 12,095 ASes. The Conficker model is based on prior work that presented a method for detecting rendezvous points for infected hosts and monitoring bot traffic to these points~\cite{Thomas:2014:KDD:2567948.2579359}. The \textit{Blackenergy} model is a SCADA-focused botnet developed from similar techniques as presented in~\cite{chang2015measuring}. The Blackenergy botnet has a total of 310,943 bots and 4,291 ASes. We present the AS-level distribution of these botnets' IPs in the appendix, Fig.~\ref{fig:botnet-dist}. All three are relatively clustered topologically. Future work could explore Maestro's affect on differently distributed flow sources, e.g. other botnets or reflection assets.

To evaluate the effectiveness of our attack as presented in Section~\ref{attack}, we choose thousands of target link/adversary AS pairings for simulated attacks. We aim to understand link vulnerability characteristics, and show how the topological position of target/adversary affect flow density. The following sections describe our link sample sets, and how adversaries are selected.

\subsection{Link Selection}{\label{link-selection}}
For the Maestro attack, we only consider customer to provider and provider to customer links. We present a related attack for peer link targets later in this section. Our attacks are conducted on the following target sample sets:

\textbf{Random:} Our first and most straightforward link sample set is 2000 links selected uniformly at random from all links in our inferred topology. The only conditions on these links were that they were 1) not last-mile links, as these can be targeted by a traditional DDoS, and 2) not peer links.

\begin{figure}[!ht]
    \begin{minipage}[t]{0.75\columnwidth}
        \includegraphics[width=\columnwidth]{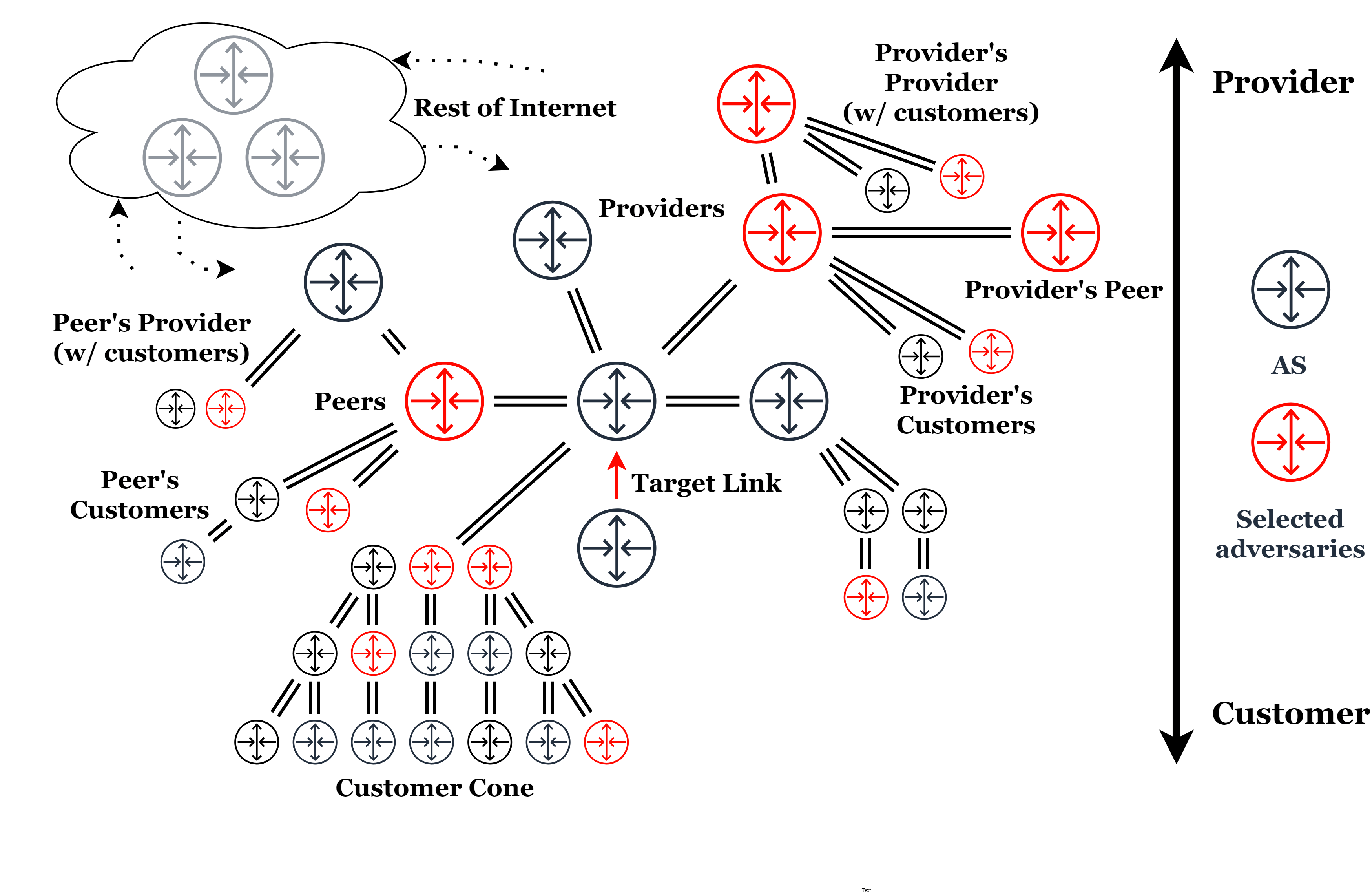}
        \captionof{figure}{Sampling adversary ASes at random along valley free paths from the To AS, within 3 topological hops, with adversaries reached from To AS's peers, customers, and providers included.}
        \label{fig:general_attack}
        \vspace{10pt}
    \end{minipage}%
    \hfill
    \begin{minipage}[t]{0.75\columnwidth}
        \centering
        \includegraphics[width=\columnwidth]{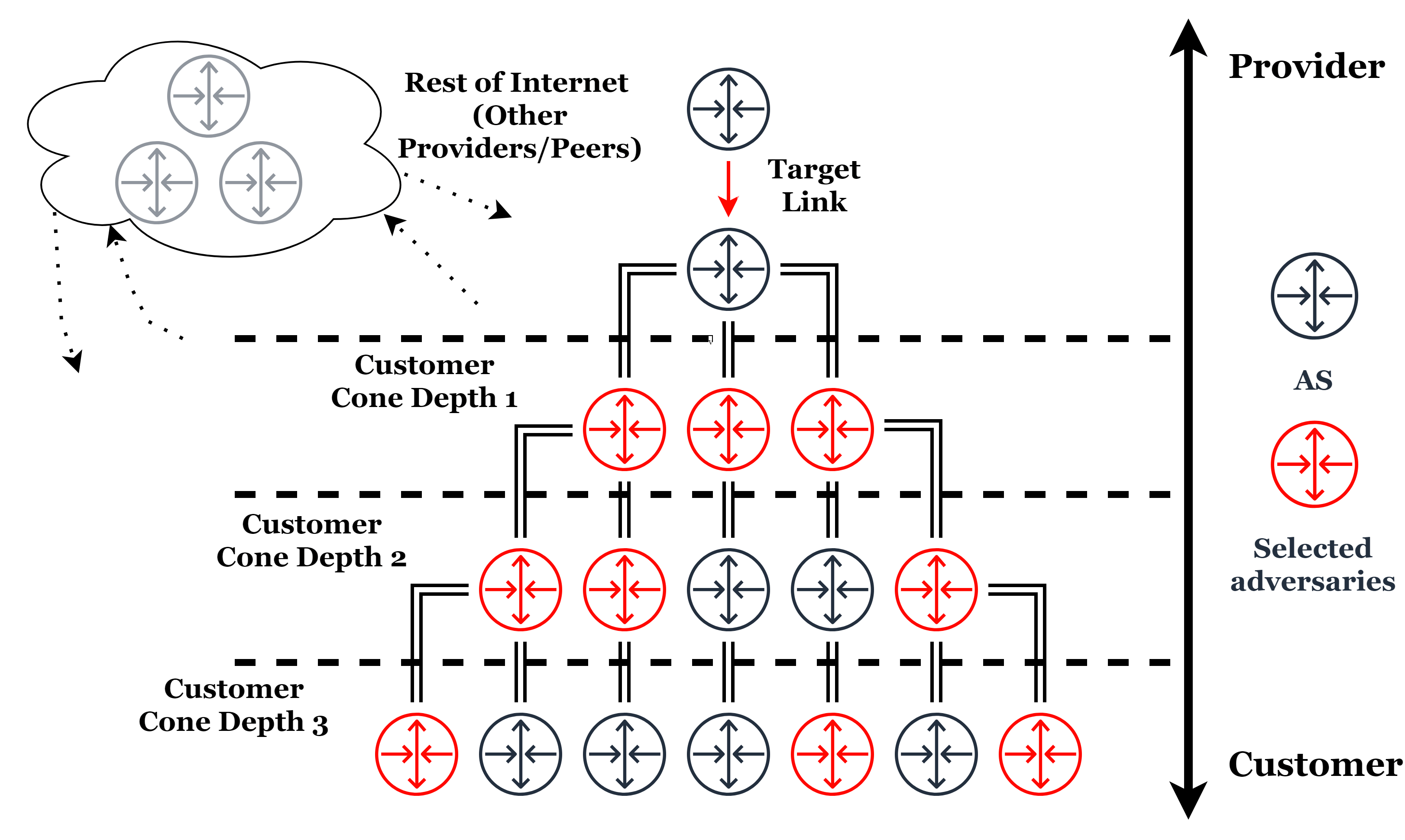}
    	\captionof{figure}{Sampling adversary ASes from multiple depths into the To AS customer cone.}
    	\label{fig:cust_attack}
    \end{minipage}
    \vspace{-10pt}
\end{figure}


\textbf{By link betweenness:} An important insight of the Crossfire attack is that degrading links in the dense core of the Internet would create broad disruption~\cite{studer2009coremelt}. These links are characterized by high \textit{betweenness}, where betweenness is quantified by the number of times a link appears on best paths between all ASes in the pre-attack inferred topology. So, for our second sample set, we divide links in the CAIDA AS relationship dataset~\cite{CAIDA} by their betweenness decile, and sample 100 links each from 1) below the 1st decile (fringe links), 2) between the 5th and 6th decile (moderately utilized links), and above the 9th decile (core links). This will allow link vulnerability comparison based on path usage.

\textbf{By bot-to-bot flow density:} Our third target link set is also sampled from low, middle, and high decile ranges, but is based on pre-attack bot-to-bot flow density rather than betweenness. For each of our three botnet models, we sample 100 links each from the low, middle, and high decile ranges described in the previous scheme. This will highlight how effective the attack is in both improving the flow density for links with moderate pre-attack exposure, and in exposing links that were previously unreachable by the botmaster. 

\textbf{By bot-to-any flow density:} Finally, we build a bot-to-any flow density link sample set, which is constructed exactly as the previous set, but using \textit{bot-to-any} rather than \textit{bot-to-bot} flow density.  


\subsection{Adversary Selection}{\label{adv-selection}}
We must also select the adversary ASes that will be used to issue poisoned advertisements. Intuitively, we expect that an AS's ability to steer traffic onto a selected link will dissipate with increased topological distance from the link. So, we constrain our adversary selection to ASes that are within 3 topological hops of the target link - roughly the average BGP path length~\cite{path-length}. To establish how distance affects attack success, we sample adversary ASes from one, two, and three hops distant from the target link.

\begin{figure}[!ht]
	\centering
	\includegraphics[width=0.60\columnwidth]{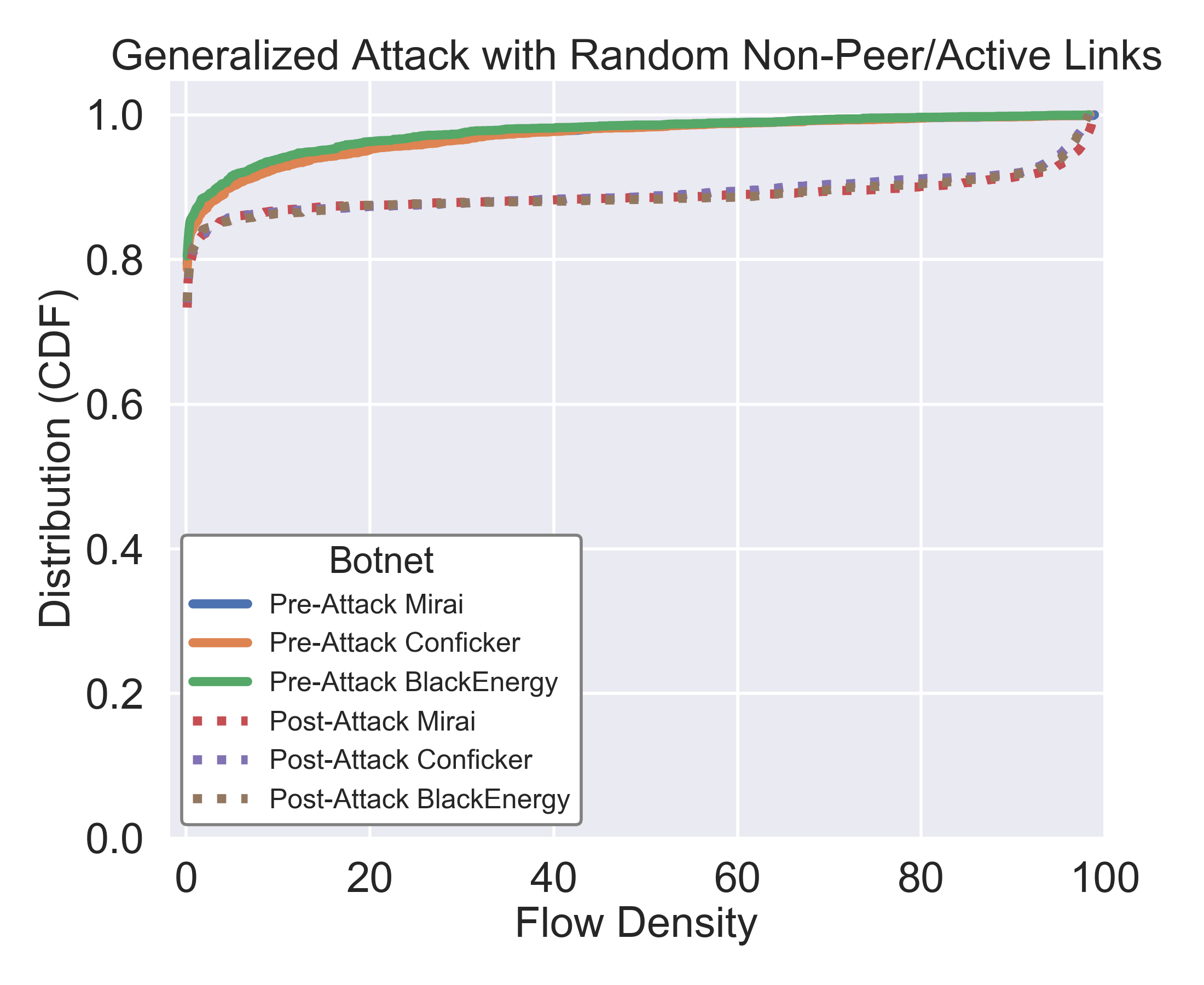}
	\vspace{-5pt}
	\caption{Pre- and post-attack bot-to-bot flow density CDF, random link selection/general adversary selection}
	{\label{fig:maestro-success-random-mirai}}
	\vspace{-10pt}
\end{figure}

\begin{figure*}[t]
    \centering
    \subfloat[Betweenness-based link selection]{\label{fig:maestro-success-lci-bet-mirai}\includegraphics[width=0.29\textwidth]{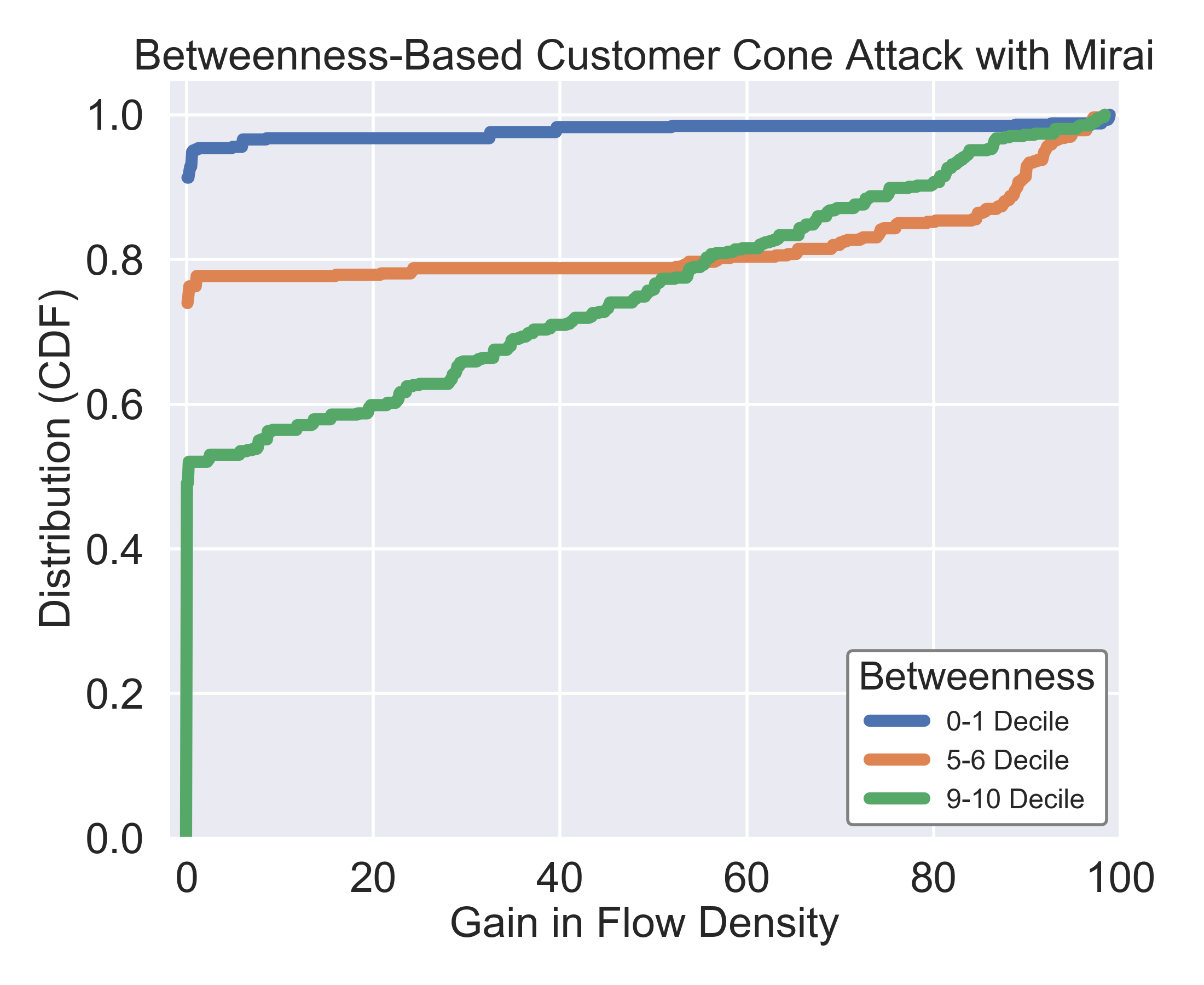}}
    \hspace{0.20cm}
    \subfloat[Bot-to-bot flow density selection]{\label{fig:maestro-success-b2b-mirai}\includegraphics[width=0.29\textwidth]{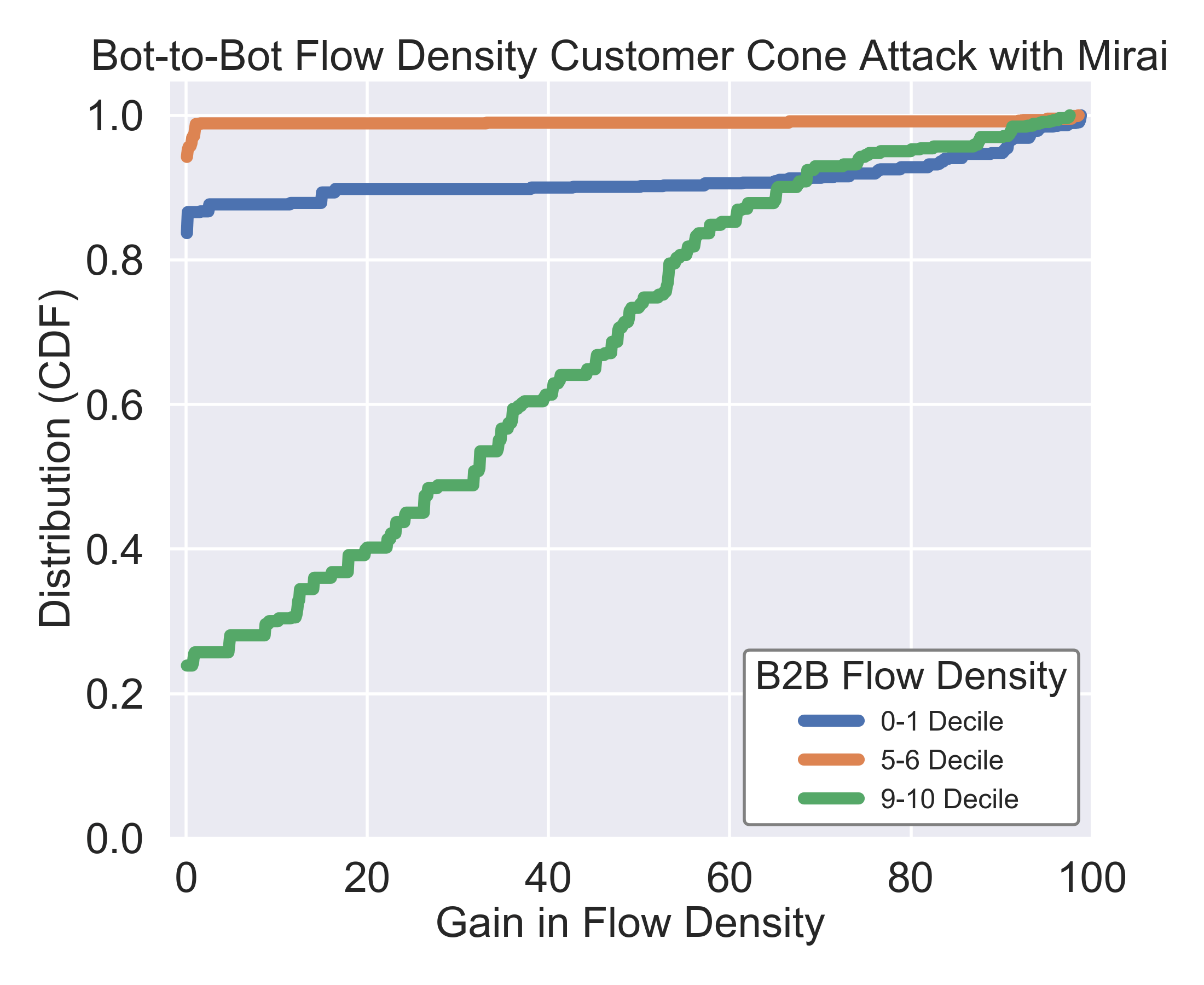}}
     \hspace{0.20cm}
    \subfloat[Bot-to-any flow density selection]{\label{fig:maestro-success-b2a-mirai}\includegraphics[width=0.29\textwidth]{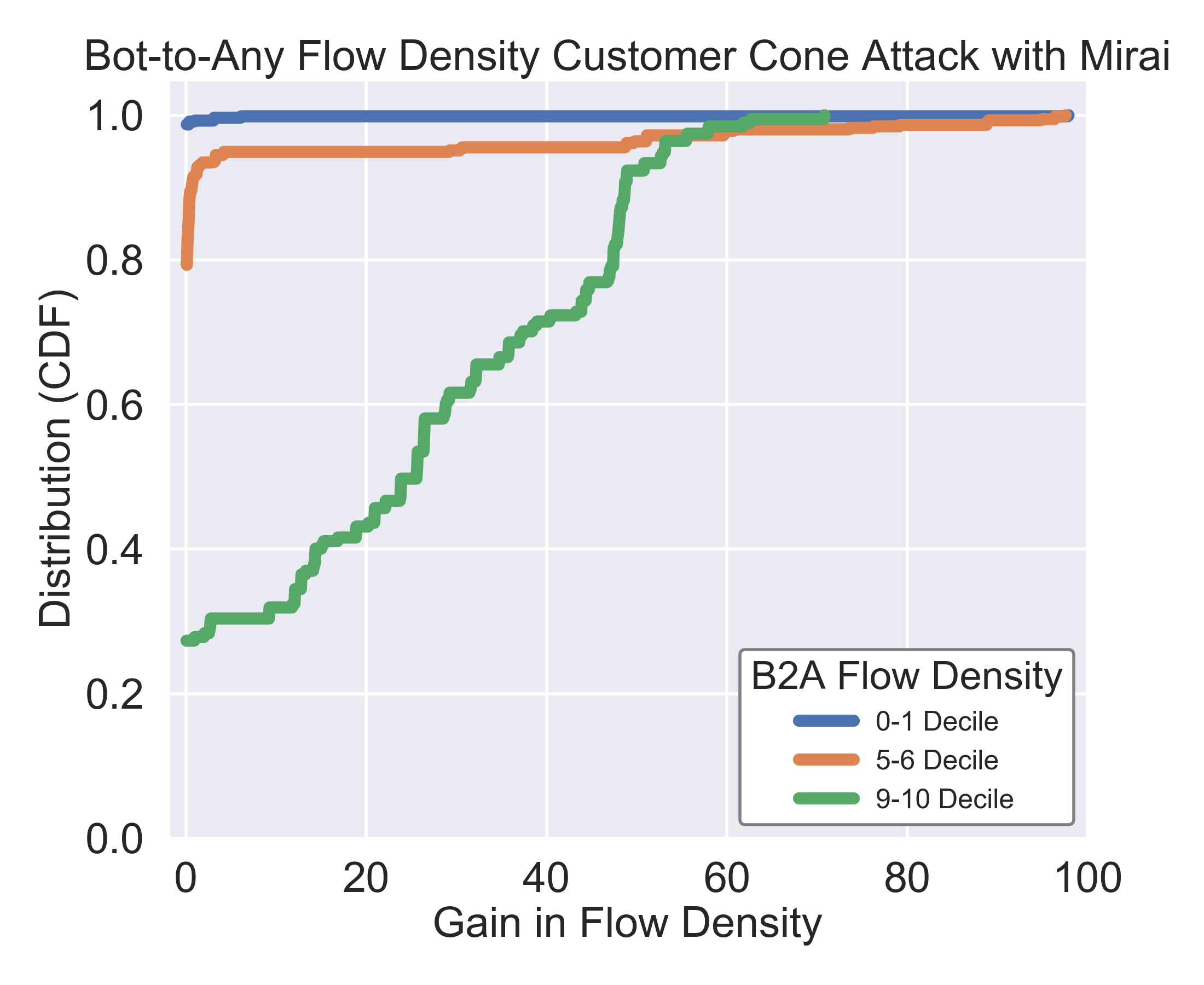}}
    \caption{Flow density gain results (post-attack density - pre-attack density) by link selection strategy, \textbf{Mirai} botnet model}
    \label{fig:maestro-success-all-link-sampling}
    	\vspace{-10pt}
\end{figure*}

\begin{figure*}[t]
    \centering
    \subfloat[Flow density CDF, provider to customer only]{\label{fig:pre-post-no-cust-prov}\includegraphics[width=0.29\textwidth]{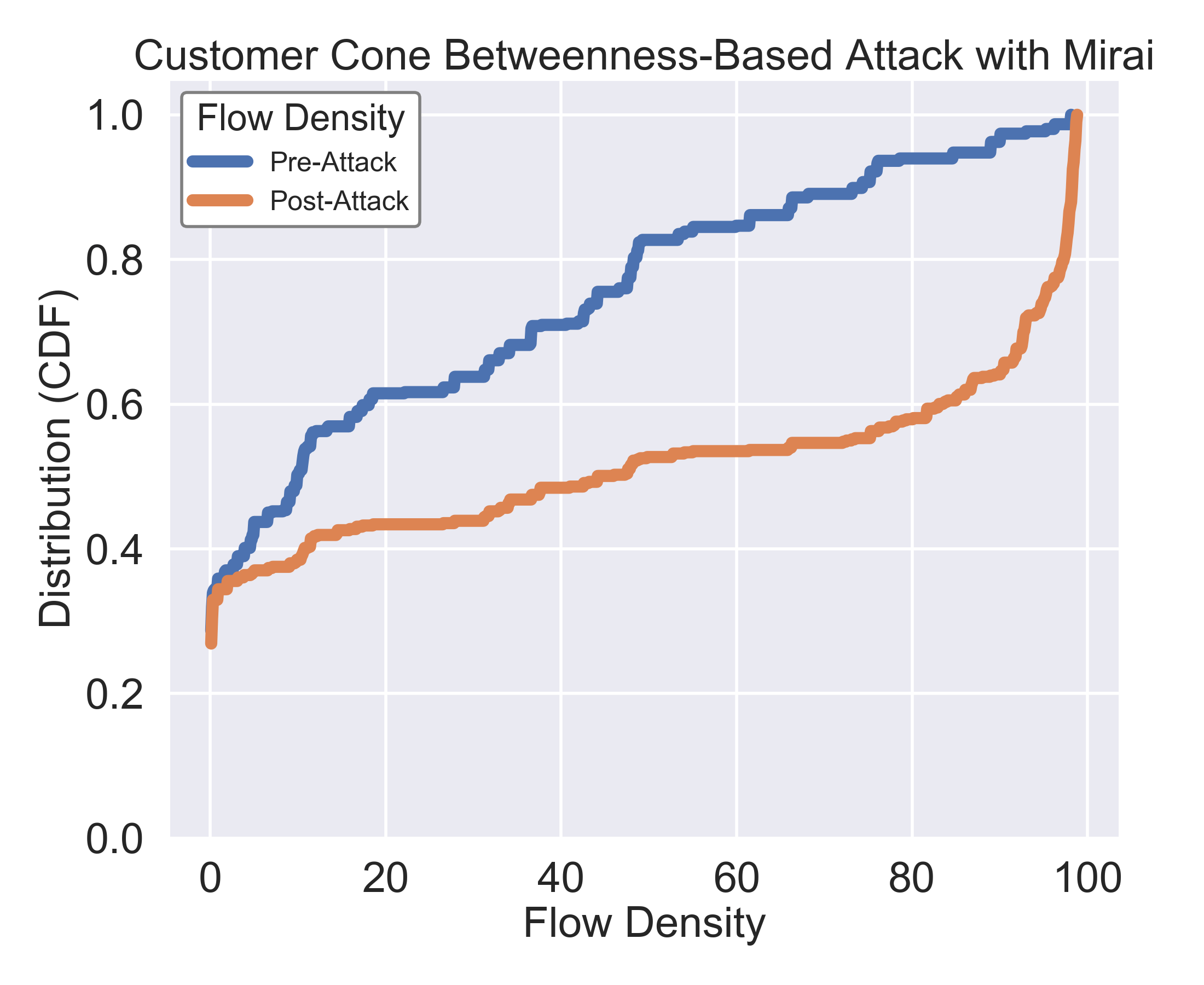}}
    \hspace{0.05cm}
    \subfloat[Heatmap of success at varying customer cone depths sample]{\label{fig:depth-heatmap}\includegraphics[width=0.29\textwidth]{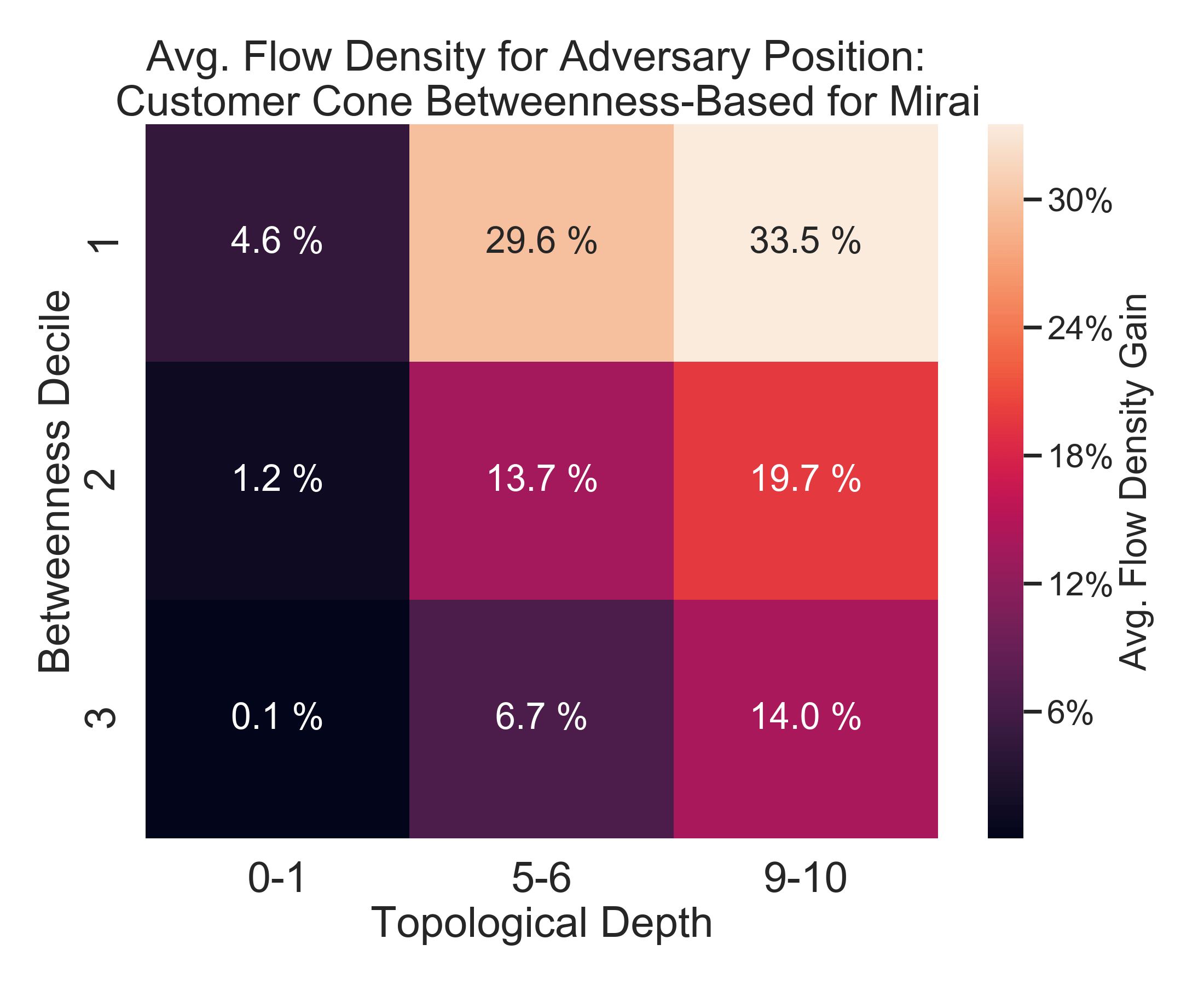}}
    \hspace{0.05cm}
    \subfloat[Distribution of success by link relationship]{\label{fig:link-rel-dist}\includegraphics[width=0.29\textwidth]{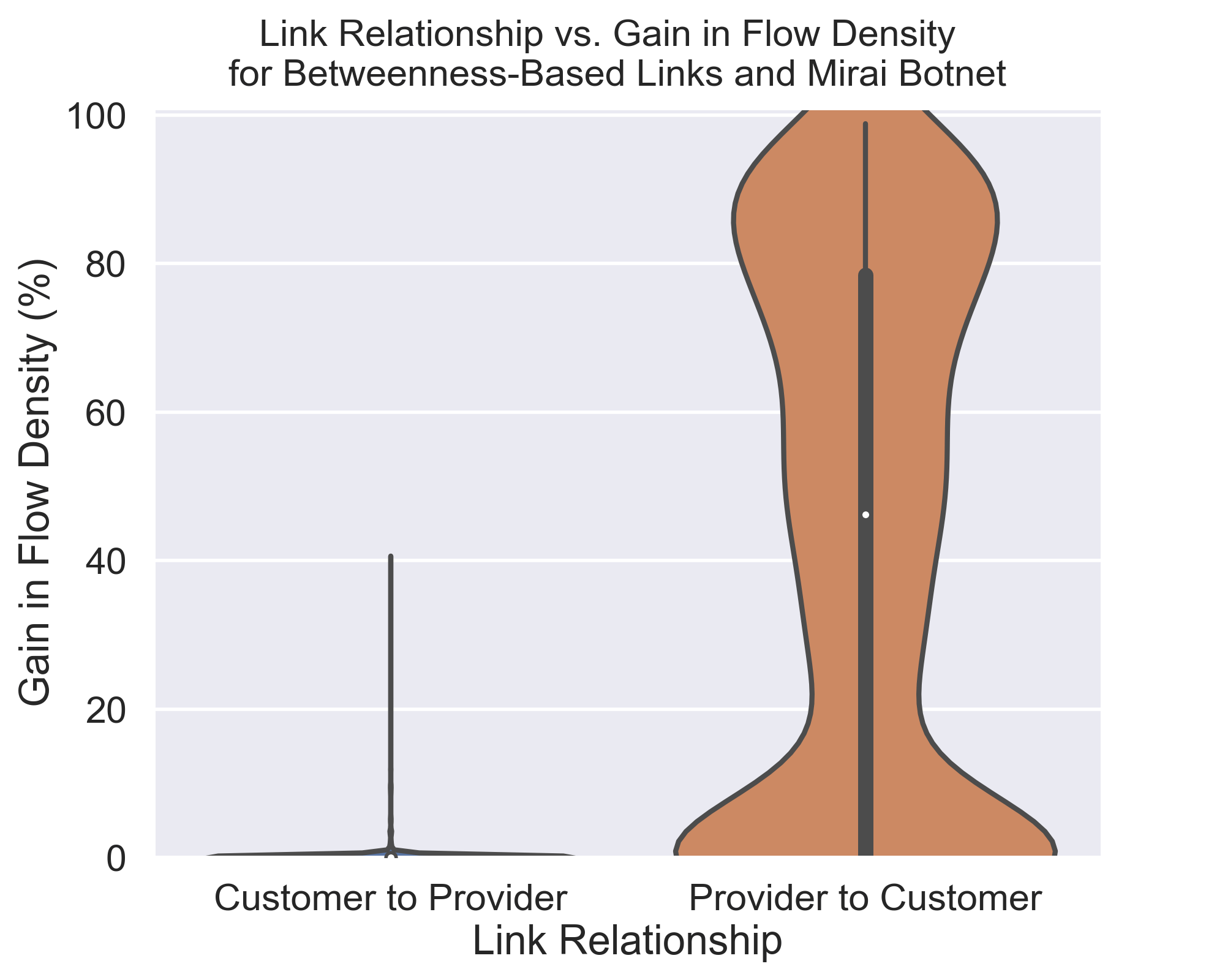}}
    \caption{Deeper look at the customer attack success, betweenness link sample}
    \label{fig:customer-attack-lci-mirai}
    	\vspace{-10pt}
\end{figure*}

\textbf{General selection:} In this method, we only constrain adversary selection to those ASes that lie within 3 hops along valley-free paths from the To AS. Because path export rules are different for providers, customers, and peers (see Section~\ref{bgp}), the prevalence of \textit{attack paths} may be affected by AS relationships. Recall that attack paths are available, valley-free paths from malicious flow sources (like bot-infested ASes) to the adversary AS that cross the target link. Note that these are not necessarily currently used best-paths, but simply some valid path that sources could choose to reach the adversary. To explore these dynamics, we ensure that ASes connected to the To AS via customers, peers, and providers (ASes in the customer, peer, and provider \textit{adversarial regions}) are represented in the sampling. Figure~\ref{fig:general_attack} shows an example sampling respecting these considerations. Note that sampling for a customer to provider link is depicted; only the customer region is available for provider to customer targets due to BGP path export rules.

\textbf{Customer-only selection:} The customer cone of an AS has the highest possible visibility of routes exported from the AS; naturally, the AS seeks to provide its customers with all of its known best paths in hopes of transiting customer traffic to the maximum number of destinations. For customer-only selection, we limit adversary sampling to those within 3 hops in the To AS customer cone. We expect that these ASes will have the maximum number of \textit{attack paths} among all potential attackers. For a depiction of this type of sampling, see~\ref{fig:cust_attack}. Note that this selection type produces a subset of the adversaries selected by the general selection method.  

The above methods for sampling adversaries and links will be combined in the following experiments to elucidate success conditions for a Maestro attacker.

\subsection{Random Attacks}{\label{random}}

In this first experiment, we set out to discover patterns in successful attacker/target pairings to guide further experimentation. We simulated attacks on 2000 links sampled from the topology according to the random selection scheme from Section~\ref{link-selection}. Adversaries were selected via general adversary selection as described in Section~\ref{adv-selection}, with three adversaries sampled at each depth 1-3 from each adversarial region (ASes reached from customers, providers, or peers of the To AS). This yields a total of about 27 attackers per link. Note that there will be fewer attackers when the To AS has a limited number of customers/providers/peers from which to sample. The results are shown in Fig.~\ref{fig:maestro-success-random-mirai}.

We make two observations about this initial experiment. First, we see that results from each of our three botnet models indicate similar steering behavior despite their differing topological distributions of infected hosts. This dynamic is consistent across all of our experiments, so we will henceforth only display results for the Mirai model. Graphs for the other botnets will be included in the appendix for completeness.

Secondly, we see that these results are frankly underwhelming. For more than 80\% of sampled targets, \textit{no} improvement was seen in flow density after the attack. An analysis of the few successful cases, however, revealed some important common factors. Successful adversaries were almost always \textit{close} to (within 3 topological hops of) the target link, confirming our suspicion that distance was likely to play a major role in moving traffic. 

Interestingly, successful attackers were almost universally located in the \textit{customer cone} of the target link destination (the To AS). This is because path export rules are most generous for customers. By locating the adversary in the customer cone of the target endpoint, we maximize attack path viability on the adversary's side of the link - because all links from the target to a direct/indirect customer adversary are provider to customer links, paths from the To AS to that adversary are naturally valley-free. Our next experiment is designed to focus our effort on adversaries in the customer cone.

\subsection{Customer Cone Attack}{\label{customer}}

In this section, we present results for betweenness, bot-to-bot flow density, and bot-to-any flow density target link selection with customer only adversary selection as described in Sections~\ref{link-selection} and~\ref{adv-selection}. By sampling 100 links each with relatively low, intermediate, and high betweenness/flow-density, we seek to illustrate how the attack performs on targets with different characteristics. Selecting only direct/indirect customers of the To AS as adversaries allows us to maximize the number of available attack paths on the adversary's side of the link. 

As in the previous experiment, three adversaries are sampled from each depth 1-3 in the To AS customer cone. This results in about 1800 adversary/link pairings per link sample set, for about 5400 total simulated attacks. The results are shown in Figures~\ref{fig:pre-post-no-cust-prov} and ~\ref{fig:depth-heatmap}. The adversary's expected success in this case is dramatically improved; for direct customers of high betweenness links, the average flow density gain is \textbf{greater than 30 percent} (Fig.~\ref{fig:depth-heatmap}). Note that this figure is not percent gain relative to existing flow density - rather, an additional 30 percent \textit{of the entire botnet} is brought on to the target. For low betweenness links, attack impact is negligible, but this is neither surprising nor particularly interesting; these links are not primary targets for LFAs. 

A deeper examination of the data yields some critical insights about the shape of a successful Maestro scenario. One of the most important is that target link relationship is critical to attack success. For an adversary in the To AS customer cone, attack paths are more available when the To AS is a customer of the From AS; that is, the target link is a \textit{provider to customer} link. This is an intuitive finding; any bot AS that must transit a provider to customer link to reach the target link \textit{cannot} then transit a customer to provider link and remain valley-free. Like locating the adversary in the To AS customer cone, targeting a provider to customer link removes a potential valley from attack paths, but \textit{at} rather than \textit{after} the target link. The importance of this dynamic is shown in Fig.~\ref{fig:link-rel-dist} - this violin plot shows the relative distribution of flow density gains by link relationship. Virtually all successful cases for this experiment were on provider to customer links. The relatively rare conditions for success in the customer to provider case are examined in the next experiment. 

Fig~\ref{fig:pre-post-no-cust-prov} displays pre vs. post attack flow density for the same betweenness link sample in Fig~\ref{fig:maestro-success-lci-bet-mirai}, filtered to include only provider to customer links. Here we see results for the ideal case for the attack: an adversary AS located in the customer cone of the To AS, when the target link is a provider to customer link. Note that the region between the curves in this figure represents the attacker's gain from executing the Maestro attack. Before the attack, most sampled links have flow densities below 10\% - that is, most link targets are vulnerable to 10\% or less of infected hosts in a Coremelt LFA. \textbf{After Maestro is executed, roughly half of sampled links have flow densities of 50\% or higher}.

While this experiment demonstrates the ideal scenario for a Maestro attack, the following studies explore how an adversary can affect other likely targets, including provider to customer and core peer links.

\begin{figure*}[!ht]
    \centering
    \subfloat[Reversing the link relationship on failed customer to provider links]{\label{fig:link-rel-swap-comparison}\includegraphics[width=0.29\textwidth]{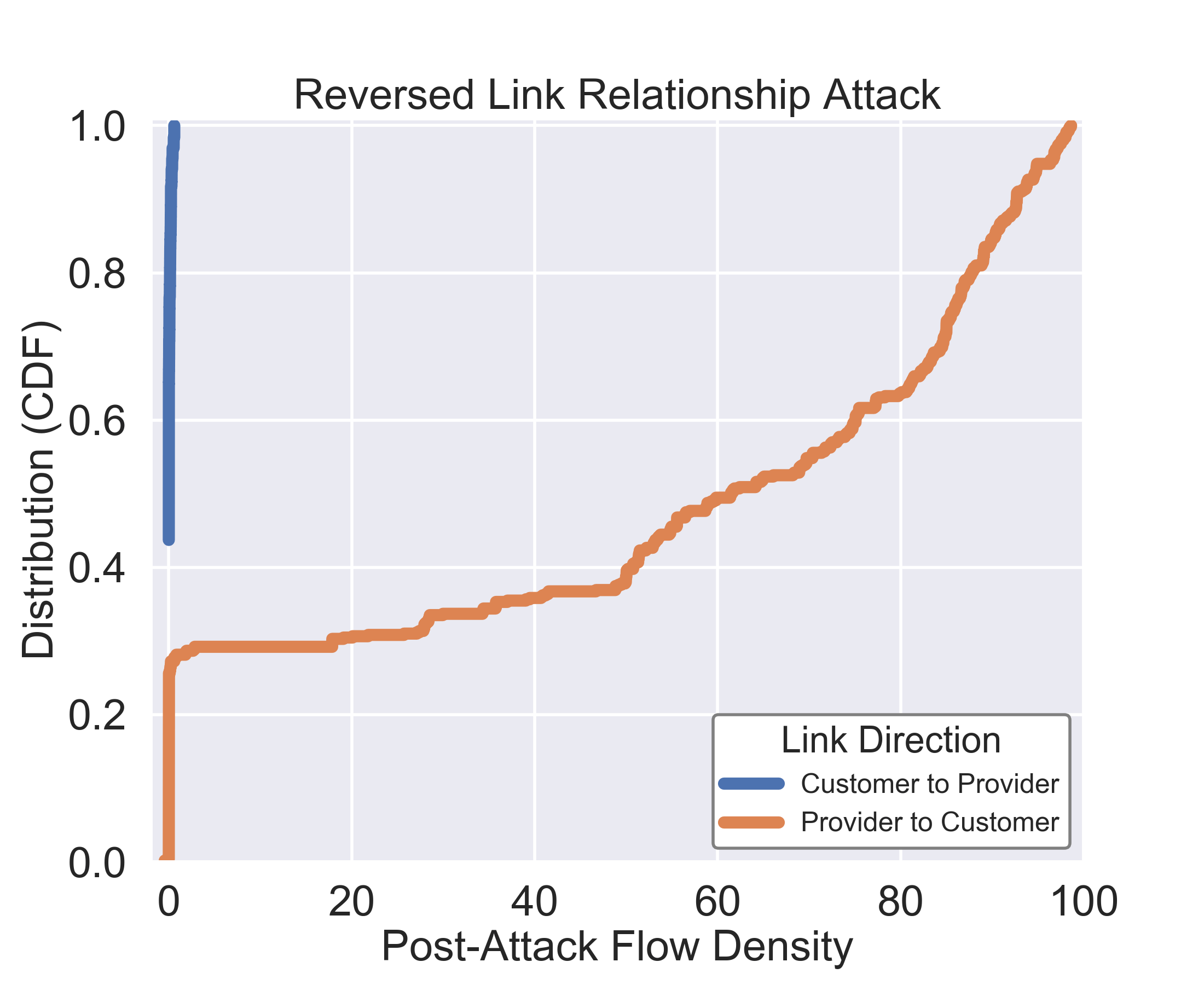}}
    \hspace{0.10cm}
    \subfloat[Attacking customer to provider links with highly infected From AS customer cones]{\label{fig:high-cust-cone-prov}\includegraphics[width=0.29\textwidth]{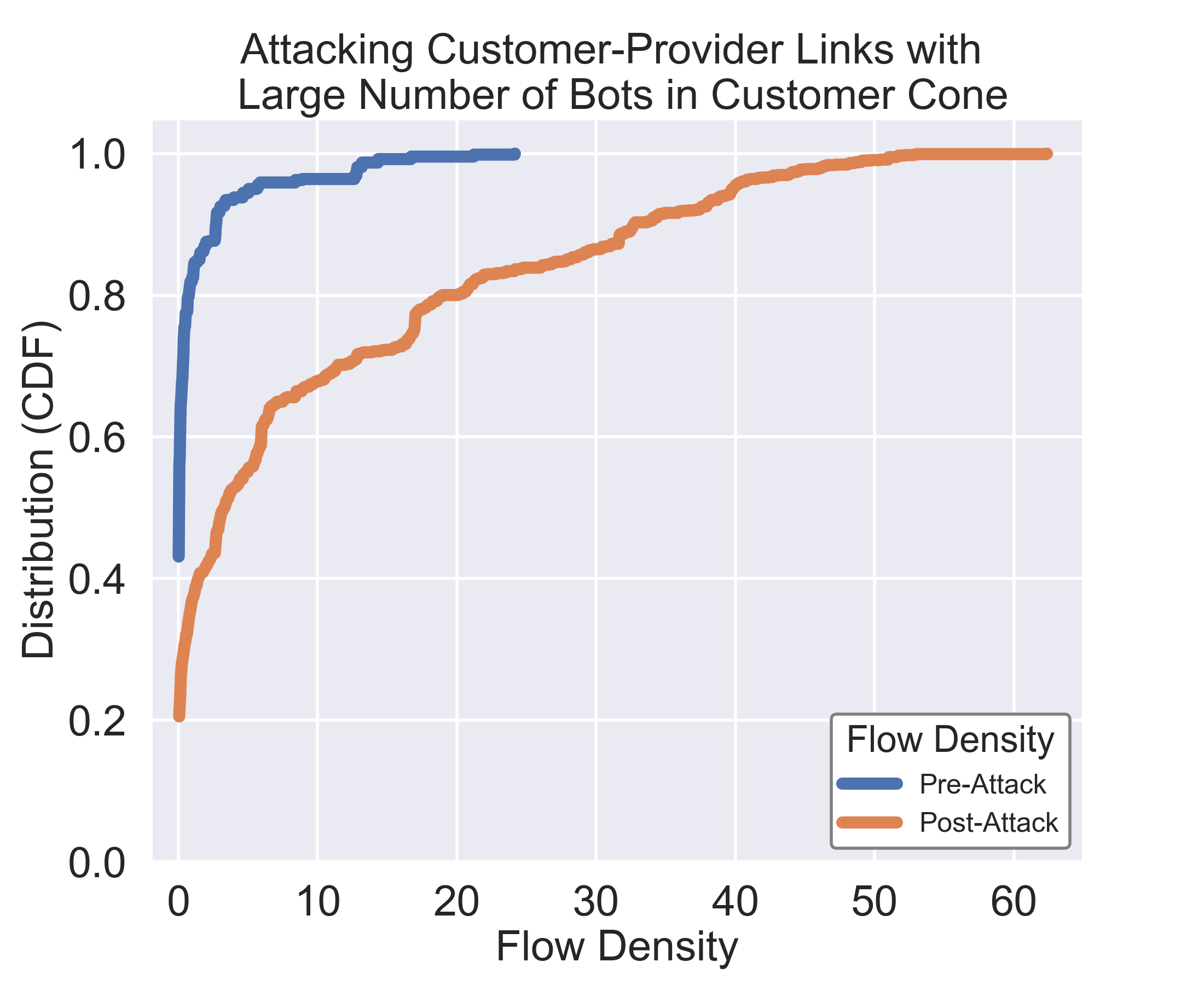}}
    \caption{Extending success to customer to provider links, Mirai botnet}
    \label{fig:reverse-link-rel-lci-mirai}
    \vspace{-10pt}
\end{figure*}

\begin{figure*}[!ht]
    \centering
    \subfloat[Topology prior to leak attack execution]{\label{fig:leak-overview1}\includegraphics[width=0.35\textwidth]{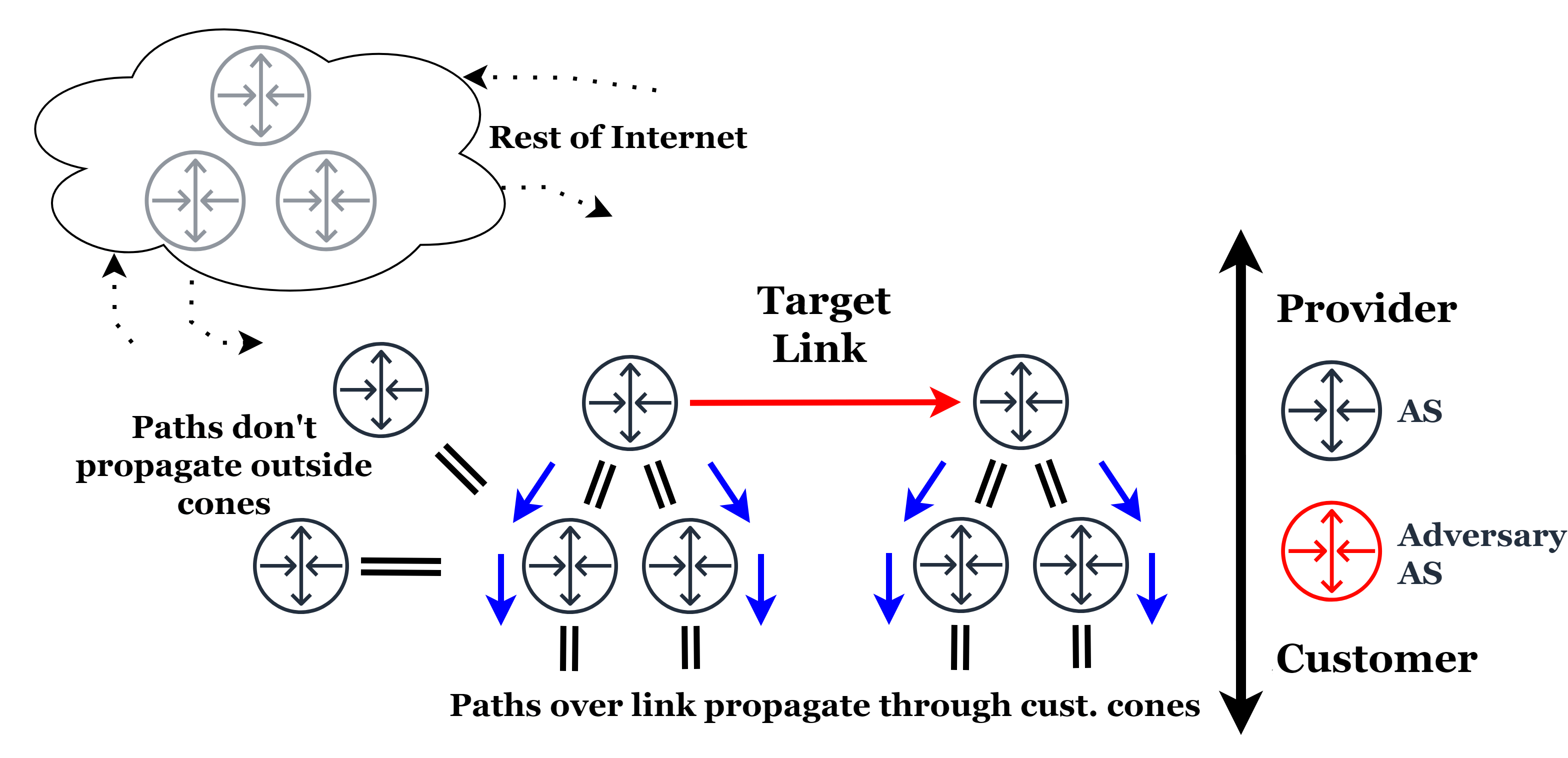}}
    \hspace{0.10cm}
    \subfloat[Post-attack topology]{\label{fig:leak-overview2}\includegraphics[width=0.35\textwidth]{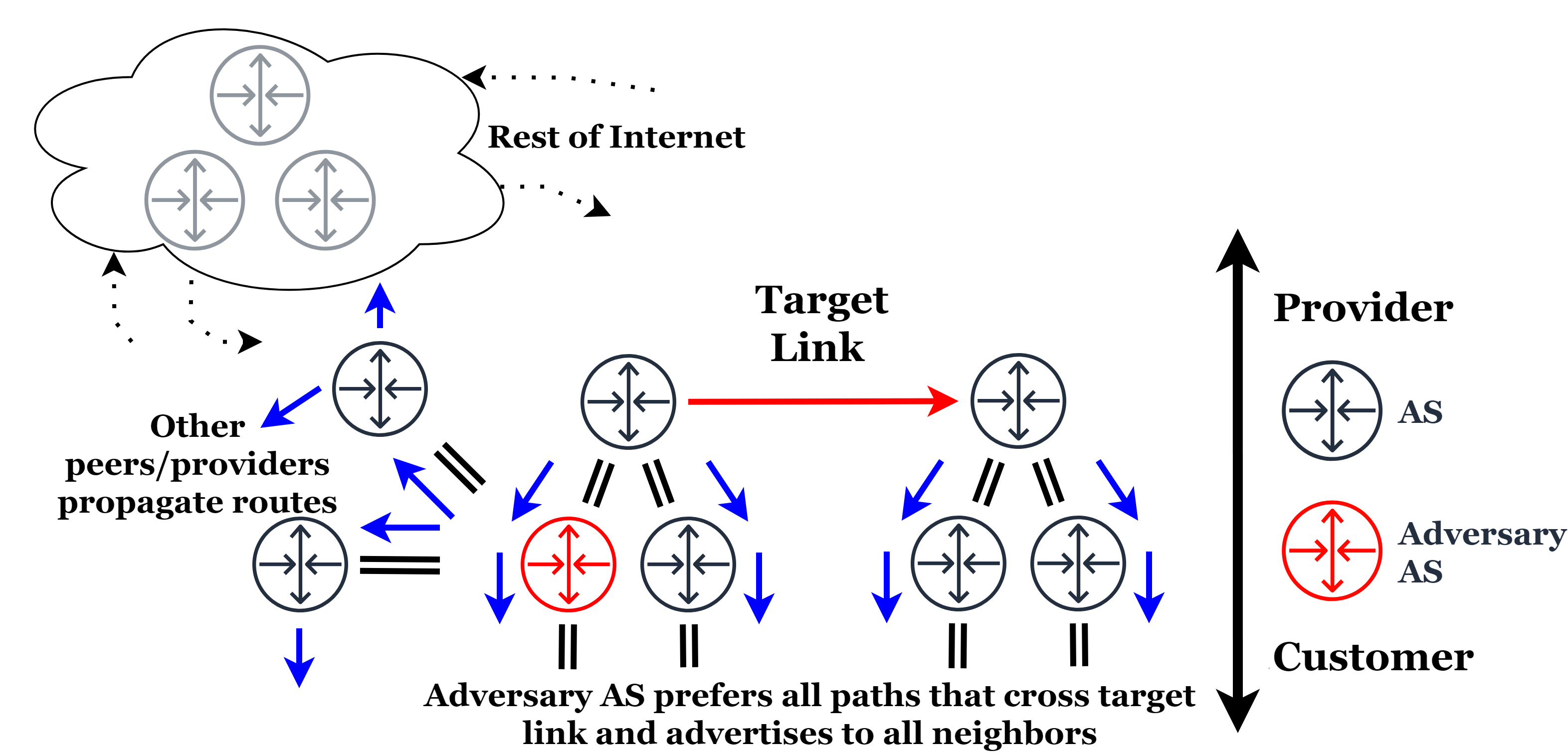}}
    \hspace{0.10cm}
    \subfloat[Flow density CDF for leak attack on core links]{\scalebox{0.75}{{\label{fig:leak-attack}\includegraphics[width=0.35\textwidth]{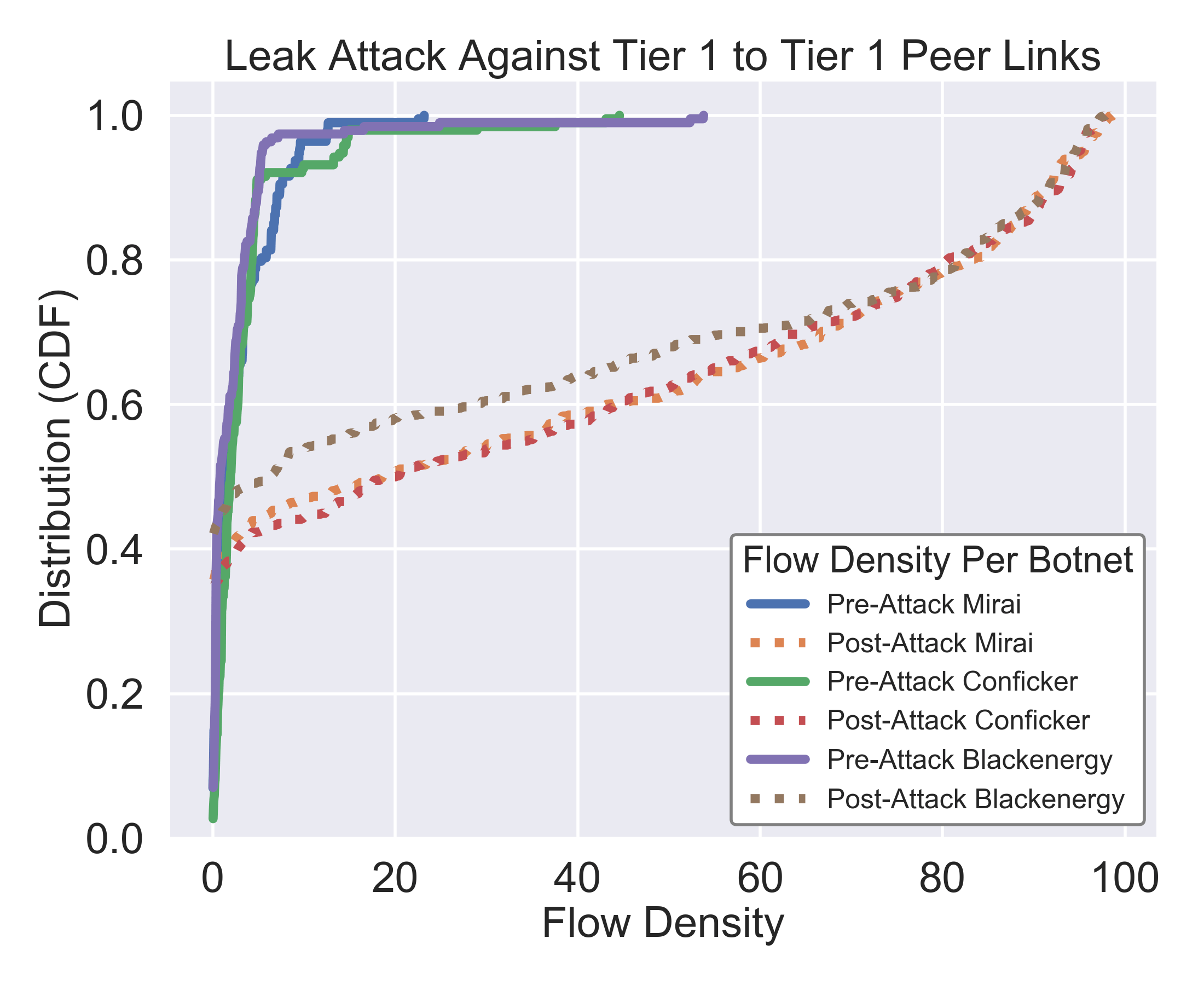}}}}
    \caption{Valley-path leak attack overview and results}
    \label{fig:leak-overview}
    	\vspace{-10pt}
\end{figure*}

\subsection{Customer to Provider Link Attacks}{\label{provider}}
In our previous experiments, we have examined the role that target link relationship and adversary position play in the prevalence of attack paths. Since most Internet services require bidirectional communication, the simplest method for attacking customer to provider links is to attack the opposite direction; i.e., target the associated provider to customer link. To confirm the viability of this method, we reversed the target link direction for all customer to provider links from our betweenness link sample set that resulted in less than 1\% post-attack flow density in the prior experiment. We re-sampled adversaries from the new To AS customer cones and simulated attacks on this target set. Fig.~\ref{fig:link-rel-swap-comparison} shows the results of these reversed attacks. Clearly, link relationship was the primary culprit preventing attack success; attacking in the reversed direction yields drastically improved flow density. For most links, we saw 50\% or greater post-attack flow density, meaning we expect to engineer most infected host paths in the botnet onto the target.

In the case where an adversary has compromised an AS in the To AS customer cone of a customer to provider link, though, the adversary may not be able to simply attack the reversed direction; this requires compromising a different AS. So, we ask now under what conditions the customer to provider direction can be successfully attacked. The only attack paths available in this case originate from within the From AS customer cone, because any flow sources \textit{not} located there must transit a peering or provider to customer link before reaching the target link. This means that potential targets are limited to those with significantly bot-infected From AS customer cones. 

To test our ability to steer bot traffic in these scenarios, we randomly sampled 300 links from between the 9th and 10th deciles for From AS customer cone infection rate from the set of all customer to provider links, simulated attacks, and measured flow density improvement. The results are shown in Fig.~\ref{fig:high-cust-cone-prov}. While these results are not as dramatic as the prior experiment, we find that we are able to exert significant steering behavior on bots in the From AS customer cone.

\subsection{Peer Link Attacks}{\label{peer}}


None of our studies have targeted peer links. In the CAIDA inferred topology, however, about 410,000 of the 660,000 total Internet links are peering links. These links play a critical role in transiting traffic between Tier 1 providers in the Internet's core.  Like customer to provider links, attack path viability in this case is limited by BGP path export rules. Unlike those targets, however, we cannot simply reverse the direction of the attack; peer link export rules are the same for both endpoints. A Maestro attacker is therefore limited to bot flow sources that exist inside the From AS customer cone. As in the previous experiment, we can often effectively steer traffic sourced from within the From AS customer cone; we leave the presentation to our appendix in Figure~\ref{fig:high-cust-cone-peer}.

\begin{table*}[!ht]
\centering
\resizebox{\textwidth}{!}{%
\begin{tabular}{|
>{\columncolor[HTML]{FFFFFF}}c |
>{\columncolor[HTML]{eeeeee}}c |
>{\columncolor[HTML]{eeeeee}}c |
>{\columncolor[HTML]{eeeeee}}c |
>{\columncolor[HTML]{dddddd}}c |
>{\columncolor[HTML]{dddddd}}c |
>{\columncolor[HTML]{dddddd}}c |
>{\columncolor[HTML]{cccccc}}c |
>{\columncolor[HTML]{cccccc}}c |
>{\columncolor[HTML]{cccccc}}c |
>{\columncolor[HTML]{bbbbbb}}c |
>{\columncolor[HTML]{bbbbbb}}c |
>{\columncolor[HTML]{bbbbbb}}c |}
\hline
 & \multicolumn{3}{c|}{\cellcolor[HTML]{FFFFFF}\textbf{Pre-Attack Vulnerability}} & \multicolumn{3}{c|}{\cellcolor[HTML]{FFFFFF}\textbf{Post-Attack Vulnerability}} & \multicolumn{3}{c|}{\cellcolor[HTML]{FFFFFF}\textbf{Avg. \# Adversaries}} &
\multicolumn{3}{c|}{\cellcolor[HTML]{FFFFFF}\textbf{Std. Dev of \# Adv.}} \\ \hline
{\color[HTML]{000000} Vulnerability Threshold} & \cellcolor[HTML]{FFFFC7}{\color[HTML]{000000} \textit{25\%}} & \cellcolor[HTML]{FFCE93}{\color[HTML]{000000} \textit{50\%}} & \cellcolor[HTML]{FFCCC9}{\color[HTML]{000000} \textit{75\%}} & \cellcolor[HTML]{FFFFC7}{\color[HTML]{000000} \textit{25\%}} & \cellcolor[HTML]{FFCE93}{\color[HTML]{000000} \textit{50\%}} & \cellcolor[HTML]{FFCCC9}{\color[HTML]{000000} \textit{75\%}} & \cellcolor[HTML]{FFFFC7}{\color[HTML]{000000} \textit{25\%}} & \cellcolor[HTML]{FFCE93}{\color[HTML]{000000} \textit{50\%}} & \cellcolor[HTML]{FFCCC9}{\color[HTML]{000000} \textit{75\%}} & \cellcolor[HTML]{FFFFC7}{\color[HTML]{000000} \textit{25\%}} & \cellcolor[HTML]{FFCE93}{\color[HTML]{000000} \textit{50\%}} & \cellcolor[HTML]{FFCCC9}{\color[HTML]{000000} \textit{75\%}} \\ \hline
\textbf{Mirai} & 0.44 & 0.18 & 0.09 & 0.97 & 0.91 & 0.87 & 135.45 & 114.64 & 91.22 & 349.69 & 312.94 & 215.85
 \\ \hline
\textbf{Conficker} & 0.44 & 0.21 & 0.08 & 0.96 & 0.91 & 0.85 & 151.25 & 100.01 & 92.35 & 360.26 & 226.39 & 216.06
 \\ \hline
\textbf{Blackenergy} & 0.38 & 0.23 & 0.10 & 1.00 & 0.94 & 0.87 & 130.61 & 103.98 & 92.48 & 335.06 & 273.92 & 217.14
  \\ \hline
\end{tabular}%
}
\caption{Maestro result summary, provider to customer betweenness link selection, customer-only attack}
\label{tab:formalizing-lci-success}
\vspace{-22pt}
\end{table*}

We observe, however, that an AS located in the From AS customer cone of a peer link target is not limited to the Maestro attack. An adversary who has compromised such an AS has a much simpler means of altering the control-plane to increase flows: leaking valley paths that cross the target link. By \textit{leaking}, we mean that the adversary AS advertises to its other upstream providers/peers valley paths from itself over the target link to bot-infected prefixes in the To AS customer cone. Normally, these paths would only be advertised to the compromised AS's customers; the compromised AS has no interest in transiting traffic between providers at its own expense. But if the AS has been compromised by an adversary, or the AS's operators value attacking the target link above the cost of the attack traffic, this technique could be an effective means of exposing the peer leak to malicious flows. Figures~\ref{fig:leak-overview1} and~\ref{fig:leak-overview2} illustrate this technique, which we term a \textit{leak attack}. While we only evaluate the leak attack in the context of peer links, an adversary could similarly leak a customer to provider link.

Unlike the Maestro attack, leak attack traffic transits the adversary AS \textit{before} crossing the target link. So, this attack is limited in flow density improvement by the capacity of the adversary AS' inbound links, as well as the capacity of links between the adversary AS and the From AS (its direct or indirect provider). Of course, the attacker does not need to consume the entire capacity of the target peer link - only to increase flow density such that normal and malicious traffic \textit{together} exhaust link capacity. Here we present results without consideration for potential pre-target bottlenecks, with an examination of relative link capacities left to future work.

We evaluate this leak attack on the Tier 1 clique; that is, all 400 peer links connecting the 20 Tier 1 providers. For each target link, we sample adversaries from 1-3 hops in the From AS customer cone, and measure bot-to-bot flow density before and after the adversary AS leaks all routes containing the target link. Fig.~\ref{fig:leak-attack} shows the pre vs. post attack CDF by flow density for these simulations. According to CAIDA's inferred topology and our real-world botnet IP mappings, these critical links in the Internet's core are almost completely isolated from potential LFA traffic before the attack. Under all botnet models, we were able to see significant improvements. For the Mirai botnet, for instance, roughly half of Tier 1 link/adversary AS pairings had post-attack flow densities of 20\% or greater. Notably, these Tier 1 to Tier 1 links have \textit{virtually no vulnerability} before the attack, and \textit{50\%+ or greater exposure to IPs in major botnets} after leak execution from a small customer.

Given the differing performance of the Maestro attack under varied link and adversary selection scenarios, we will present a summary of attack results and perform further statistical analysis to better understand link vulnerability characteristics.

\section{Attack Scope and Vulnerability}{\label{formalization}}
As shown in Section~\ref{evaluation}, the general conditions under which we can expect success as a Maestro adversary are: 1) target a provider to customer link and 2) compromise a direct/indirect customer of the To AS. We present a table to summarize the results of the Maestro attack on provider to customer links from our betweenness link sample, Table~\ref{tab:formalizing-lci-success}. This table displays the proportion of links from the sample that an adversary can achieve a targeted bot-to-bot flow density level on, called the ~\textit{vulnerability threshold}, before and after attack execution. \textbf{Observe that 50\% or less of links in this sample have 25\% or greater pre-attack flow density across all botnet models; after the attack, greater than 95\% of these links are above this threshold}.

The adversary sample set columns display, for each vulnerability threshold, the average set size for the number of potential ASes that an adversary could compromise to perform the attack. To calculate this number, we first note that adversary AS size is not a significant determinant of attack success - the absolute value of the correlation coefficient of CAIDA AS rank~\cite{CAIDA-rank} and flow density gain for these attacks was always smaller than $.01$. Success for adversaries in the customer cone is instead dominated by relative topological position. For this reason, if an AS two hops into the To AS customer cone is successful at a given threshold, we expect every other AS at the same or lower customer cone depth to also successfully execute Maestro at that threshold. Out of about 600 provider to customer attacks in this sample set, we found that fewer than 3\% violated this expectation for any botnet model. To that end, we estimate the potential adversary pool as the To AS customer cone to the depth of the deepest successful adversary. The standard deviation is also presented for these adversarial sets. Note that the high variance in set size expressed by these statistics illustrates that the pool of potential adversaries varies greatly. For target links with large ISPs as the To As (like core Internet links), the To AS customer cone can include thousands of potential adversaries. For many smaller targets, only a handful of ASes are well-positioned for the attack. 


\subsection{Statistical Analysis}{\label{ml-analysis}}

In addition to understanding the scope of the attack, we also want to understand adversary AS/target properties that influence attack success. For this purpose, we statistically analyze a set of feature vectors composed of candidate attack success predictors. Using 17,000 simulations of the Maestro attack with randomized link selection and general adversary selection, we extract the following features: BGP best path distance from the adversary to the target link, the topological depth of the adversary in the To AS customer cone, number of bots in the From AS's customer cone, and the AS ranks of the adversary, From AS, and To AS. We selected these features based on properties that can be easily determined using standard traceroutes and open datasets such as CAIDA's AS Rank and AS relationships~\cite{CAIDA-rank,CAIDA}.

\begin{figure}[!ht]
	\centering
	\includegraphics[width=0.70\columnwidth]{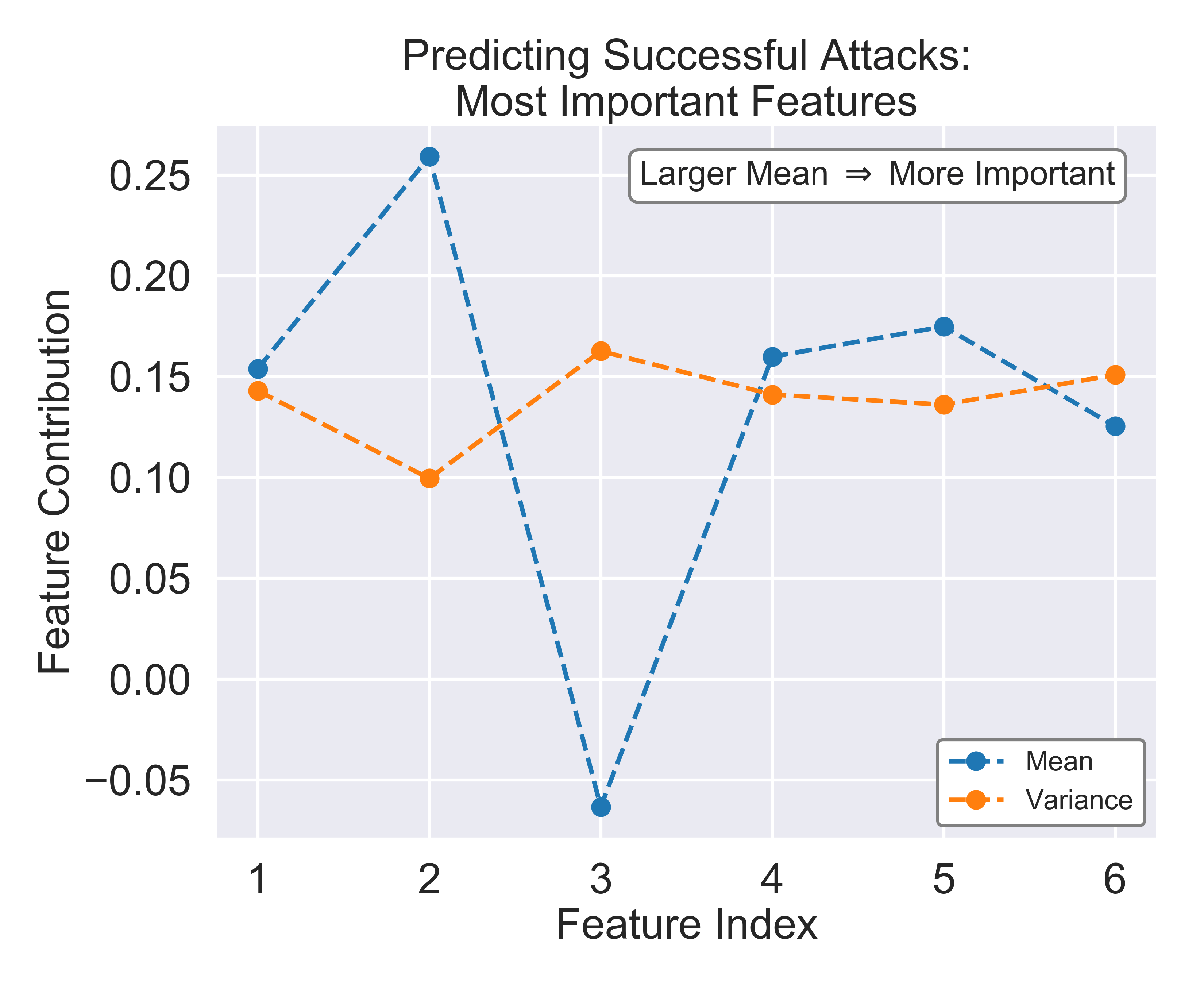}
	\caption{Most important features from Principal Component Analysis of link/adversary features for random attack instances}
	{\label{fig:ml-pca}}
	\vspace{-8pt}
\end{figure}

Fig.~\ref{fig:ml-pca} shows a Principal Component Analysis (PCA) algorithm used to rank all prior features by their mean and variance. Higher feature means suggest greater influence on attack success. We find the most important predictor to be the \textit{topological depth of the adversary}. The least important is the \textit{adversary AS rank}. Together, these findings demonstrate that topological positioning outweighs adversary AS size in determining attack success - small ASes can conduct the Maestro attack against relatively large transit AS links. The heatmap presented in Figure~\ref{fig:depth-heatmap} meshes with the above findings to illustrate how the adversary's ability to concentrate flows on the target decreases with distance. From this attack analysis, we move to a consideration and evaluation of potential mitigations.

\begin{figure*}[!ht]
    \centering
    \subfloat[Flow density CDF by number of poisons]{\label{fig:poison-details}\includegraphics[width=0.32\textwidth]{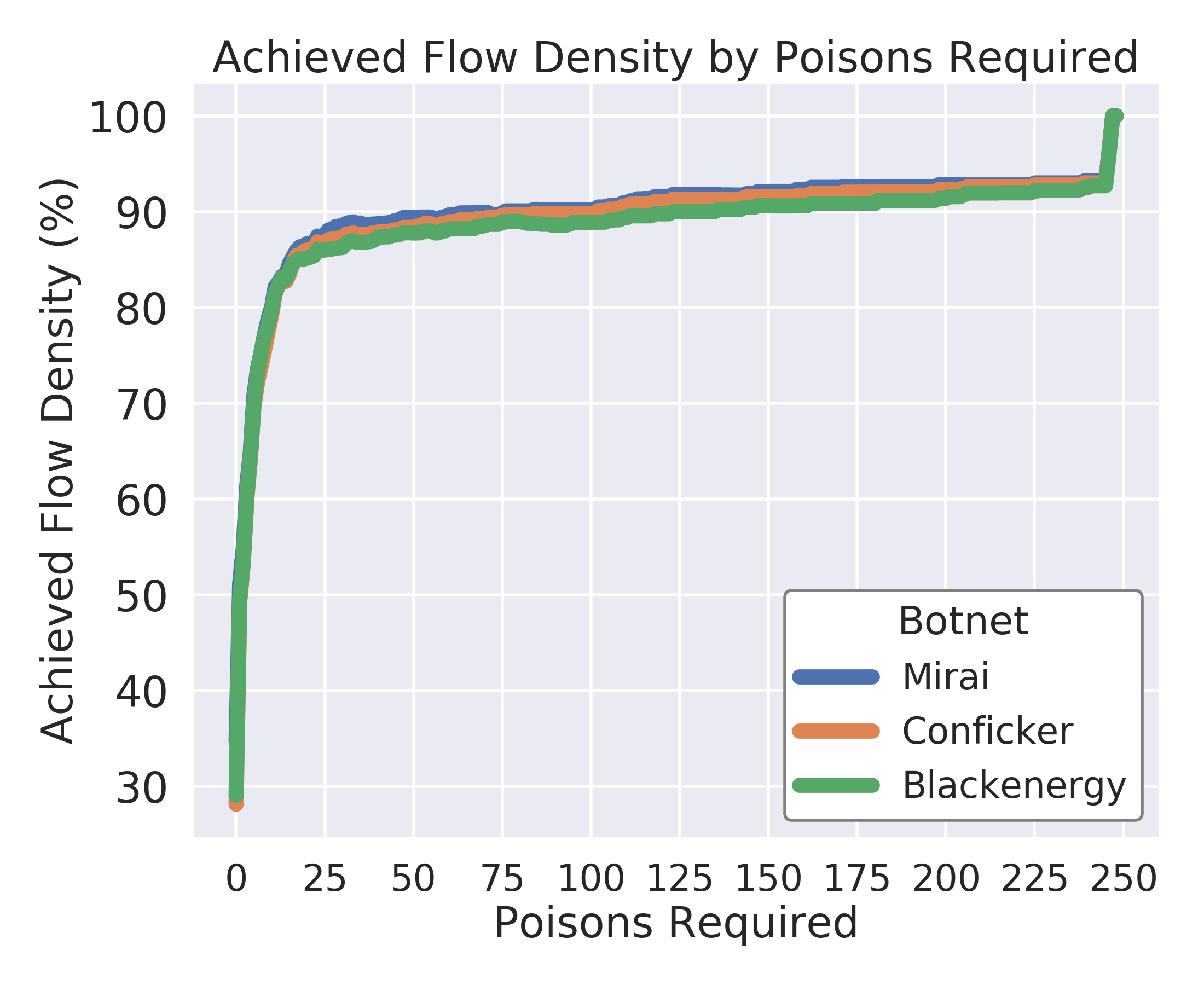}}
    \hspace{0.20cm}
    \subfloat[Flow density CDF against path filtering]{\label{fig:poison-capping}\includegraphics[width=0.32\textwidth]{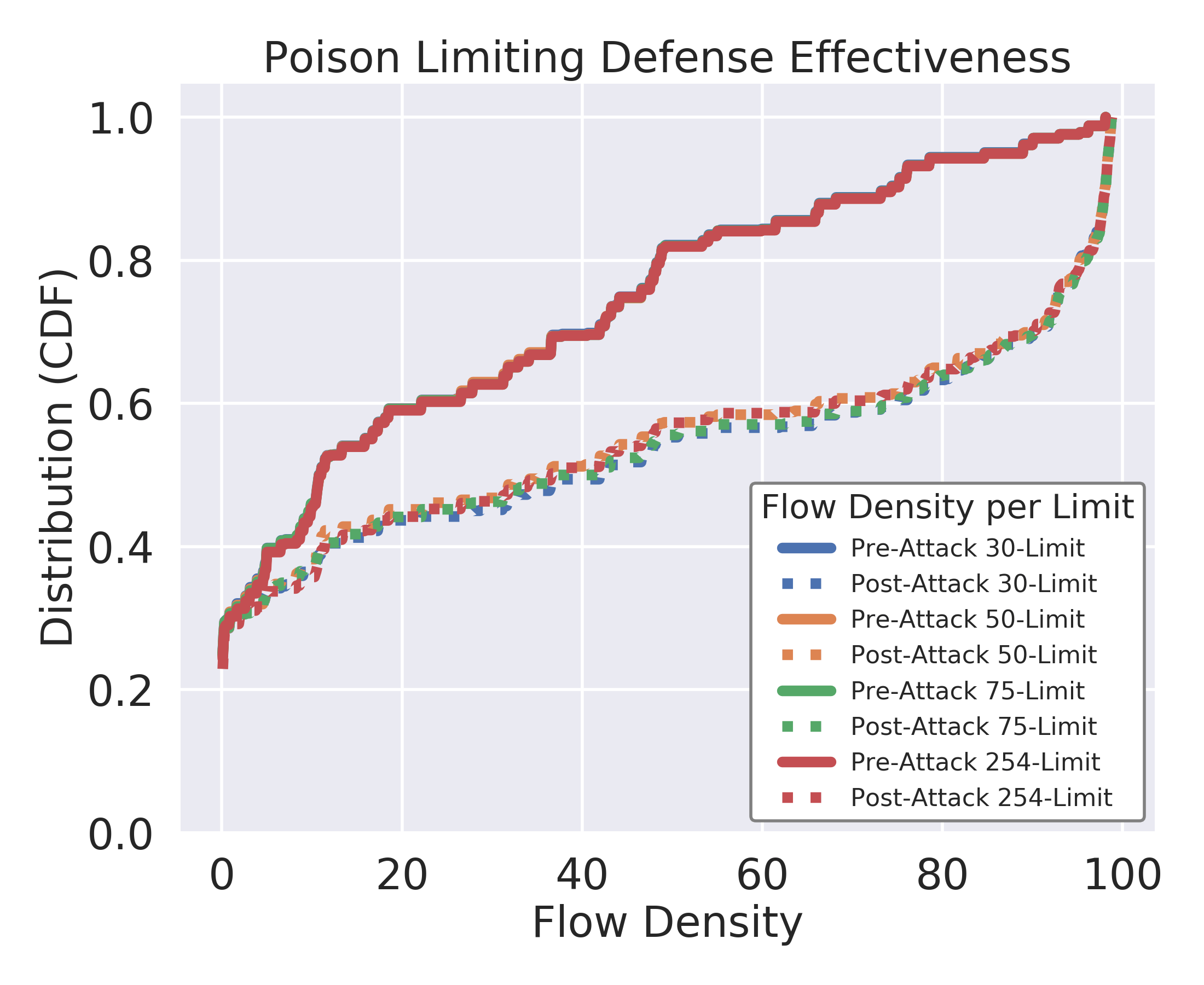}}
    \hspace{0.20cm}
    \subfloat[Flow density CDF by various AS filtering sets]{\label{fig:poison-filtering}\includegraphics[width=0.32\textwidth]{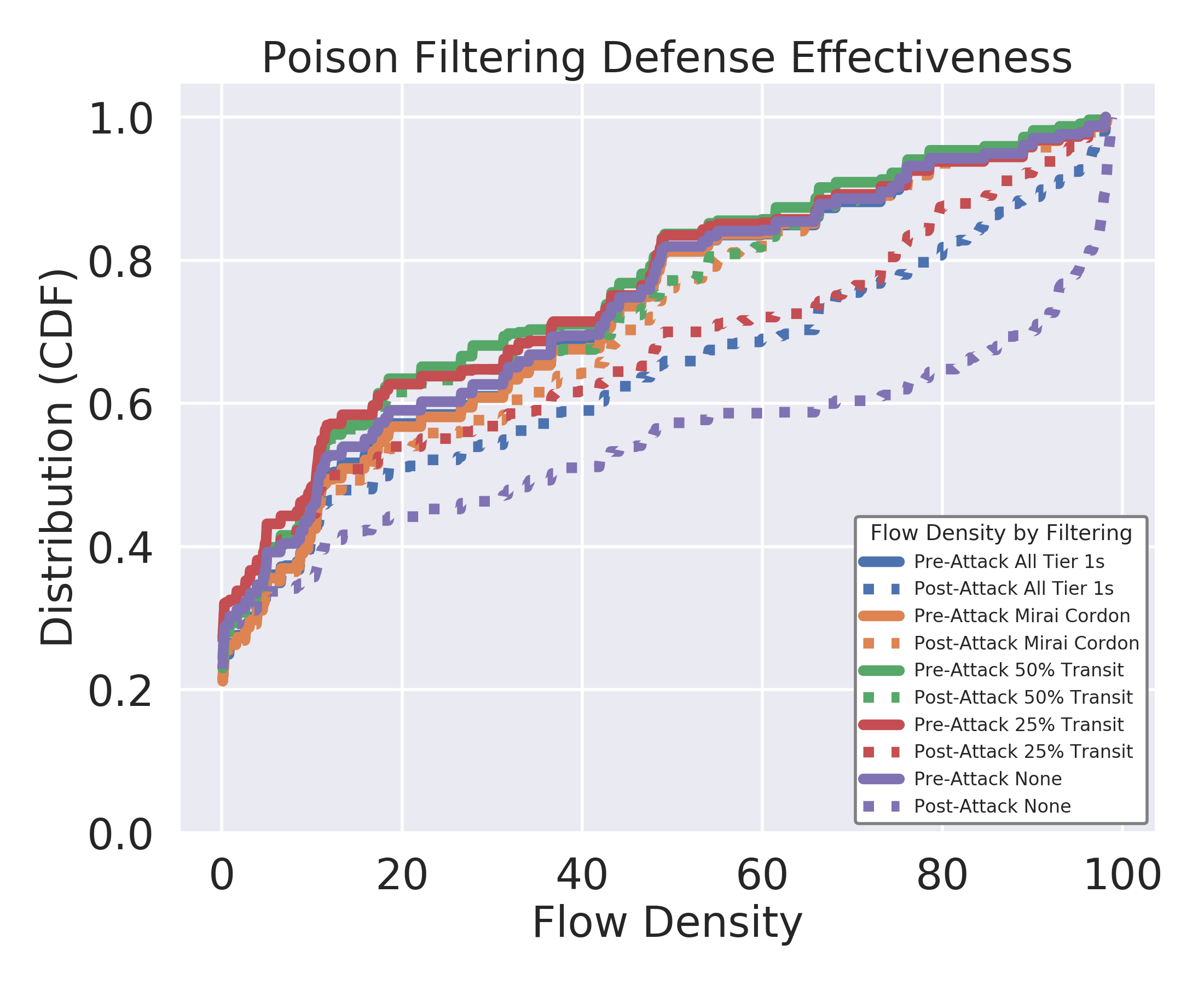}}
    \vspace{-10pt}
    \caption{Evaluation of Maestro defenses}
    \label{fig:mitigation-results}
    \vspace{-10pt}
\end{figure*}
\section{Towards Defenses}{\label{mitigation}}
\subsection{Defenses for Presented Attacks}

\subsubsection{Leak Defense}
Leak attack mitigation is a challenging problem. No current, widely deployed method exists to filter advertised valley paths. Worse, ground-truth AS relationships are not publicly available. In fact, valley paths have been observed on the Internet in practice and studied in prior work~\cite{giotsas2012valley}. The cost of transiting attack traffic over the leaked route could be prohibitive for some adversaries, though this would not dis-incentivize an attacker who has compromised (does not actually own) the AS. The surest defense for upstream providers who wish to defend themselves from this attack is \textit{monitoring advertisements from their customer cone for leaked routes}. In response to suspect advertisements, the defender could respond by disconnecting the violator.

\subsubsection{Maestro Defense}
Two broad categories exist for defense against this attack: general LFA defense solutions and solutions targeting the spread of poisoned announcements. Unfortunately, many state-of-the-art defensive options are not widely available to network operators. Some (like CoDef) require collaboration across ASes, a difficult feat in an Internet architecture that does not encourage cooperation~\cite{lee2013codef}. The next-generation architecture SCION includes mechanisms to solve this and many other routing problems~\cite{sibra2016}; unfortunately, SCION has yet to see broad deployment. Promising SDN-based LFA mitigations have been proposed, but these require infrastructure capabilities currently unavailable to most operators~\cite{kang2016spiffy, liaskos2016novel}. Nyx, a re-routing system itself based on BGP poisoning~\cite{smith2018routing} could partially mitigate Maestro for a single deployer AS and some chosen critical AS, but the target link would still be affected for all other AS pairings.

The second and more relevant category of mitigations target Maestro's poisoned advertisements. Note that Route Origin Authorization would not disrupt this attack, as the compromised AS owns the advertised destination prefixes~\cite{lepinski2012rfc}. BGPSEC, if widely deployed, could prevent this kind of AS PATH tampering; unfortunately, it is not deployed at scale nor effective with partial deployment~\cite{goldberg2014taking}. Detecting or filtering advertisements by individual network operators is the most straightforward approach to countering Maestro. However, some proposed DDoS mitigation~\cite{smith2018routing} and link failure response~\cite{katz2012lifeguard} systems rely on BGP poisoning, and network operators sometimes employ it for traffic engineering~\cite{roughan201110}. Filtering all BGP poisons, then, may have some cost. Alternatively, path length limiting could be an effective mitigation, as each poison required to steer flows increments the path length of the poisoned advertisement. In order to determine which mitigations are most appropriate, we examined these options in the Chaos simulator with the same attack mechanics presented earlier.

As with the peer link, the most effective defense across botnet models is careful monitoring of downstream advertisements. \textbf{The major threat from this attack is from within link endpoints' own customer cone}. While this is a higher cost activity than simply implementing a filtering rule, it means that ASes can act in their own self-interest to defend their own links from Maestro attacks.

\subsection{Poison-Focused Defenses Evaluation}{\label{defense-eval}}

\subsubsection{Path Length Defense}{\label{limiting-defense}}
The path length defense - rejecting advertisements above some limit - is one easy-to-implement response to this attack. Unfortunately, as shown in Fig.~\ref{fig:poison-details}, nearly all of the attack effect is achieved with 5 or fewer poisons; this is partially an artifact of relatively dense botnet distributions as discussed in the next mitigation. Prior work from Tran et al.~\cite{tran2019feasibility} observed AS PATH lengths commonly reach 30 hops in legitimate advertisements, indicating that adding 5 hops to the AS PATH is likely not sufficient to distinguish attack advertisements. So, a defense based strictly on path length is probably not feasible. Fig.~\ref{fig:poison-capping} shows the effect if all ASes limited advertisement AS PATH length at various levels, including those observed by prior work (30 and 75) and implemented in common routing hardware (254)~\cite{cisco}. In short, these limits \textit{do not decrease the achieved flow density}. 

\subsubsection{Poison Filtering Defense}{\label{filtering-defense}}
ASes could also prevent this attack by filtering advertisements with poisons. This is a feasible approach, because poisoned advertisements have a clear signature - AS PATHs that include a list of poisoned ASes between copies of the originator's ASN. If all ASes on the Internet filtered poisons, the Maestro attack would have no effect. However, as previously discussed, this could have some cost on benign traffic engineering. ASes are also generally under unique administrative control, so operator outreach and deployment is a challenge.

Fig.~\ref{fig:poison-filtering} shows how our betweenness-based link selection/customer only adversary selection experiment responds to different sets of filtering ASes. For this mitigation trial, we ran our attack against four filtering sets. The first two sets are composed of 25\% and 50\% of all transit ASes (those with one or more customers) filtering all poisoned advertisements. The next two sets are smaller but more strategically targeted. We include a set of all 20 Tier 1 ASes to quantify how well the largest and most influential providers can protect the Internet as a whole.

Finally, we explore a botnet-specific defense designed from the observation that the bulk of infected hosts in all three botnet models are concentrated in relatively few ASes (for detail on botnet distributions, see Fig.~\ref{fig:botnet-dist} in the appendix). In the Mirai model, for example, 64\% of the botnet is hosted in 30 ASes. So, for our final filtering set, we include all providers (58) for these 30 ASes. In effect, those 58 providers form a poisoned advertisement \textit{cordon} or barrier around the botnet's core to prevent path steering for the majority of infected hosts. While cordoning is a botnet-specific defense, it illustrates how a handful of well-positioned filterers is orders of magnitude more effective per AS deployer than many randomly distributed ones. \textbf{Both the 50\% transit and Mirai cordon filtering sets eliminate virtually all Maestro flow density gains}. Interestingly, filtering at all Tier 1s (20 ASes) provides greater overall reduction in flow density than filtering at 25\% of all transit providers (2441 ASes). Unfortunately, prior work indicates that filtering short poisoned announcements under lengths of 50 ASes rarely occurs in practice~\cite{smith2019internet,tran2019feasibility}.

\subsection{Ethics and Operator Engagement}{\label{operator-feedback}}
As described in Section~\ref{evaluation}, all experiments were performed in simulation; no production networks were adversely affected. Additionally, we submitted a preprint to the NANOG (North American Network Operators Group) mailing list to solicit feedback on the attack and disseminate mitigations. Responses indicated that operators had not seen any similar attack executed in practice. Some operators suggested that the "peer lock" mechanism, widely deployed by the largest providers, could provide some protection from the attack~\cite{peerlock}. Hurricane Electric lists peer locking as an upcoming feature in their filtering algorithm~\cite{hepeerlock}. Peer locking validates advertisements against known relationships, an intuitive step in averting path manipulations like those used in Maestro. However, this requires periodic out-of-band information exchange between ASes, and it is unclear how far this feature has penetrated beyond Tier 1 ASes. We leave to future work a full quantification of this tool's impact on Maestro, though note that our Tier 1 filtering examination from Fig.~\ref{fig:poison-filtering} should provide a similar result if peer locking deployment is limited mostly to Tier 1 providers.
\section{Related Work}{\label{related}}
The Coremelt~\cite{studer2009coremelt} and Crossfire~\cite{kang2013crossfire} attacks are discussed in detail in the background, Section~\ref{lfas}. Classifying links by BGP betweenness is a technique employed by in Schuchard et al.'s Losing Control of the Internet (LCI) attack on the BGP control-plane~\cite{schuchard2010losing}. Interestingly, the LCI attack used the control-plane to attack the data-plane; here, we leverage the control-plane to augment a data-plane attack. LFA mitigation work that applies to this attack is presented earlier during the mitigation discussion, Section~\ref{mitigation}.

Other uses of BGP poisoning include LIFEGUARD from Katz-Bassett et al.~\cite{KatzBassett:2010wp,katz2012lifeguard} as well as Anwar et al.'s policy exploration~\cite{Anwar:2015tk}. Nyx~\cite{smith2018routing} from Smith et al. employs BGP poisoning for DDoS mitigation. In~\cite{colitti2006internet}, Colitti et al. use poisoning for route discovery similar to Anwar et al.~\cite{Anwar:2015tk}. The propagation of poisoned advertisements throughout the Internet is actively measured in~\cite{smith2019internet}, which informs us regarding how BGP operators respond to such advertisements in practice.
\section{Conclusion}{\label{conclusion}}
In this work we explored both the limitations of LFAs launched by botnets, and approaches adversaries have to overcome these limitations with control of a BGP speaker.  Our experiments showed that contrary to assumptions in previous literature, botnet-sourced LFAs cannot target arbitrary links with full force in practice. In fact, many core Internet links can be reached by just a fraction of infected hosts in all three of our botnet models. Our simulations show that the Maestro attack can partially overcome these reach restrictions. Most troublingly, high betweenness links (core) are most vulnerable to this attack, and the rank of AS adversaries plays little role in attack success. Provider to customer targets are far more vulnerable to a Maestro attack, but customer to provider and peer links with significantly infected From AS customer cones can be affected by Maestro, as well. Additionally, both could be subject to leak attacks that expose them to malicious flows.  Our exploration of defenses suggests poison filtering at topologically strategic points drastically reduces vulnerability. \\

\noindent \textbf{Future Work}: \textit{1)} Improvements to our poison scoring heuristic could be explored. The version we employed for experiments weighs all source ASes equally when making poisoning decisions; infected hosts, on the other hand, are not uniformly distributed throughout the Internet topology. \textit{2)} The use of multiple prefixes independently could improve adversarial success by eliminating poison set conflicts, and could be used to attacking multiple links for a more sophisticated isolating LFA. \textit{3)} Botnets are not the only sources of malicious flows. Maestro can be used to steer any attack flows directed to the adversary AS; this could include normal inbound traffic for a large adversary AS, or reflection-generated flows~\cite{majkowski2018memcrashed}.


\appendix

\section{Optimal Poison Choice}{\label{npcomplete-formalization}}
We do not utilize this method in practice because it has high runtime complexity and does not exploit the specific structure of our problem.

We can solve for the optimal poison set by re-framing the problem as MAX-SAT, a generalization of boolean satisfiability (SAT) where we seek to assign truth values to variables in order to maximize the number of satisfied clauses rather than achieve complete boolean formula satisfaction~\cite{vazirani2013approximation}. Consider that each source AS has a set of poison sets $S$ that map to resulting paths $P$ over the target link to the adversary AS, where each set $s$ $\in$ $S$ corresponds to a resulting path to the poisoning prefix $p$ $\in$ $P$ (that is, $s$ $\mapsto$ $p$ for that source AS). This signifies that if the adversary chooses a poison set that contains all of the ASes in $s$ and none of the ASes in $p$, the source AS will shift onto path $p$. Note that, depending on the adversary AS and target link position relative to the source AS, $S$ and $P$ may be empty; in that case, there is no way to steer the source AS onto the target link.

In the boolean formula for a problem instance, the variables will be the ASes in the topology. We can define the structure of the boolean formula by building a clause for each source AS thusly: an AS appearing in a poison $s$ is represented by its AS variable, and the ASes in the resulting path $p$ are represented by the inversion of their AS variables. These variables are joined conjunctively, along with the source AS itself; if it is poisoned, it will of course not have a path to the poisoned prefix. We disjunctively join each source AS's conjunctive clauses, one for each $s$ $\mapsto$ $p$, to form a clause in disjunctively normal form for that source AS. This clause is the boolean representation of the poison choices we must make to bring the source AS onto the target link. 

Finally, we join the all source AS clauses by conjunction, and we have defined a boolean formula that describes how our poison choices will affect the paths of ASes containing bots to the adversary. While we do not reformulate the problem in conjunctive normal form, it is always possible to do so~\cite{kalish1964logic}.

Unfortunately, MAX SAT is APX-complete; no efficient algorithm can solve it, and no polynomial time approximation scheme can be devised (unless $P=NP$)~\cite{cohen2004complete}. This is problematic because exploring how the relative topological position of the adversary, target, and flow sources requires the simulation of thousands of attacks. We also note that while we can perfectly map poison choices to resultant paths in the Chaos simulator, AS policy choices are not publicly known and can be changed at any time. Without this information, the "ideal" poison set cannot be computed. So, this optimal choice algorithm is unused in our experiments.

\begin{figure*}[]
	\centering
	\includegraphics[width=0.65\textwidth]{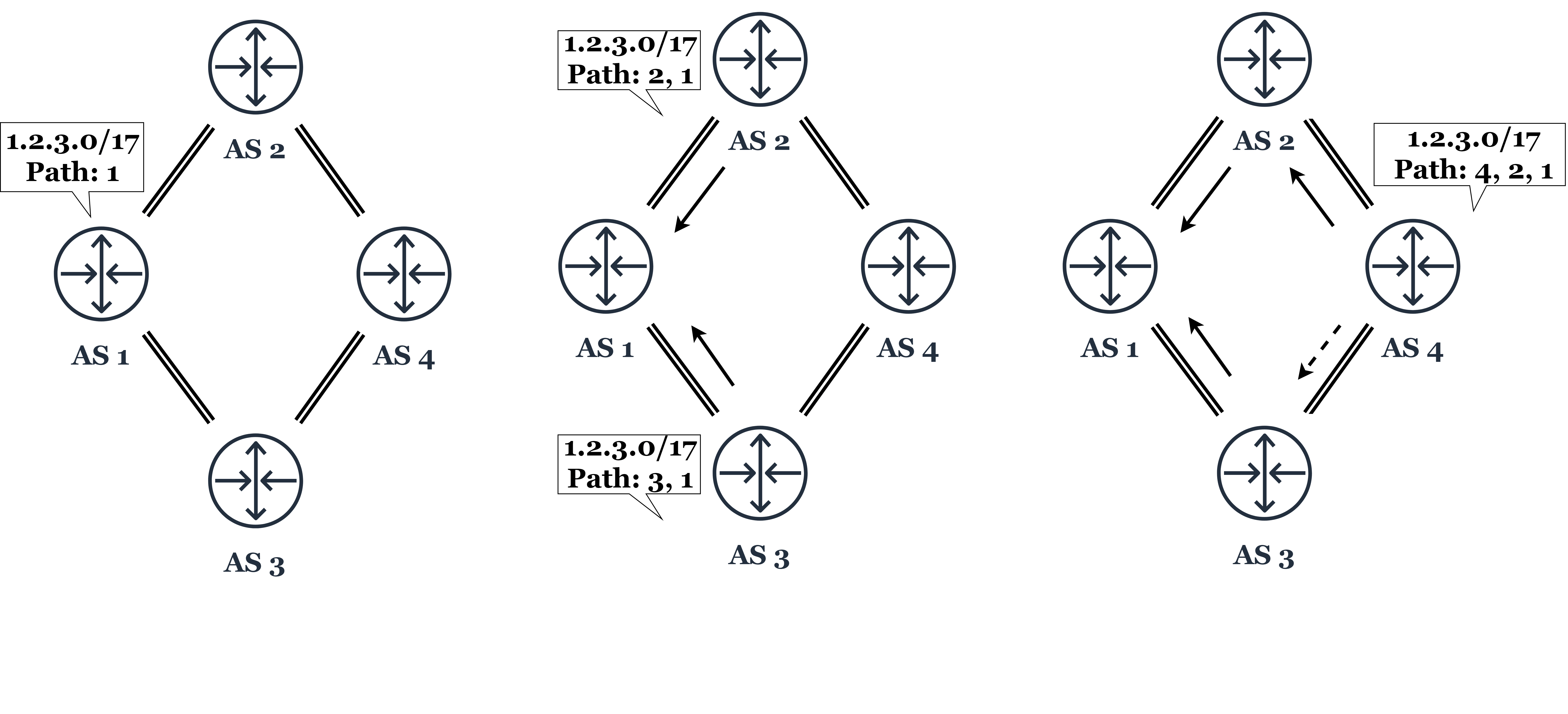}
	\caption{BGP routes built iteratively as they are propagated by neighboring ASes. Since 4 chooses path \{2, 1\} to reach 1, path \{1, 3, 4\} is not exported.}{\label{fig:simple_bgp}}
\end{figure*}

\begin{figure}[]
	\centering
	\includegraphics[width=0.70\columnwidth]{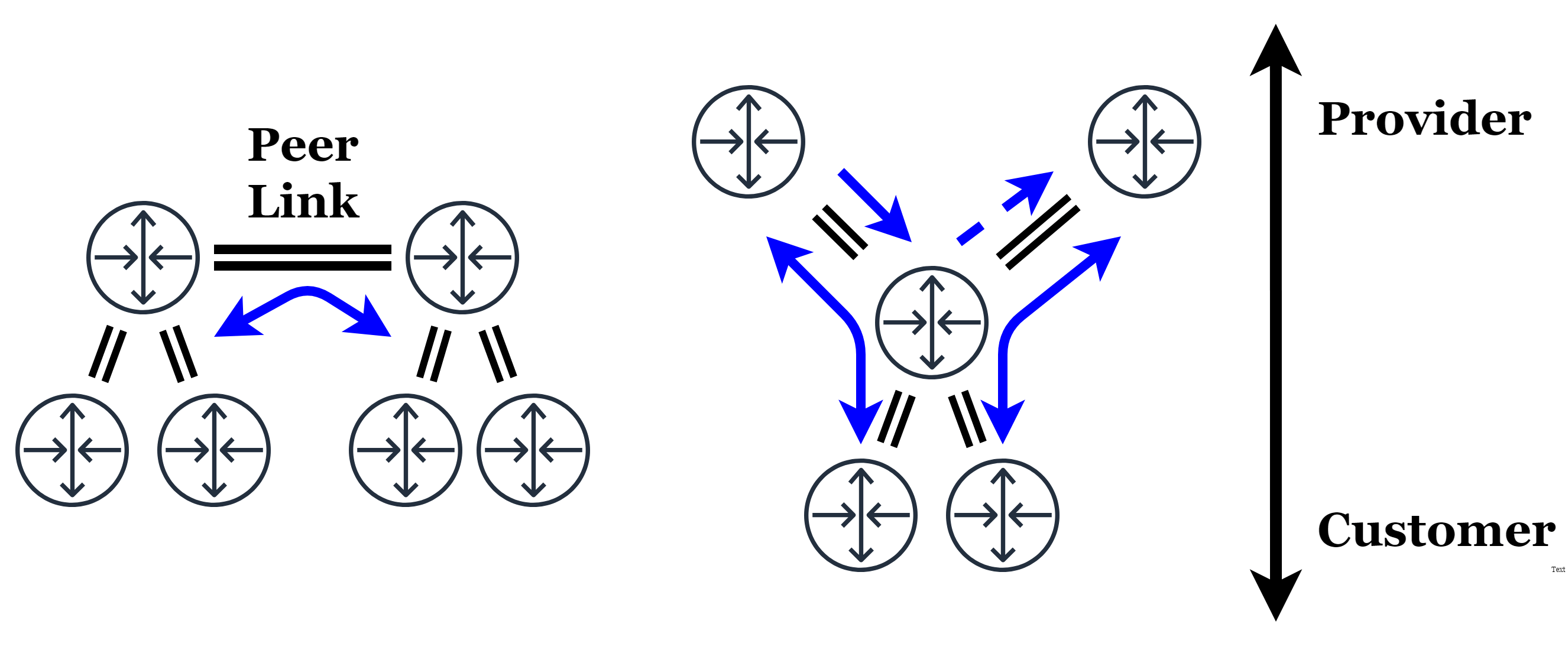}
	\caption{Valley free routing: ASes inform customers of all paths, but do not transit traffic for providers}{\label{fig:bgp_rel}}
\end{figure}

\begin{figure}[]
    \centering
    \includegraphics[width=0.70\columnwidth]{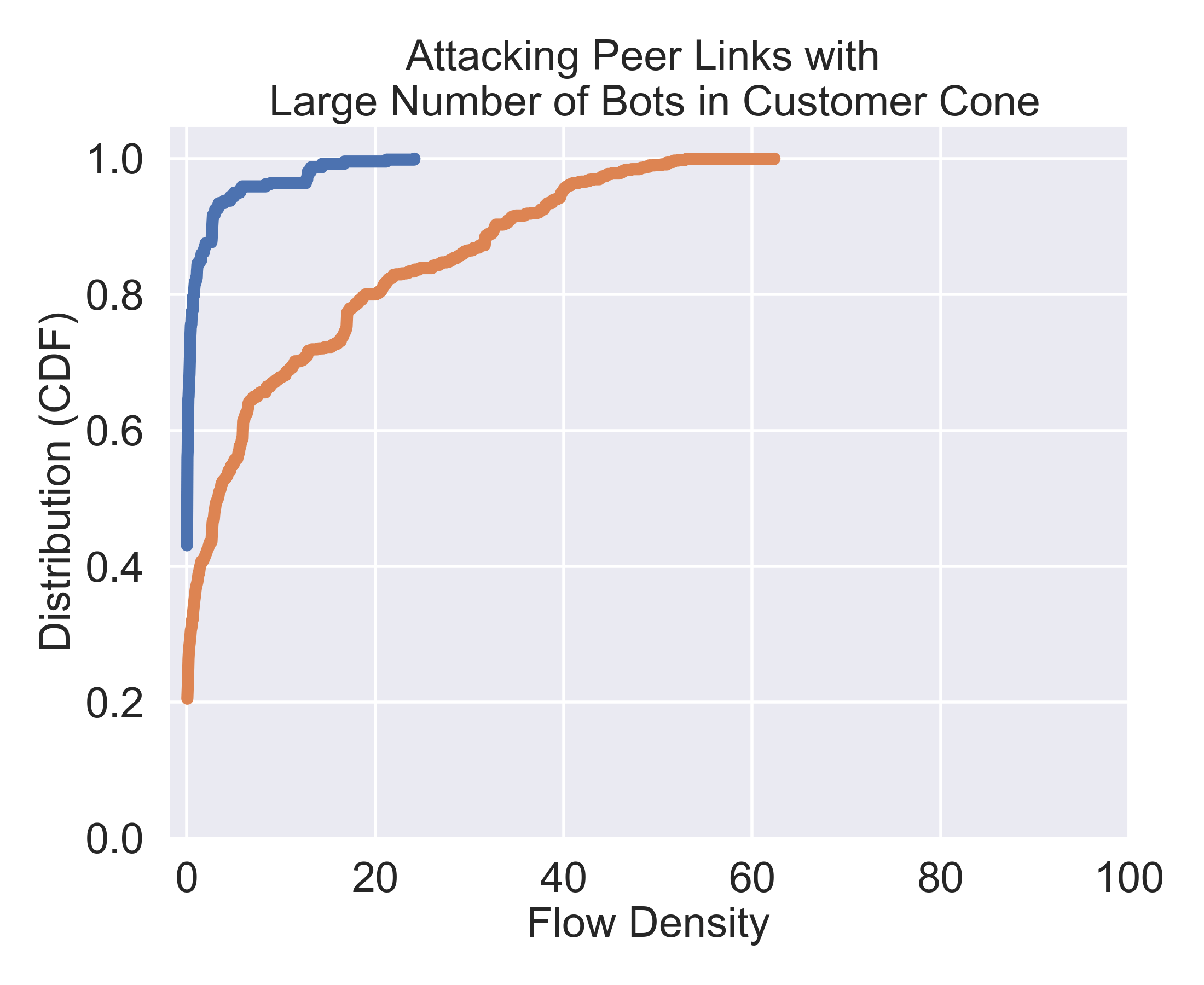}
    \caption{Attacking Peers with Bot-Focused Topological Positioning}
    \label{fig:high-cust-cone-peer}
\end{figure}

\begin{figure}[]
    \centering
    \includegraphics[width=0.70\columnwidth]{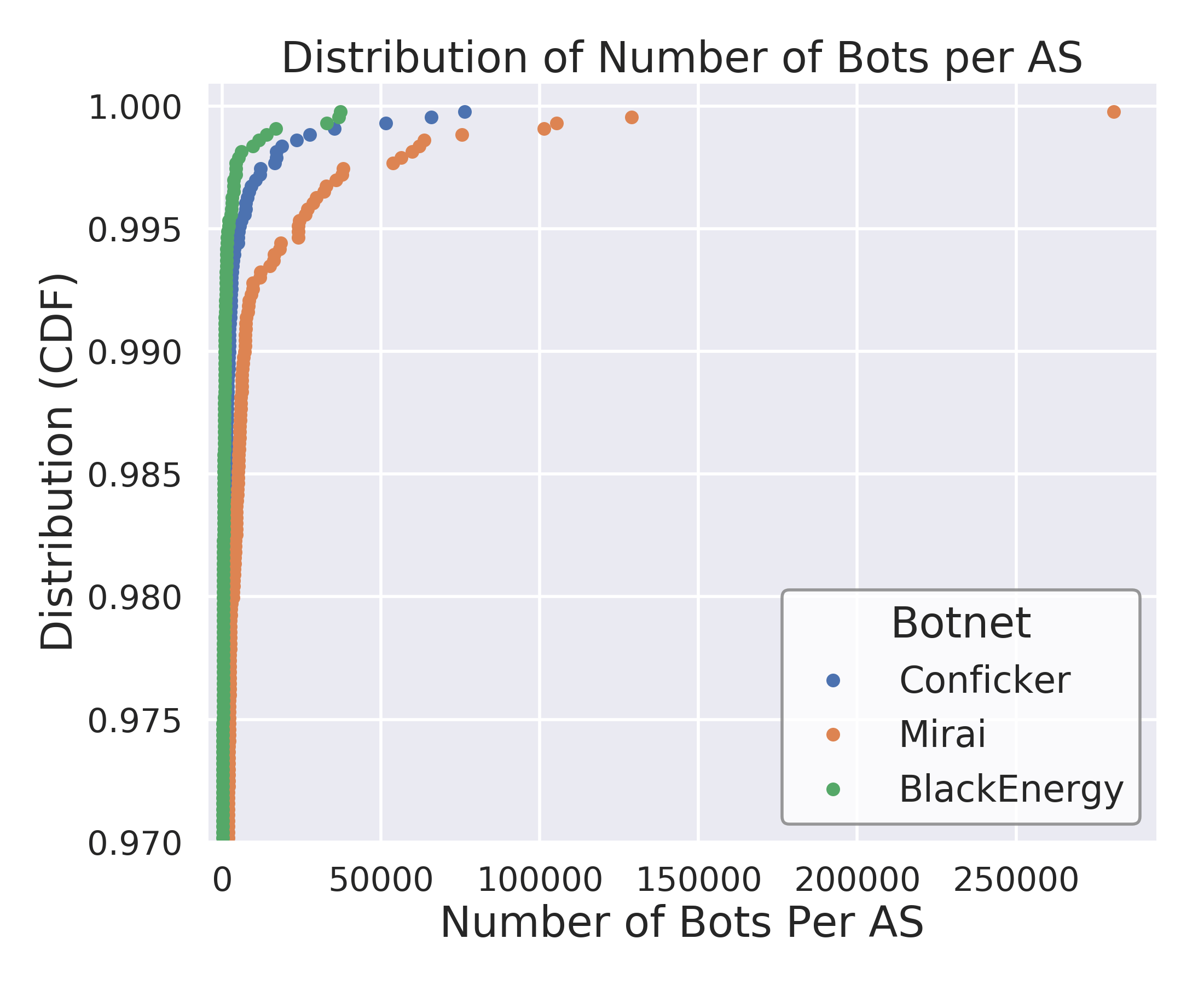}
    \caption{Botnet distributions: Mirai, Conficker, and BlackEnergy}
    \label{fig:botnet-dist}
\end{figure}

\section{Example Attack}
\begin{figure*}[h]
	\centering
	\subfloat[Topology at start with target link 22 $\rightarrow$ 31 and adversary 60 in red.]{\label{fig:example1}\includegraphics[width=0.31\textwidth]{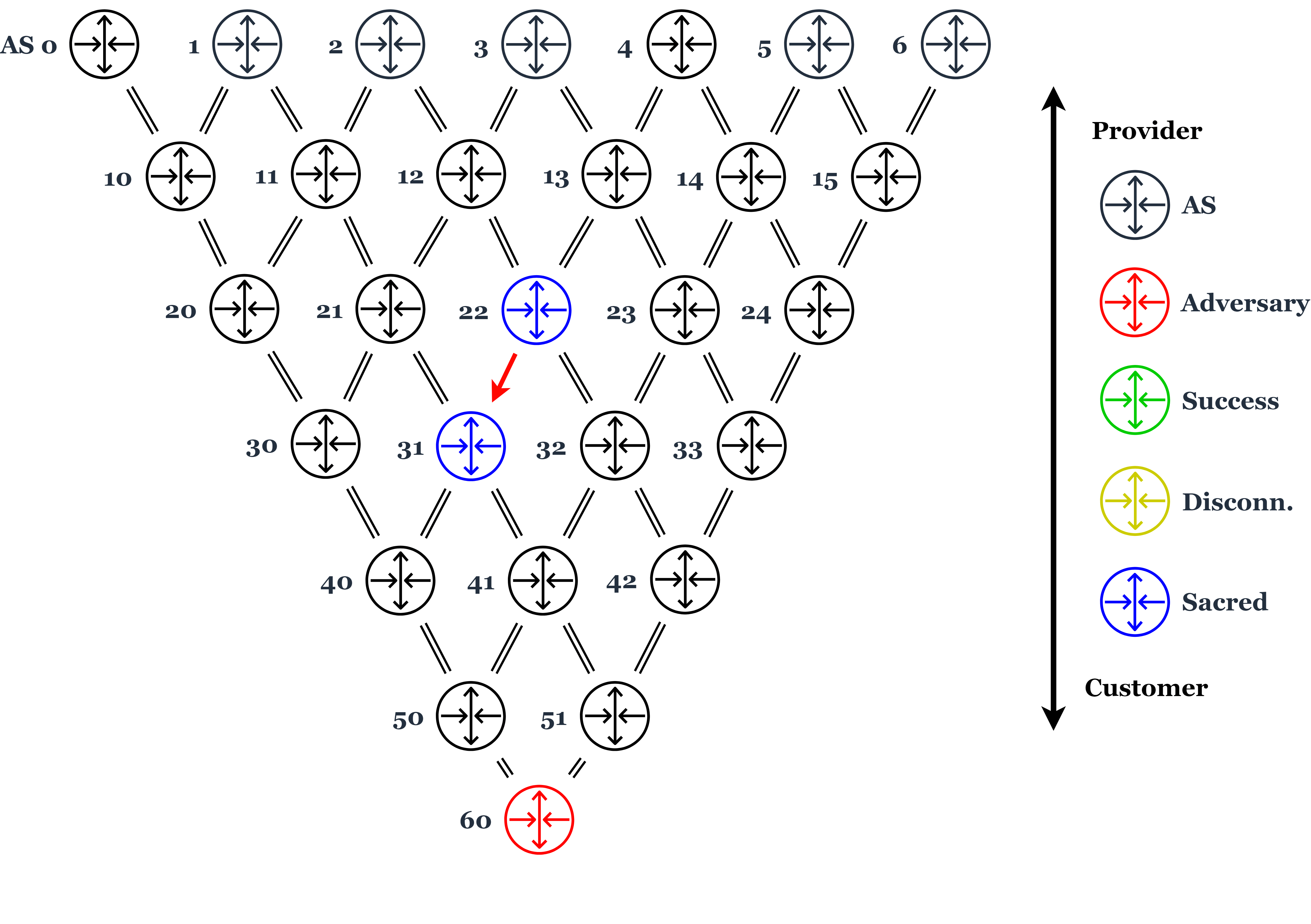}}
	\hspace{0.50cm}
	\subfloat[ASes prefer customer routes, shorter paths, and lower ASNs. \{4, 13, 22\} on link at start; 50 marked sacred, 31 not advertised a path to 60 without it. Select \textbf{40} to poison; most left side ASes transit it to 60.]{\label{fig:example2}\includegraphics[width=0.31\textwidth]{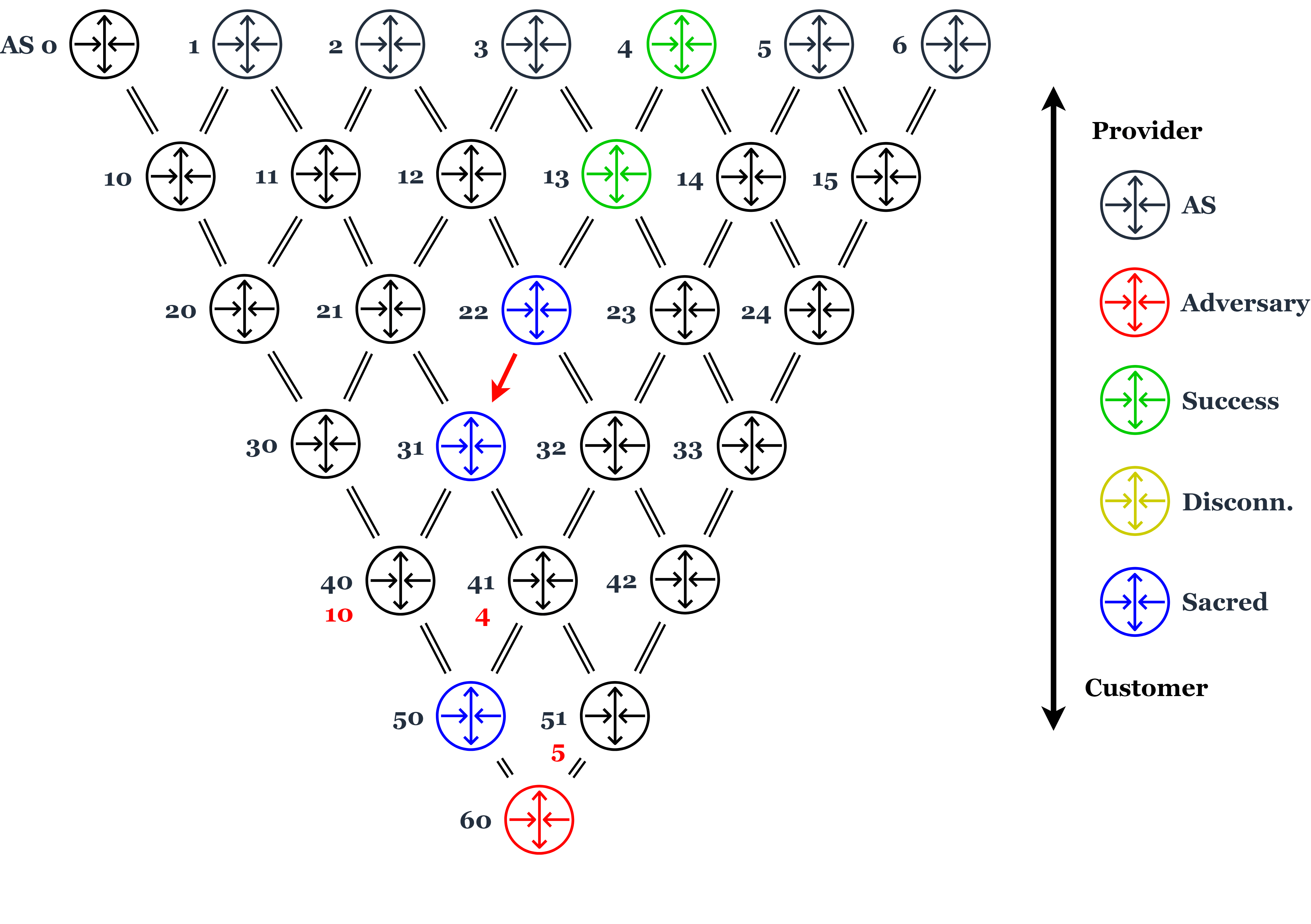}}
    \hspace{0.50cm}
	\subfloat[This isolates 0; 0 has no valley free path to 60 without 40. No ASes move onto link after first poison. Most of top left now channeled through 21. Add \textbf{21} to poison set: \textbf{\{40, 21\}}. ]{\label{fig:example3}\includegraphics[width=0.31\textwidth]{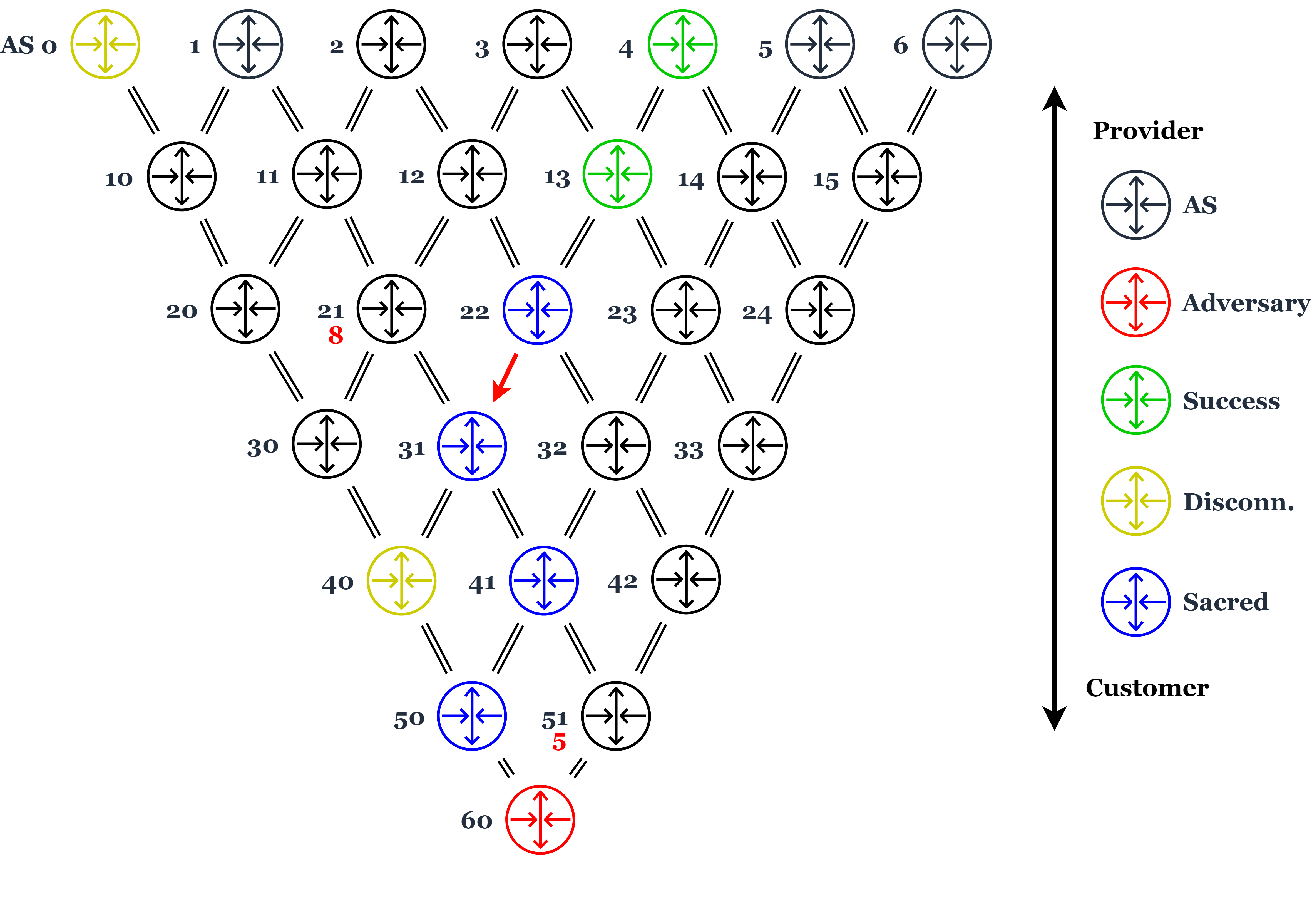}}
	\\
	\subfloat[Left side ASes on link or disconnected. Poison: \textbf{51}. Poison set: \textbf{\{40, 21, 51\}}.]{\label{fig:example4}\includegraphics[width=0.31\textwidth]{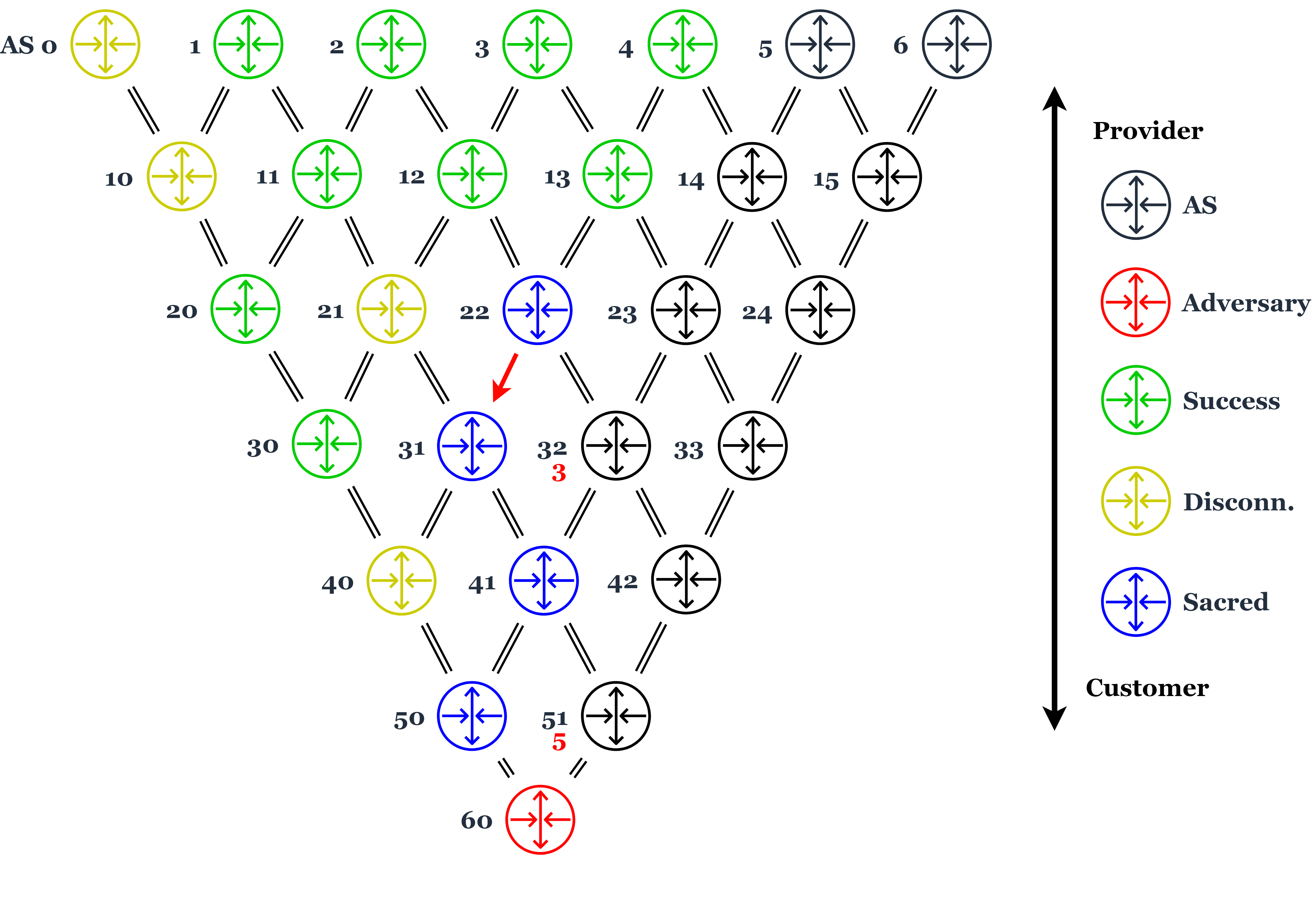}}
    \hspace{0.50cm}
	\subfloat[32 is the next poison choice. \textbf{\{40, 21, 51, 32\}}.]{\label{fig:example5}\includegraphics[width=0.31\textwidth]{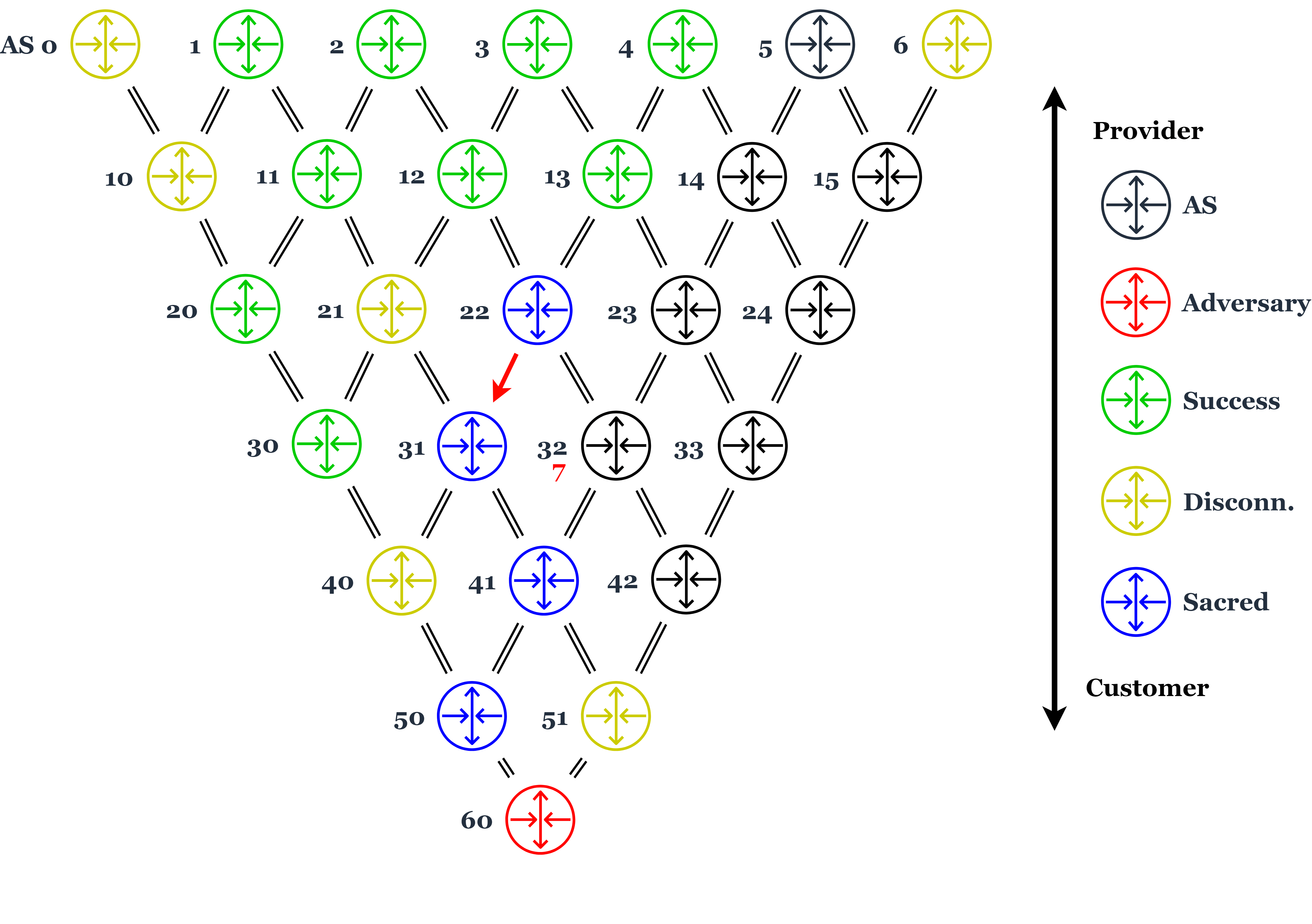}}
	\hspace{0.50cm}
	\subfloat[We reach the termination condition: all ASes either transit link to adversary, or are disconnected/sacred. \textbf{14} ASes' traffic on target, compared to \textbf{3} initially.]{\label{fig:example6}\includegraphics[width=0.31\textwidth]{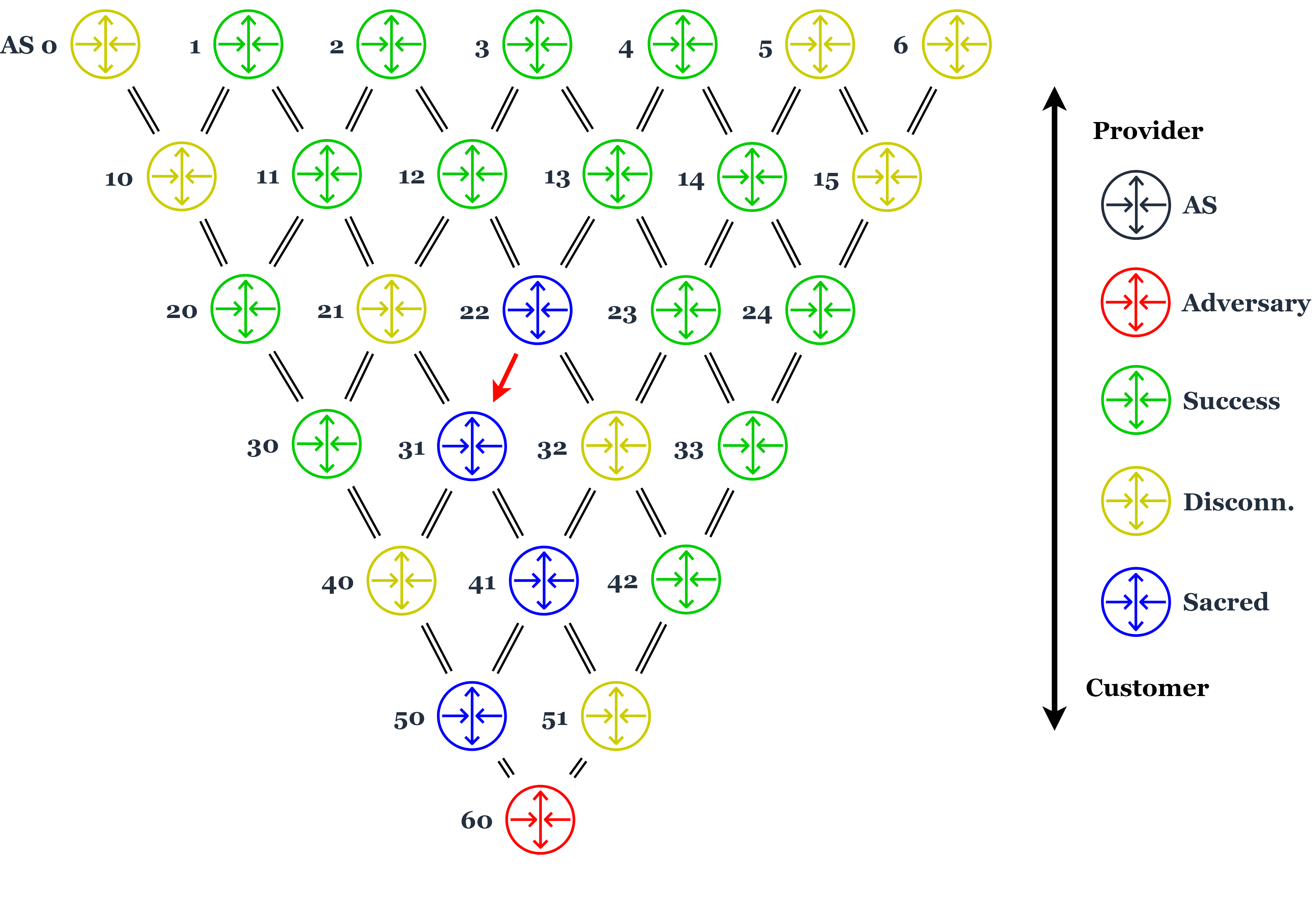}}
	\begin{center}
		\caption{Example Poison Scoring for Attack}
		{\label{fig:frrp}}
	\end{center}
\end{figure*}

{\label{example-subsection}}
To further elucidate our heuristic approach, we present an example attack on a small toy topology in Fig~\ref{fig:frrp}. ASes above (closer to the top) of the figure provide for the linked ASes below them. Each subfigure displays the results at one iteration (after the initial depiction of the topology), with the betweeenness-based poison score displayed in red for select high score ASes. 

\section{Supplementary Evaluations}
Here we present similar experiments to Section~\ref{evaluation} for other botnet models - Conficker and Blackenergy. These results generally exhibit the same patterns as those found in that section.

\begin{figure*}[p]
    \centering
    \subfloat[Betweenness-based link selection for Conficker]{\label{fig:conficker-lci-all-deciles}\includegraphics[width=0.26\textwidth]{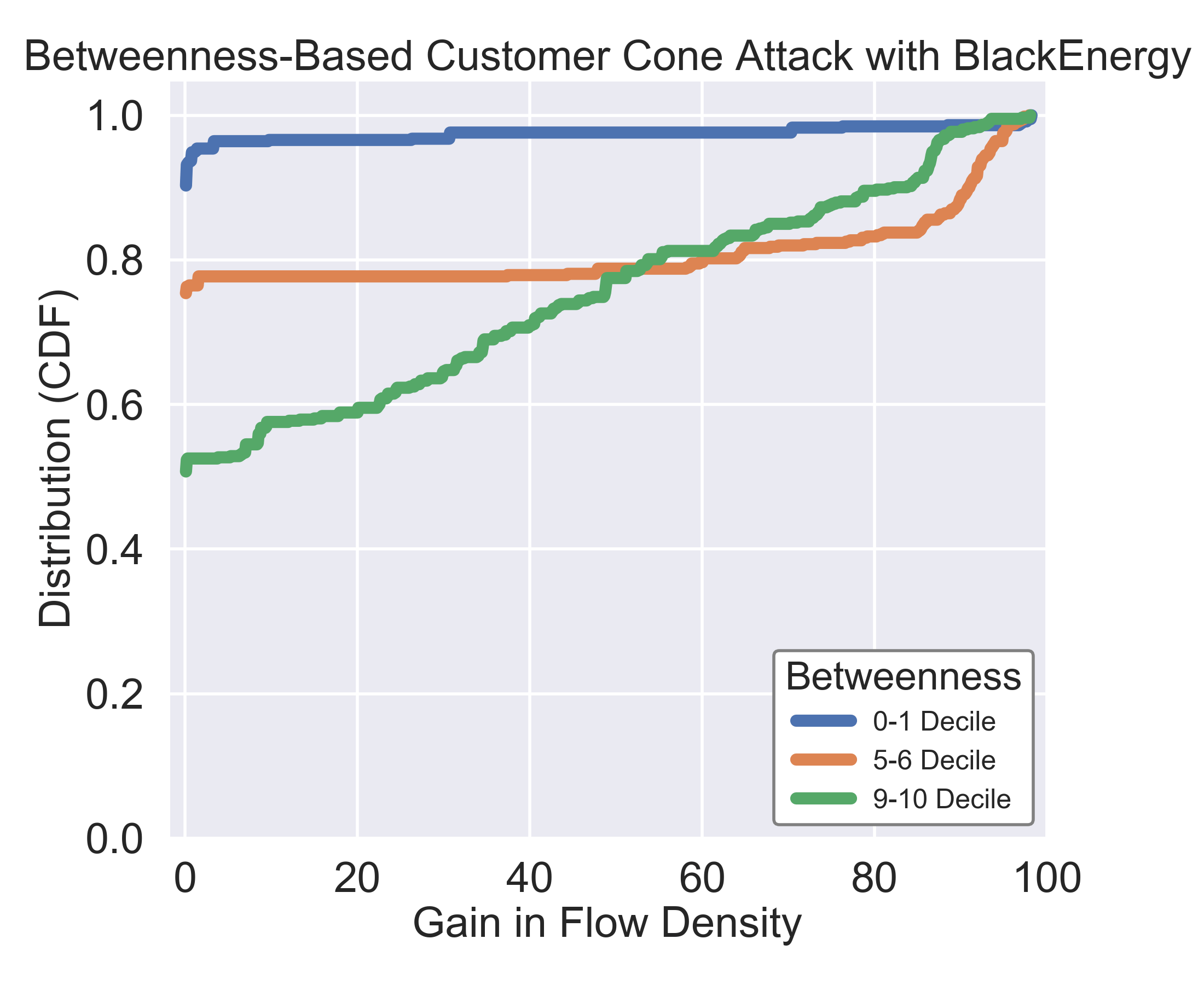}}
    \hspace{0.50cm}
    \subfloat[Bot-to-bot (Coremelt) link selection for Conficker]{\label{fig:conficker-core-all-deciles}\includegraphics[width=0.26\textwidth]{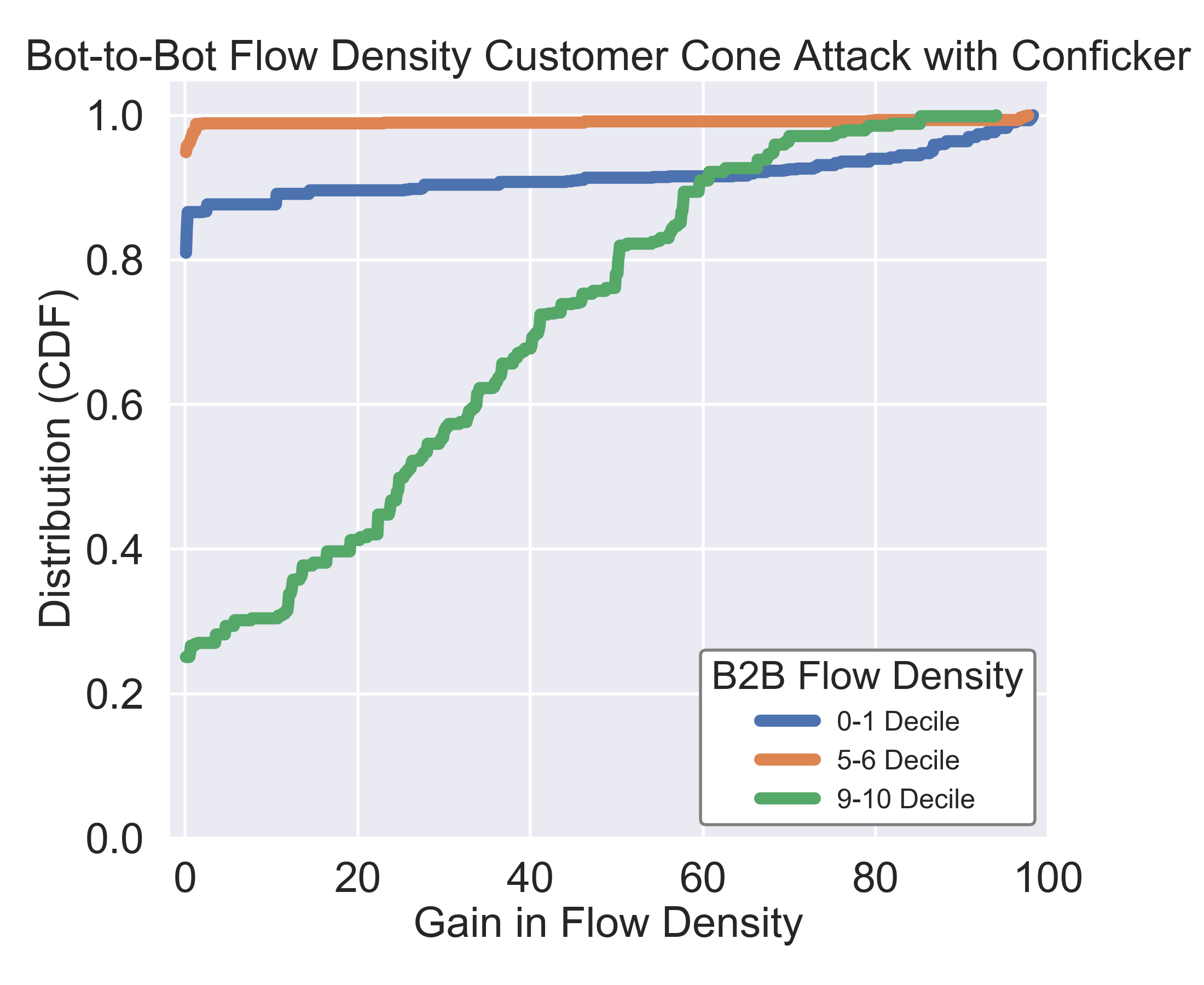}}
    \hspace{0.50cm}
    \subfloat[Bot-to-any (Crossfire) link selection for Conficker]{\label{fig:conficker-cross-all-deciles}\includegraphics[width=0.26\textwidth]{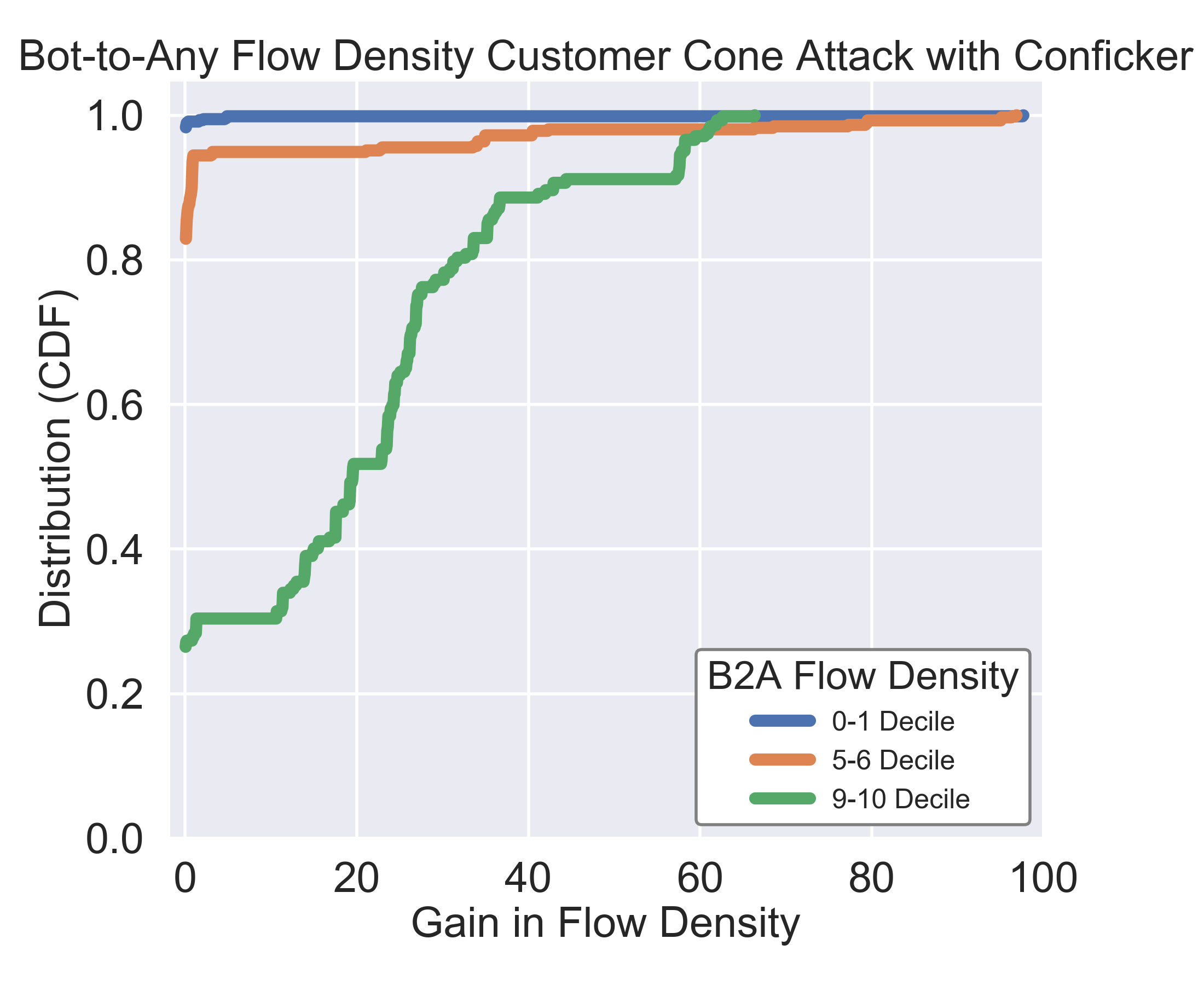}}
    \\
    \subfloat[Betweenness-based link selection for Blackenergy]{\label{fig:blackenergy-lci-all-deciles}\includegraphics[width=0.26\textwidth]{graphs/appendix/cust/customer-region-lci-blackenergy-flow-density-decile_cdf.png}}
    \hspace{0.50cm}
    \subfloat[Bot-to-bot (Coremelt) link selection for Blackenergy]{\label{fig:blackenergy-core-all-deciles}\includegraphics[width=0.26\textwidth]{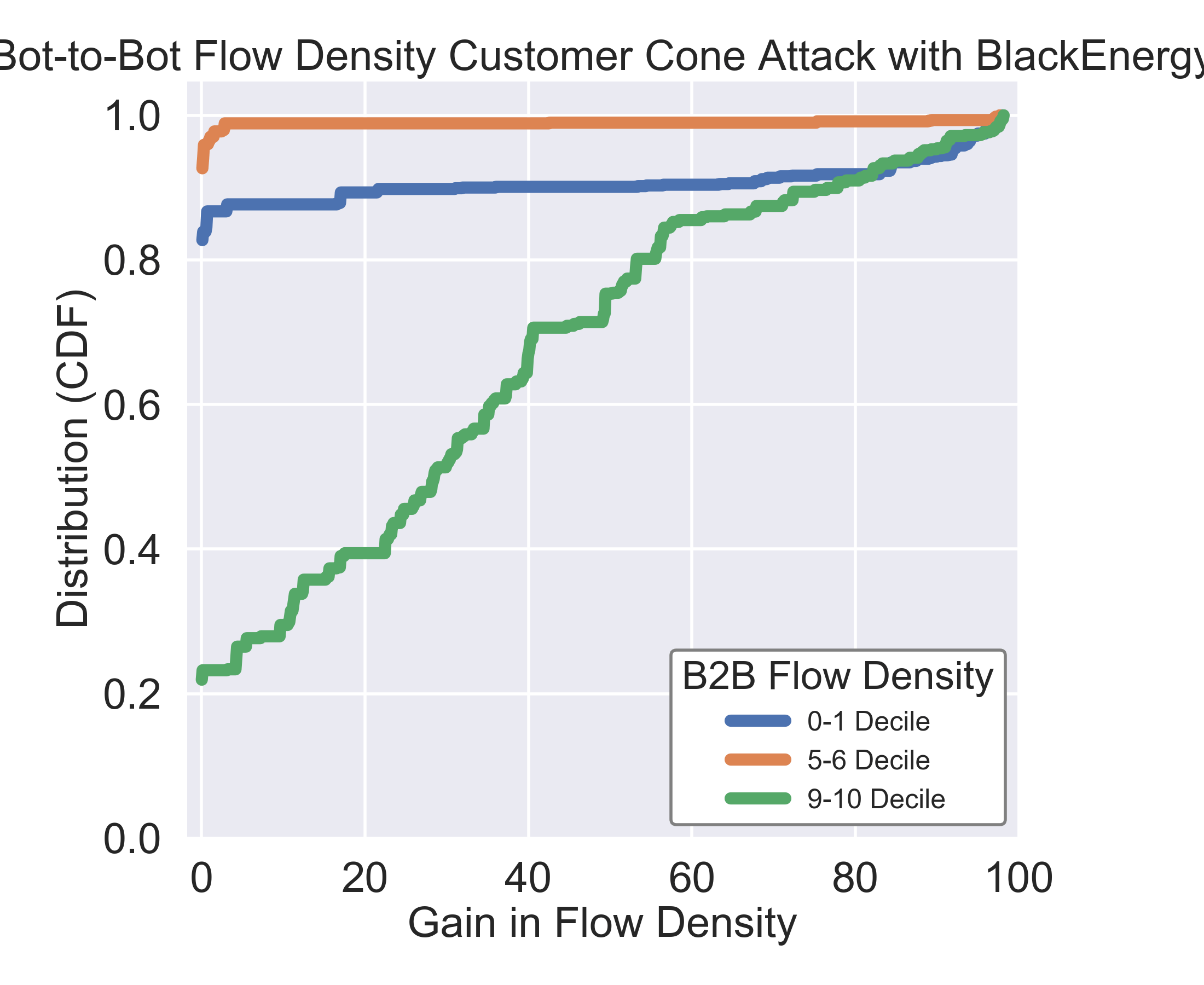}}
    \hspace{0.50cm}
    \subfloat[Bot-to-any (Crossfire) link selection for BlackEnergy]{\label{fig:blackenergy-cross-all-deciles}\includegraphics[width=0.26\textwidth]{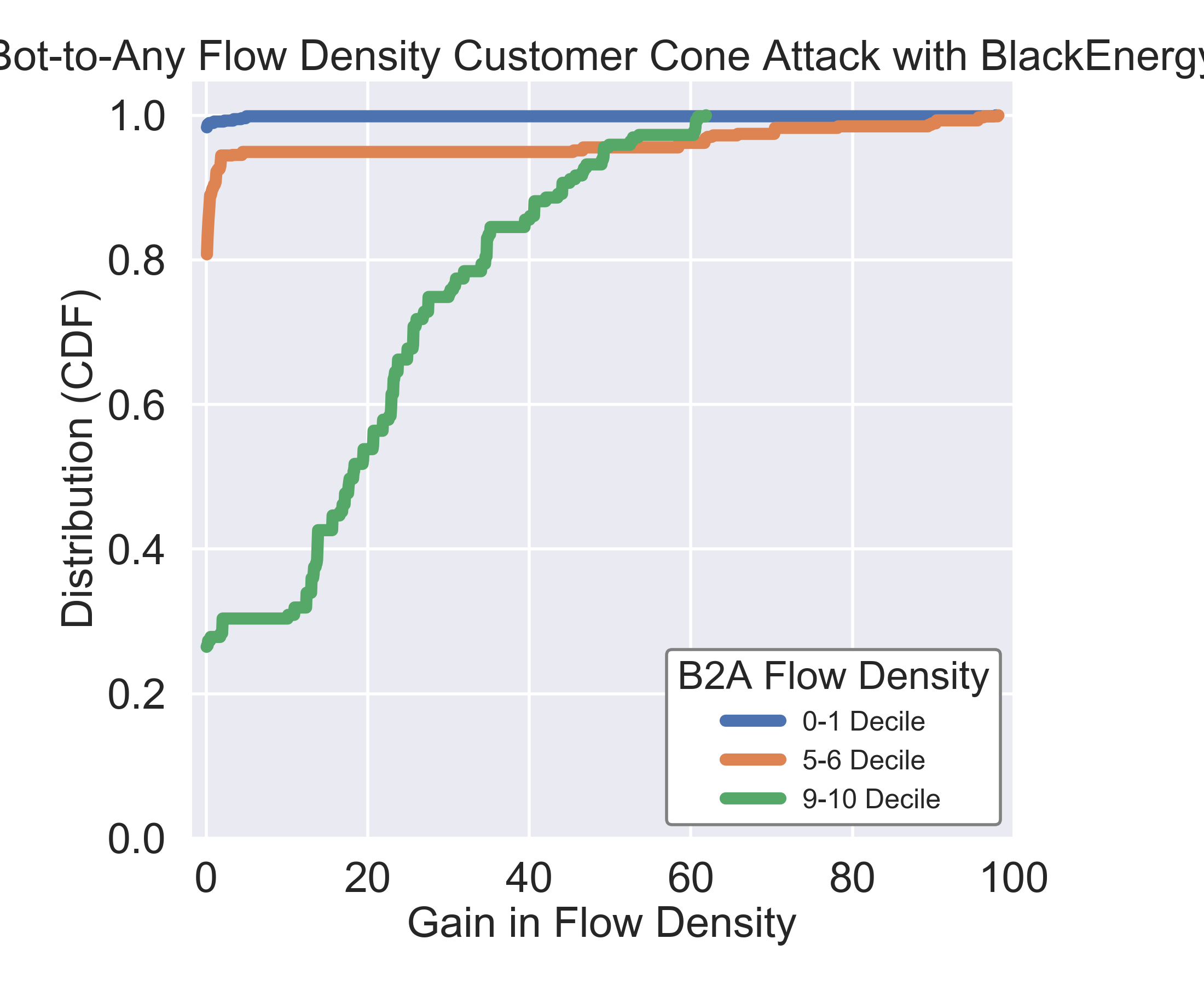}}
    \begin{center}
        \caption{Maestro Success in Flow Density Gain (post flow density - pre flow density) for Conficker and BlackEnergy Botnet Models for all three link-selection strategies (across all deciles)}
        {\label{fig:other-botnets-all-deciles}}
    \end{center}
    \vspace{-20pt}
\end{figure*}

\begin{figure*}[p]
    \centering
    \subfloat[Betweenness-based link selection for Conficker]{\label{fig:conficker-lci-prepost}\includegraphics[width=0.26\textwidth]{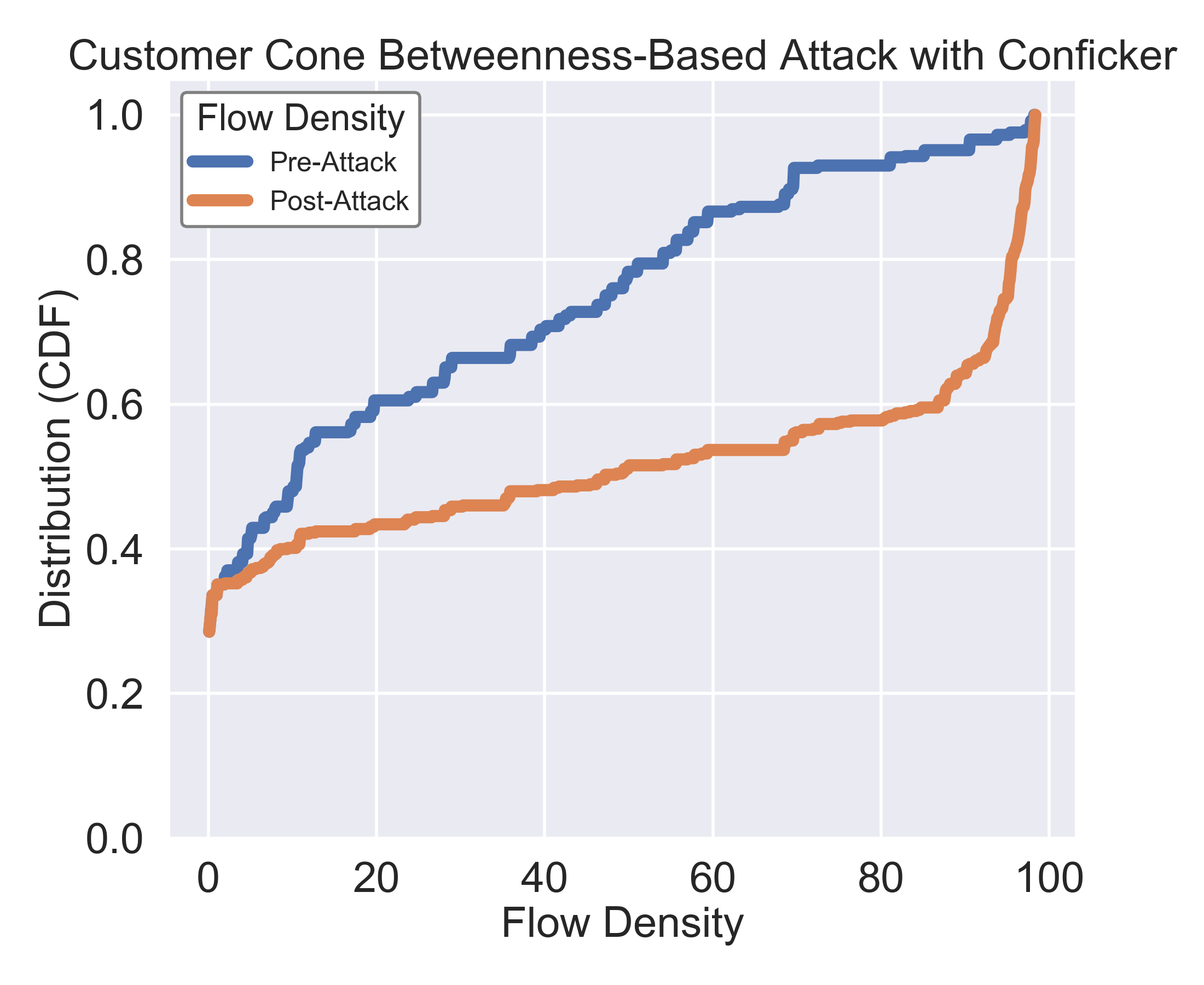}}
    \hspace{0.50cm}
    \subfloat[Bot-to-bot (Coremelt) link selection for Conficker]{\label{fig:conficker-core-prepost}\includegraphics[width=0.26\textwidth]{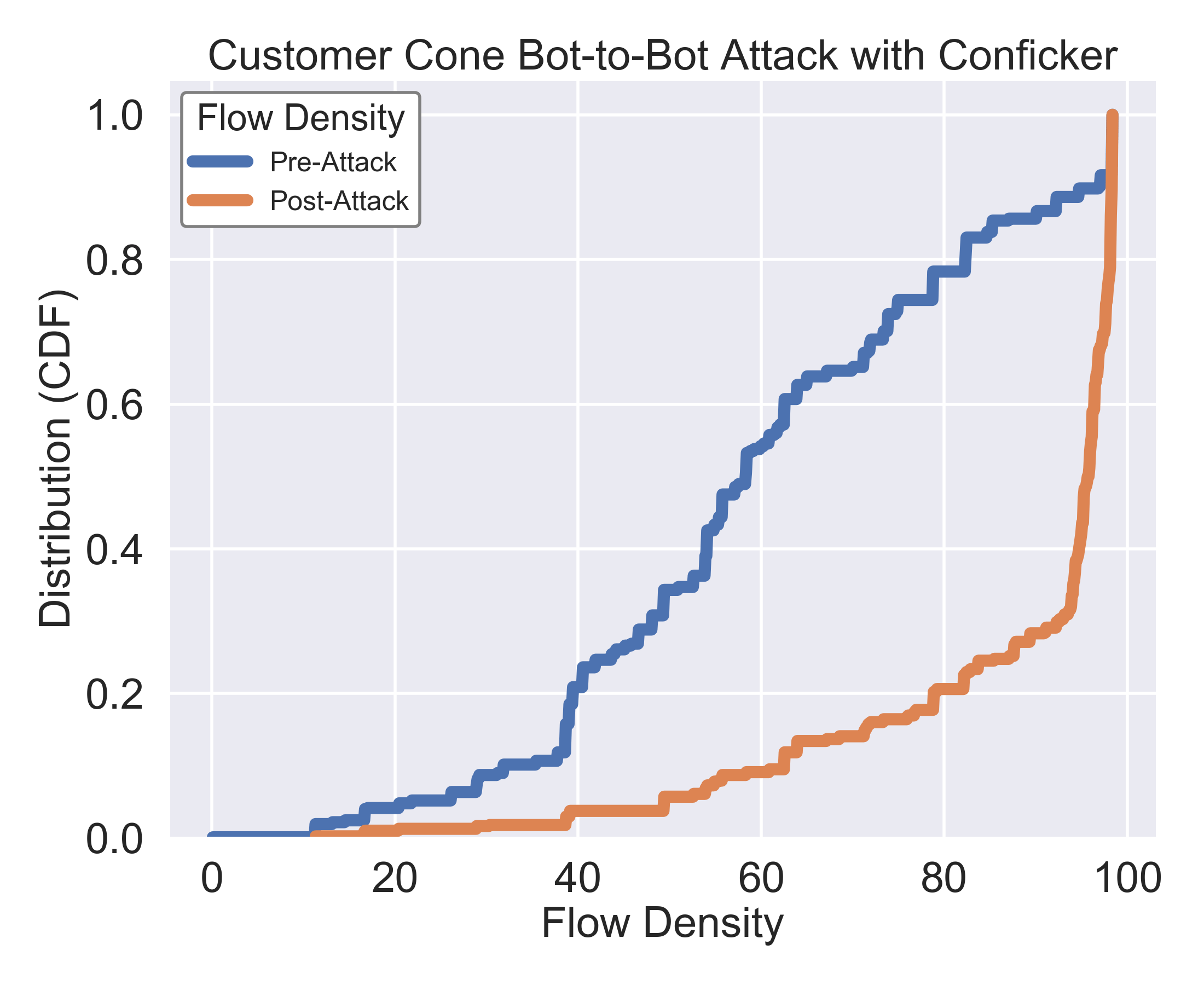}}
    \hspace{0.50cm}
    \subfloat[Bot-to-any (Crossfire) link selection for Conficker]{\label{fig:conficker-cross-prepost}\includegraphics[width=0.26\textwidth]{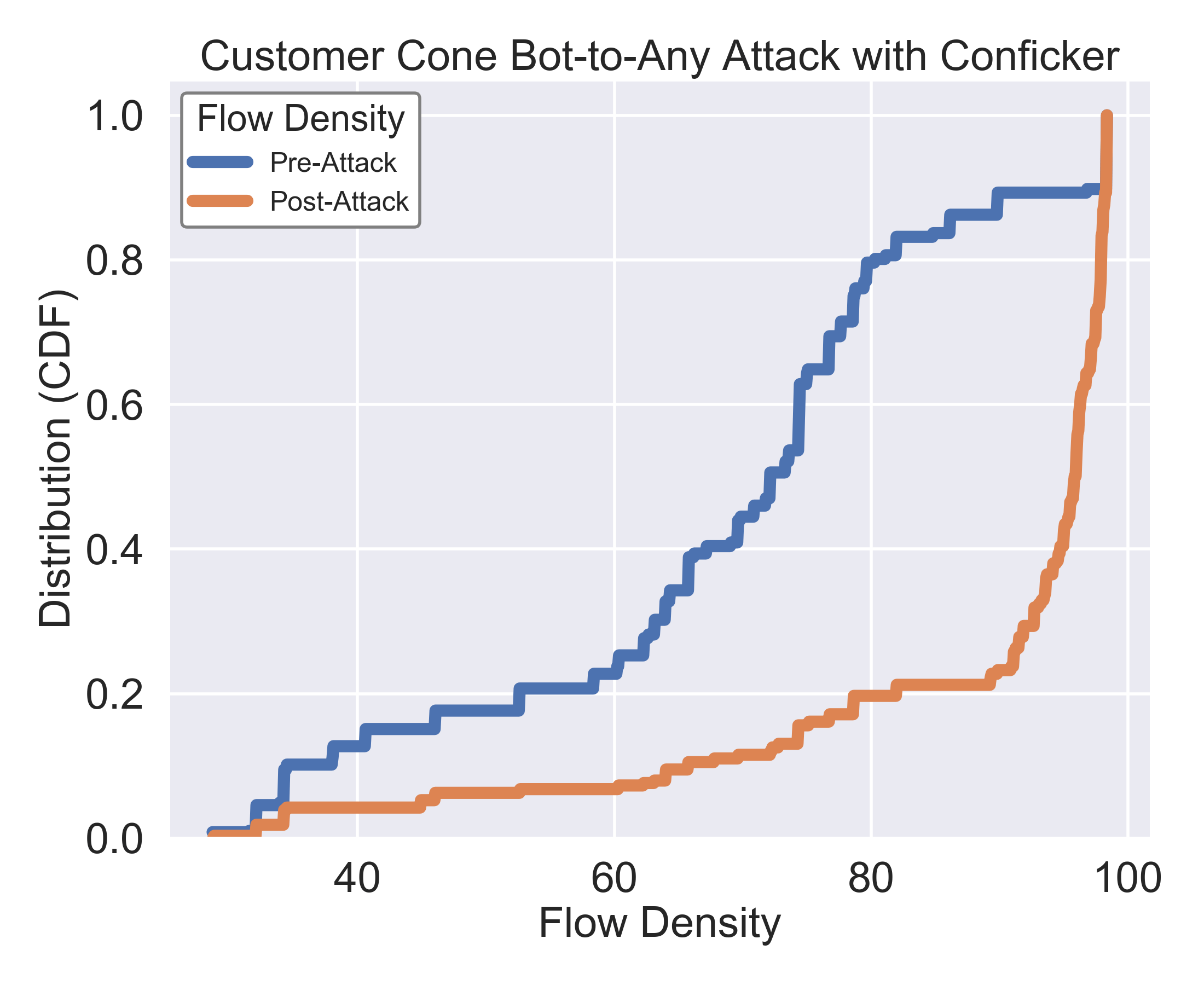}}
    \\
    \subfloat[Betweenness-based link selection for Blackenergy]{\label{fig:blackenergy-lci-prepost}\includegraphics[width=0.26\textwidth]{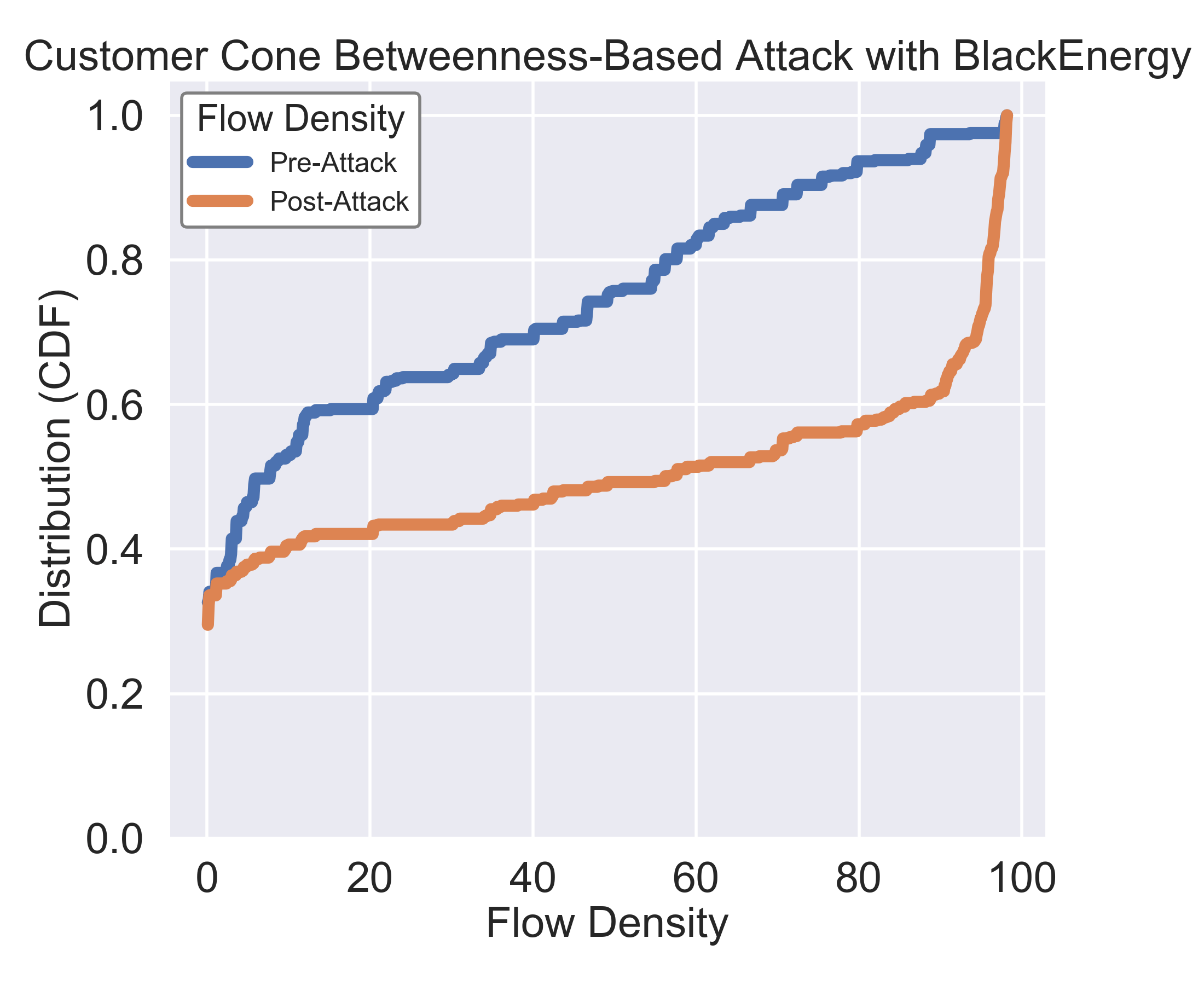}}
    \hspace{0.50cm}
    \subfloat[Bot-to-bot (Coremelt) link selection for Blackenergy]{\label{fig:blackenergy-core-prepost}\includegraphics[width=0.26\textwidth]{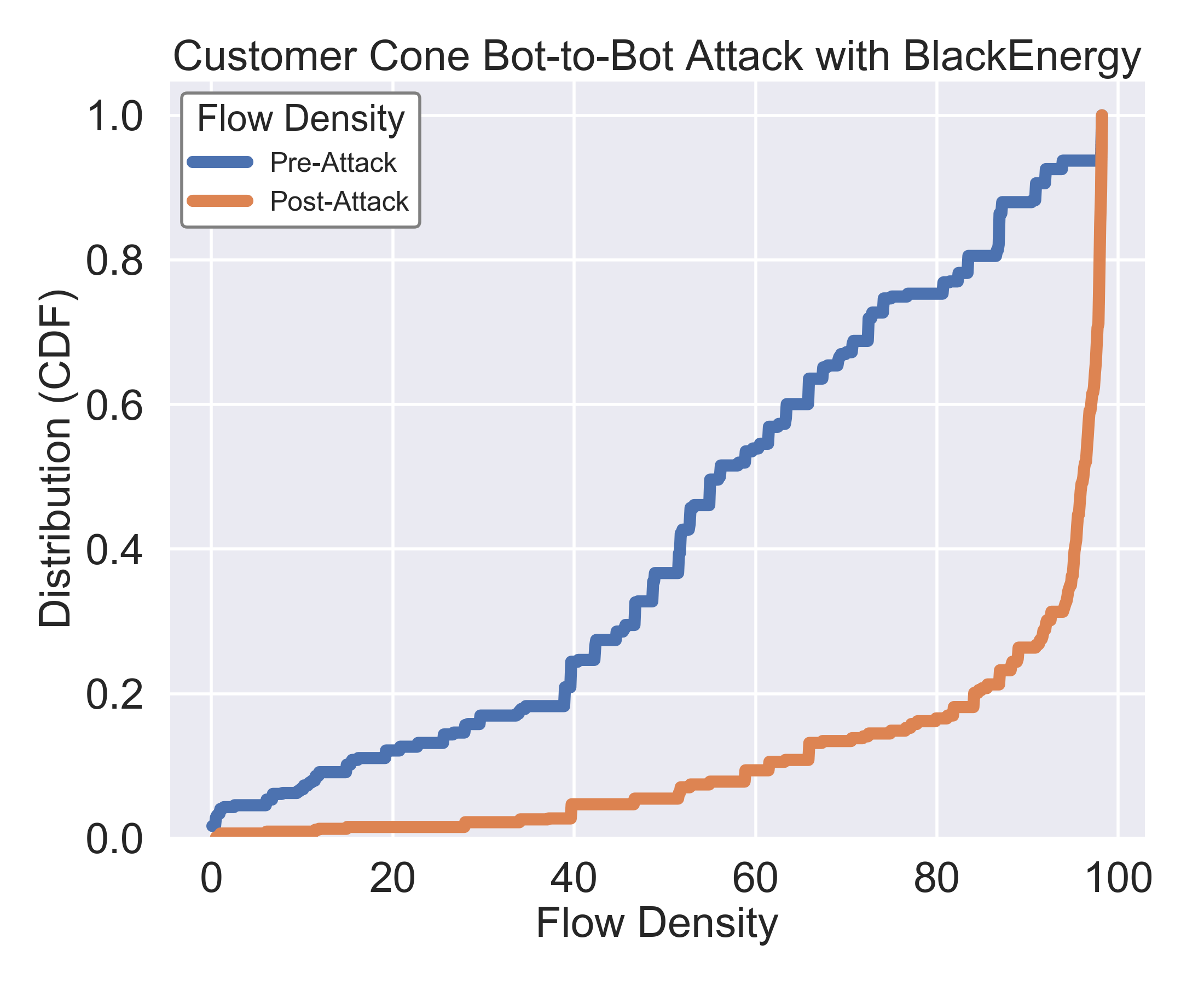}}
    \hspace{0.50cm}
    \subfloat[Bot-to-any (Crossfire) link selection for BlackEnergy]{\label{fig:blackenergy-cross-prepost}\includegraphics[width=0.26\textwidth]{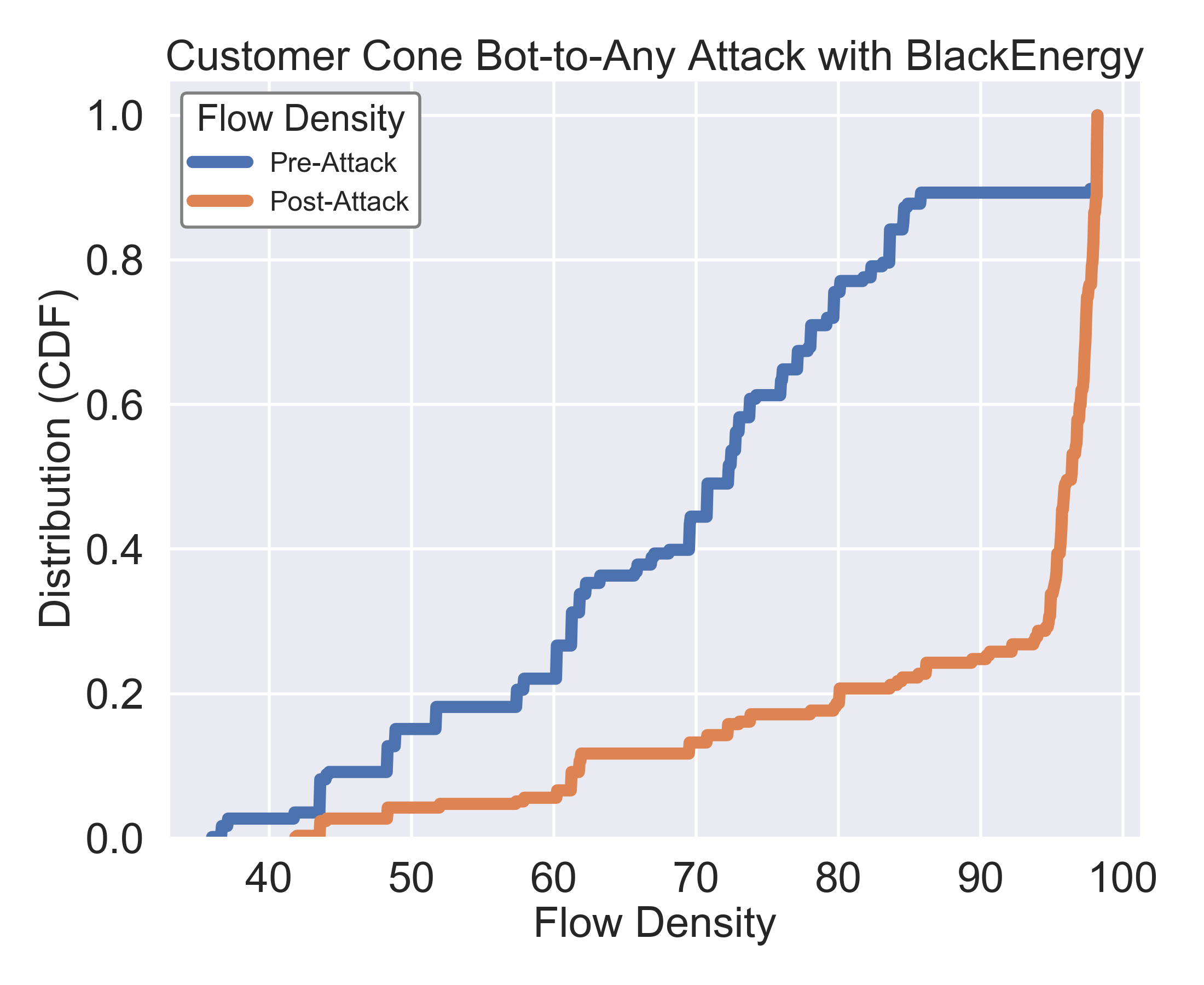}}
    \begin{center}
        \caption{Maestro Success in Pre- vs. Post-attack Flow Density for Conficker and BlackEnergy Botnet Models for all three link-selection strategies for highest decile}
        {\label{fig:other-botnets-pre-post}}
    \end{center}
    \vspace{-20pt}
\end{figure*}

\end{document}